\documentclass[onecolumn]{aastex62}
\submitjournal{The Astrophysical Journal -- Accepted Feb 14, 2019}
\shorttitle{}
\shortauthors{}

\begin{document}


\title{Mapping the Interstellar Reddening and Extinction towards Baade's Window Using Minimum Light Colors of ab-type RR Lyrae Stars. Revelations from the De-reddened Color-Magnitude Diagrams}


\correspondingauthor{Abhijit Saha}
\affil{National Optical Astronomy Observatory, 950 N. Cherry Ave, Tucson, AZ 85719}
\email{saha@noao.edu  \& asaha56@earthlink.net}

\author[0000-0002-0278-1600]{Abhijit Saha}
\affil{National Optical Astronomy Observatory, 950 N. Cherry Ave, Tucson, AZ 85719}

\author[0000-0003-4341-6172]{A. Katherina Vivas}
\affil{Cerro Tololo Inter-American Observatory, National Optical Astronomy Observatory, Casilla 603, La Serena, Chile}
%\email{kvivas@ctio.noao.edu}

\author{Edward W. Olszewski}
\affil{Steward Observatory, The University of Arizona, Tucson, AZ 85721}
%\email{eolszewski@as.arizona.edu}

\author{Verne Smith}
\affil{National Optical Astronomy Observatory, 950 N. Cherry Ave, Tucson, AZ 85719}
%\email{vsmith@email.noao.edu}

\author{Knut Olsen}
\affil{National Optical Astronomy Observatory, 950 N. Cherry Ave, Tucson, AZ 85719}
%\email{kolsen@noao.edu}

\author{Robert Blum}
\affil{National Optical Astronomy Observatory, 950 N. Cherry Ave, Tucson, AZ 85719}
\affil{Large Synoptic Survey Telescope, 950 N. Cherry Ave, Tucson, AZ 85719}
%\email{rblum@noao.edu}

\author{Francisco Valdes}
\affil{National Optical Astronomy Observatory, 950 N. Cherry Ave, Tucson, AZ 85719}
%\email{fvaldes@noao.edu}

\author{Jenna Claver}
\affil{National Optical Astronomy Observatory, 950 N. Cherry Ave, Tucson, AZ 85719}
%\email{jenna@noao.edu}

\author{Annalisa Calamida}
\affil{National Optical Astronomy Observatory, 950 N. Cherry Ave, Tucson, AZ 85719}
\affil{Space Telescope Science Institute, 3700 San Martin Drive, Baltimore, MD 21218}
%\email{calamida@stsci.edu}

\author[0000-0002-7123-8943]{Alistair R. Walker}
\affil{Cerro Tololo Inter-American Observatory, National Optical Astronomy Observatory, Casilla 603, La Serena, Chile}
%\email{awalker@ctio.noao.edu}

\author[0000-0001-6685-0479]{Thomas Matheson}
\affil{National Optical Astronomy Observatory, 950 N. Cherry Ave, Tucson, AZ 85719}
%\email{saha@noao.edu}

\author{Gautham Narayan}
\affil{National Optical Astronomy Observatory, 950 N. Cherry Ave, Tucson, AZ 85719}
\affil{Space Telescope Science Institute, 3700 San Martin Drive, Baltimore, MD 21218}
%\email{gnarayan@stsci.edu}

\author{Monika Soraisam}
\affil{National Optical Astronomy Observatory, 950 N. Cherry Ave, Tucson, AZ 85719}
%\email{soraisam@noao.edu}

\author{Katia Cunha}
\affil{Observatorio Nacional/MCTI, Rua Gen. Jose Cristino, 77 20921-400, Rio de Janeiro, Brazil}  
\affil{Steward Observatory, The University of Arizona, Tucson, AZ 85721}

\author{T. Axelrod}
\affil{Steward Observatory, The University of Arizona, Tucson, AZ 85721}
%\email{taxelrod@gmail.com}

\author{Joshua S. Bloom}
\affil{Department of Astronomy, University of California, Berkeley, CA 94720-3411}
%\email{joshbloom@berkeley.edu}

\author{S. Bradley Cenko}
\affil{NASA Goddard Space Flight Center, Mail Code 661, Greenbelt, MD 20771}
\affil{Joint Space-Science Institute, University of Maryland, College Park, MD 20742}
%\email{brad.cenko@nasa.gov}

\author{Brenda Frye}
\affil{Steward Observatory, The University of Arizona, Tucson, AZ 85721}
%\email{brendafrye@gmail.com}

\author{Mario Juric}
\affil{DIRAC Institute, Department of Astronomy, University of Washington, 3910 15th Avenue NE, Seattle, WA 98195}
%\email{mjuric@uw.edu}

\author{Catherine Kaleida}
\affil{Cerro Tololo Inter-American Observatory, National Optical Astronomy Observatory, Casilla 603, La Serena, Chile}
\affil{Space Telescope Science Institute, 3700 San Martin Drive, Baltimore, MD 21218}
%\email{ckaleida@stsci.edu}

\author{Andrea Kunder}
\affil{Cerro Tololo Inter-American Observatory, National Optical Astronomy Observatory, Casilla 603, La Serena, Chile}
\affil{St. Martin's University, 5000 Abbey Way SE, Lacey WA 98503}
%\email{akunder@stmartin.edu}

\author{Adam Miller}
\affil{Center for Interdisciplinary Exploration and Research in Astrophysics and Department of Physics and Astronomy, Northwestern University, 2145 Sheridan Road, Evanston, IL 60208}
%\email{amiller@northwestern.edu}

\author{David Nidever}
\affil{National Optical Astronomy Observatory, 950 N. Cherry Ave, Tucson, AZ 85719}
%\email{dnidever@email.noao.edu}

\author{Stephen Ridgway}
\affil{National Optical Astronomy Observatory, 950 N. Cherry Ave, Tucson, AZ 85719}
%\email{sridgway@noao.edu}

\begin{abstract}

We have obtained repeated images of 6 fields towards the Galactic
bulge in 5 passbands ($u,g,r,i,z$) with the DECam imager on
the Blanco 4m telescope at CTIO. From over 1.6 billion individual
photometric measurements in the field centered on Baade's window, we
have detected 4877 putative variable stars.  474 of these have been
confirmed as fundamental mode RR~Lyrae stars, whose colors at minimum
light yield line-of-sight reddening determinations as well as a
reddening law towards the Galactic Bulge which differs significantly
from the standard $R_{V} = 3.1$ formulation. Assuming that the stellar
mix is invariant over the 3 square-degree field, we are able to derive
a line-of-sight reddening map with sub-arcminute resolution, enabling
us to obtain de-reddened and extinction corrected color-magnitude
diagrams (CMD's) of this bulge field using up to 2.5 million
well-measured stars. The corrected CMD's show unprecedented detail and
expose sparsely populated sequences: e.g., delineation of the very wide
red giant branch, structure within the red giant clump, the full
extent of the horizontal branch, and a surprising bright feature which
is likely due to stars with ages younger than 1~Gyr. We use the
RR~Lyrae stars to trace the spatial structure of the ancient stars,
and find an exponential decline in density with Galactocentric
distance. We discuss ways in which our data products can be
used to explore the age and metallicity properties of the bulge, and how
our larger list of all variables is useful for learning to interpret future
LSST alerts.
~\\
\end{abstract}


\section{Introduction}
\label{sec:intro}

Observationally the Galactic bulge is a concentration of stars towards
the galactic center with chemistry, age distribution, and dynamics
that set it apart from the disk and halo. A comprehensive review with 
leads into the extensive literature is given by 
\citet{barbuy18}.  By combining what we know about our bulge with those 
in other galaxies we are led to understand that bulges come in two
forms, classical bulges and pseudo-bulges \citep{kormendy04}. 
Modern observations of the Milky Way bulge indicate that it has
a bar \citep{dwek95} with some characteristics
of a classical bulge and some of a pseudo-bulge. While the majority of
Bulge stars seem to be old, there is still debate about the
percentage of younger stars, a debate that can be informed by 
the inspection and analysis of color-magnitude diagrams from which 
a) the line-of-sight reddening and extinction are removed, and b) contamination by foreground stars 
is identified and eliminated on the basis of proper motions.  

Thus, in addition to the complications of performing accurate photometry 
in severely over-crowded fields, the construction of suitable color-magnitude diagrams involves 
removing reddening on the finest possible angular scales.  The color of the red clump (RC) stars 
just off the giant branch has been used as a standard color-marker (or standard crayon) in 
many studies, most notably by \citet{Nataf13, Nataf16} and references therein. They found that not only does the standard reddening law predict the line-of-sight reddening to the bulge incorrectly, but that the true reddening law in these directions varies on angular scales of a few degrees.

%However, the 
%color and brightness of the clump depend on the age, metallicity and possibly helium abundance of the parent stellar population, 
%rendering the process somewhat circular if the end goal is to construct a color-magnitude diagram to 
%investigate the very nature of the same population of stars.  

Removal of foreground stars using proper motions up to 19th mag over wide fields of 
view is possible with \emph{Gaia}, though we may have to wait for the mission to complete to do this 
comprehensively. It may well be that due to the high stellar density in these areas, \emph{Gaia}'s selection of stars 
in this part of the sky is incomplete. Over time, the VVV survey \citep{minniti10} and its followup provide both the time base and 
object completeness, which are likely required to complete the task. From the analysis of asymptotic giant branch and cool supergiant stars near the Galactic center, 
\citet{blum03} implied that about 25\% of the stars in the central few parsecs are younger than 5 Gyr. However this may not be representative of the bulge as a whole.
The \emph{Hubble Space Telescope} ($HST$) has already been used to carry this out 
for small fields of view in the bulge \citep[e.g.,][]{clark08, cala14}, with ensuing cleaned color-magnitude diagrams such as 
by \citet{brown09}, and  more recently by \citet{bernard18}. The latter work goes on to derive star formation histories in different bulge fields from their CMDs, and report 
that up to 20 or 25\% of the most metal rich stars are younger than 5~Gyr.  
The drawback is that rare(r) stars can only be
seen as populations in larger-area studies than possible with \emph{HST}, and reddening and extinction 
corrections used in these studies involve adopting the standard Galactic extinction law, 
which \citet{Nataf13, Nataf16} show to be invalid.

In this paper we explore an alternative route to deriving reddening and extinction 
following the precepts enunciated by \cite{sturch66} about the constancy and universality of the 
colors of fundamental mode RR~Lyrae stars while they are 
in the pulsation phase corresponding to near minimum light. 
The potential advantage of this approach is that since RR~Lyrae are also standard candles, they 
can be used to investigate not only the reddening, but also the ratio of total to selective extinction.
In our experiment, we have obtained and analyzed multi-band, multi-epoch wide field bulge images
to construct light curves of the RR~Lyraes, and employ them to examine the intervening dust reddening 
and extinction. The emphasis is on avoiding any prior assumptions about the bulge's stellar population make up.

RR~Lyrae stars are also probes of ancient stellar populations, and their distribution in the bulge traces 
that of the oldest stars. Recent searches for these stars in the near infrared 
through the very obscured inner regions of the bulge by the VVV survey \citep{minniti10} indicate 
that these stars do not follow the bar like structure, but have a smoother distribution \citep{minniti17}. This is contrary to an older 
result based on OGLE data \citep{Piet12}, who claim that the RR~Lyrae spatial distribution is elongated along 
the Galactic bar. It is quite possible that the accuracy in the adopted reddening and total to selective reddening laws 
impact such findings.

\begin{figure*}[htb!]
\plotone{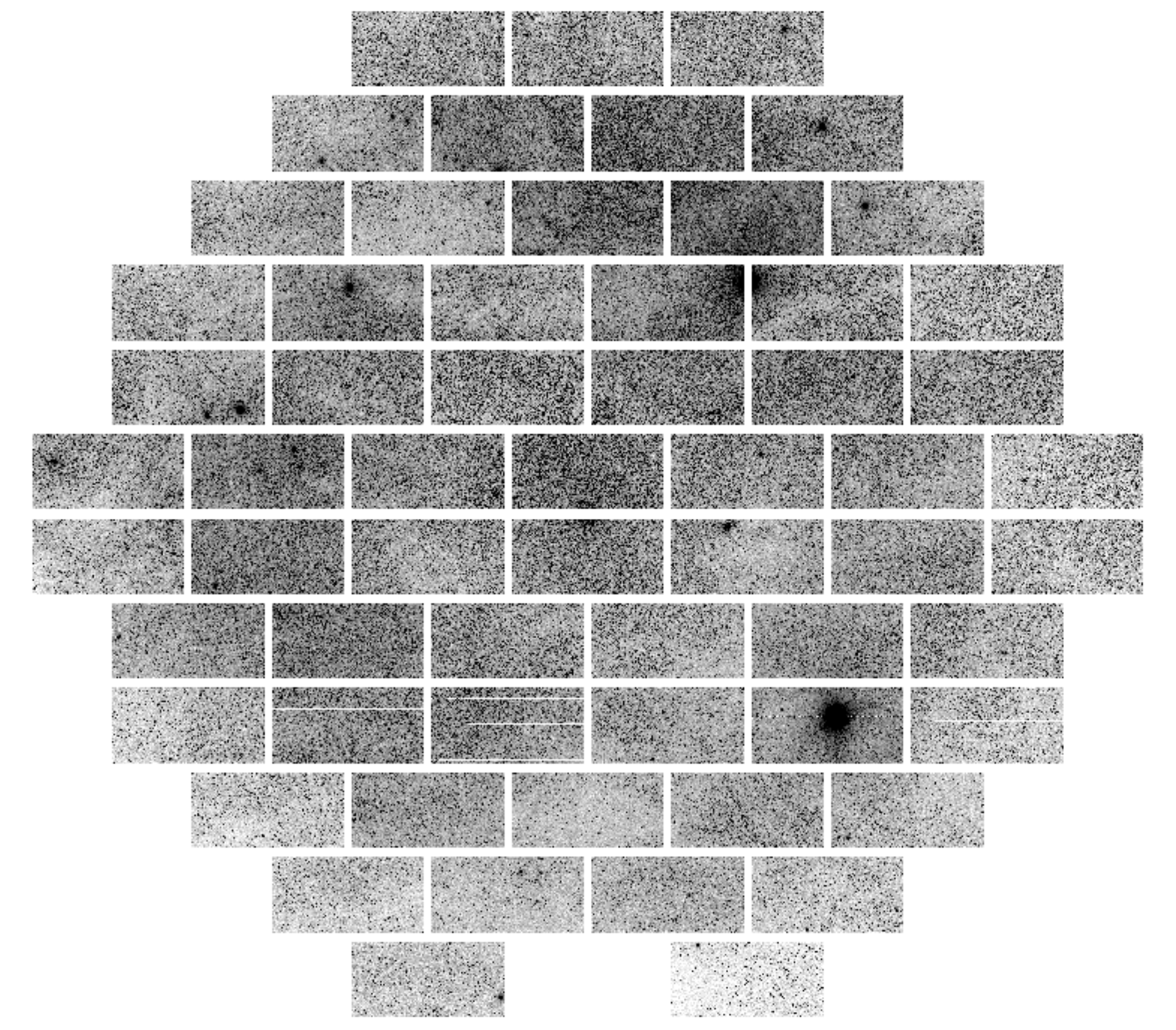}
\caption{Panoramic image of the B1 field showing the full DECam field (with radius $\sim 1^{\circ}$) from a 300s exposure in the $u$-band.  North is up, and East to the left.  The patchiness in extinction is apparent, with structure on scales of less than an arc-minute (the individual chips have dimensions of $ 9 \times 18 $ arc-minutes on the sky.}
\label{fig:panorama} 
\end{figure*}

We obtained images of 6 select fields towards the general direction of the
Galactic center with the DECam imager \citep{flaugher15}
over multiple epochs in 5 different passbands $u,g,r,i,z$. The chosen
fields are shown in Table~\ref{tab_fields}, and named B1 through B6.
B1 is centered on the well known ``Baade's Window,'' and gets close
to the direction of the Galactic center while remaining relatively
transparent.  The footprint of the DECam field is significantly larger
than the original area considered by Baade, and has patches of
reddening much higher than the value of $ E(B-V) \sim 0.7 $ often
ascribed to it. Figure~\ref{fig:panorama} shows an image of the field in 
the $u$ passband, which highlights the patchiness in extinction that must be 
dealt with.  B2 is an adjacent field midway between 2 fields found
by \citet{blanco92} and \citet{blancos97}, with lower and less uneven extinction
than B1, but slightly farther from the direction of the Galactic
center. There is a small intentional overlap between B1 and B2 for the
purpose of verifying photometric accuracy in our data. B5 is set $\sim
10^{\circ}$ south of the Galactic Center, and is intended as a probe of the region off
the Galactic plane, but within the bulge. B3 and B4 are fields at
similar Galactic latitude as B1, but $\sim 10^{\circ}$ and $\sim 5^{\circ}$ away in
longitude respectively in the direction of the near side of the bar,
while B6 is $10^{\circ}$ away on the far side of the bar. These field
choices sample the run of stellar populations along and across the
Galactic disk. The exact placement of the fields was made to have
minimal extinction compared to their surroundings using the dust maps
by \citet{sfd98}\footnote{http://irsa.ipac.caltech.edu/applications/DUST/}. 
This paper deals only with field B1, but also details the
analysis methodology that will be used for the remaining fields.

\begin{deluxetable}{ccccc} 
\tabletypesize{\scriptsize}
\tablewidth{0pt}
\tablecolumns{5}
\tablecaption{Field Positions \label{tab_fields} }
\tablehead{\colhead{Field Name} &
  \colhead{RA (J2000)} &
  \colhead{DEC (J2000)} &
  \colhead{Galactic long.} &
  \colhead{Galactic lat.} \\ 
  \colhead{} &
  \colhead{( h : m : s )} &
  \colhead{( $^{\circ}$ : \arcmin : \arcsec ) } &
  \colhead{(degrees)} &
  \colhead{(degrees)}  \\
}
\startdata 
 B1   & 18:03:34.0  &  $-$30:02:02  &   1.02 &  $-$3.92  \\
 B2   & 18:09:24.4  &  $-$31:26:06  &   0.40 &  $-$5.70  \\
 B3   & 18:26:41.9  &  $-$22:39:21  &  10.00 &  $-$5.00  \\
 B4   & 18:14:23.3  &  $-$27:56:49  &   4.00 &  $-$5.00  \\
 B5   & 18:26:41.8  &  $-$33:45:24  &   0.00 & $-$10.00 \\
 B6   & 17:48:10.8  &  $-$37:08:15  & 353.25 &   $-$4.70  \\
\enddata
\end{deluxetable}

The organization of this paper is as
follows. \S~\ref{sec:observations} describes the
observations. \S~\ref{sec:photometry} describes the processing of the
data, including photometry and calibration onto an absolute flux based
magnitude scale for the native DECam
passbands. \S~\ref{sec:variability} deals with the detection of
variable stars, followup analysis including period determination to
identify the RR~Lyrae stars, template light curve fitting,
measurement of minimum light brightness in each passband, and
determination of completeness. \S~\ref{sec:minlightreddening}
describes the derivation of reddening to the individual fundamental
mode RR~Lyrae stars, and utilization of the differential reddening and
extinction of the ensemble of these stars to independently derive the
total to selective absorption ratios for the line-of-sight encompassed
by the field B1.  We present a comparison with the standard extinction
law. In \S~\ref{sec:CMDs} the observed colors and magnitudes of all
stars in the field are used in conjunction with the RR~Lyrae reddening
values to correct the observed color-magnitude diagrams (CMDs) for
extinction, and the prominent features in the corrected CMD are
discussed.  We show a reddening map in \S~\ref{sec:redmap} with angular bins of $0.5 \times 0.5$ arc-minutes. The implications of our analysis of reddening for tracing the geometry of the Galactic bulge are presented in \S~\ref{sec:geom}. In \S~\ref{sec:discussion} we summarize our findings, and suggest how the data-set presented in this paper may be profitably used in future analysis and investigations.

An ancillary benefit of the data set is that we have light curve data in many LSST-like passbands for a 
cornucopia of variable stars (and possibly transients) that begins to inform us about how to interpret the variability 
alert stream from LSST when it begins operation.

\section{Observations}
\label{sec:observations}

The journal of observations is given in Table~\ref{tab_journal}. There
were three dark runs in 2013, in May, June and August, each of 3 to 4
nights. All 6 fields were visited 3 to 4 times on each of these nights
(weather permitting), and exposed in all 5 bands $u.g,r,i,z$
successively. Consecutive visits to the same field on any night were spaced about 2
hours apart. The exposure times (typically 300s in $u$, 100s in
$g,r,i,z$) are long enough to detect F stars to $r
\sim 24$ mag in dark skies and arcsec or better seeing in \emph{uncrowded} 
fields in the absence of reddening/extinction. For these
fields, and particularly for the B1 field, the crowding is extreme and
reddening is significant, so the actual detection limit is
substantially brighter. In the best (and deepest) images, saturation sets in at $r
\sim 15$ mag. Some of the images were taken in poorer seeing (up to
1.5 arcsec), and also on occasion through light clouds. Some
additional epochs in the $i$, and $z$ bands were obtained in June
2013, during bright time. In addition, we obatained a set of much shorter exposures in March of 2015. These provide some measure of
longer term sampling of the brighter stars relative to the 2013 data,
and also provide data for stars of interest that may have been
inadvertently saturated in the longer exposures of 2013. 

\begin{deluxetable}{cccc} 
\tabletypesize{\scriptsize}
\tablewidth{0pt}
\tablecolumns{4}
\tablecaption{Journal of Observations for field B1 in all bands \label{tab_journal} }
\tablehead{\colhead{HJD$-$2400000.0} &
  \colhead{Passband} &
  \colhead{DECam Exposure No.} &
  \colhead{Exposure time} \\
  \colhead{} &
  \colhead{} &
  \colhead{} &
  \colhead{(seconds)} \\
}
\startdata
56423.659625  &  $r$  &     205418  &   100.  \\
56423.661128  &  $i$  &     205419  &   100.  \\
56423.662623  &  $z$  &     205420  &   100.  \\
56423.664117  &  $g$  &     205421  &   100.  \\
56423.665618  &  $u$  &     205422  &   300.  \\
56423.808243  &  $r$ &     205489  &   100.  \\
56423.809737  &  $i$  &     205490  &   100.  \\
56423.811255  &  $z$  &     205491  &   100.  \\
56423.812743  &  $g$  &     205492  &   100.  \\
56423.814222  &  $u$  &     205493  &   300.  \\
\enddata
\tablecomments{Table~\ref{tab_journal} is published in its entirety in the electronic edition of the journal. A portion is shown here for guidance regarding its form and content.}
\end{deluxetable}
The images were processed through the NOAO DECam pipeline \citep{Valdes14}, for bias and flat-field correction, bad pixel masking and WCS (world coordinate system) fitting.  Reduced images are available publicly through the NOAO Science Archive\footnote{ https://archive.noao.edu}.  We intend to make catalog data available through the Community
NOAO Data Lab\footnote{https://datalab.noao.edu}. \\

\section{Photometry, Cross-matching, and Calibration}
\label{sec:photometry}

We measured photometry in these very crowded fields using a variant of the DoPHOT program \citep{schechter93} maintained by one of us (Saha).  The procedures and considerations for optimizing the DoPHOT parameters and evaluating aperture corrections done using a bespoke procedure written in IDL are fully described in \S~3.2
of \citet{saha10} and need not be repeated here. The only differences, mentioned also in \citet{vivas17}, are that unlike as in \citet{saha10}, where the 8 individual chips were combined into a single image by applyng a gnomonic projection, 
in the present case the individual chips were processed independently, and the output photometry lists from the individual chips are concatenated 
into one single file for the whole image. The requirement for this is that the photometry across all chips (for a given image) be on the  same footing. The justification for this premise has been previously given in \S~2 of \citet{vivas17}.

We thus created independent photometry lists for each image. In
addition to the aperture corrected instrumental magnitudes and
associated error estimates, each object carries its {\it RA} and {\it
  DEC} positions, as well as the chip on which it was detected, and
the pixel coordinates within that chip for each image/epoch. Each
object on each image also carries the object type code assigned by
DoPHOT.  These codes distinguish well fitted bona-fide single stars
(type 1), multiple star blends (type 3), other extended objects (type
2), cosmic rays (type 8), image pathologies (types 4, 5, 8 and 9), and
objects too faint to disambiguate between stars and extended objects
or blends (type 7) ), and the fitted background or ``sky.''  In
addition the following attributes were also evaluated and recorded
(for each object on each image):
\begin{enumerate}

\item 
Whether the object lies within 50 pixels of the chip's edge.

\item Whether there were two or more cosmic-ray (or other pathologies)
  detected within a radial distance of 1 full width at half-maximum
  (FWHM) of the stellar point-spread-function (PSF) as measured along
  the major-axis of the PSF.

\item
Whether there were any bona-fide objects detected within a 1 FWHM
radius footprint around the object as described above, and if so, the
cumulative flux from those objects expressed in magnitudes relative to
the flux of the object in question. We denote this by $m_{neighbors}$.

\item
High and low  percentile values for the distribution of fitted sky values of all objects for the entire image were also evaluated and recorded for each image.

\item
For all stellar objects on a given image, the total reported error for individual objects was fitted as a function of reported magnitude.  The fitted value $err_{exp}$ at a given measured brightness is a good expectation of what the measurement error should be for an object of that measured brightness. Reported errors much higher than the expected value for that brightness are suspect. The value $err_{lim} = 2 \times err_{exp} + 0.05 $ for each object for each epoch was evaluated and recorded as an attribute.

\end{enumerate}

The photometry list from the best deep image (best seeing in photometric conditions) in each passband was assigned
as the deep template object list in that band, and a similarly
suitable short exposure image in each band was assigned as the shallow
template object list. For each band, the deep and shallow template
lists were merged by matching to a coordinate tolerance of 0\farcs3 (eliminating all multiple matches within this matching
tolerance), The instrumental magnitude difference of the matched
objects was used to adjust the instrumental mag system of the shallow
template to that of the deep one.  This process allows the objects
saturated in the deep template to be represented in the eventual
object list in each passband, and at this point there is a ``grand'
template list in each of the 5 passbands that spans the full dynamic
range of magnitudes spanned by both the deepest as well as shallow
exposures,  with instrumental mags on the system of the deep template.
Finally, the $r$ band ``grand'' template was adopted as the {\emph
  {master}}-template, containing the master-list of all objects.  In the subsequent processing,  the numerical ID's of objects on this master-list serve as the final object ID's for all objects in this field.  Any objects that are not on this list (for whatever reason) are not considered further.  

A particular detail for preparing the template lists before matching and combining into the ``grand'' templates is worth mentioning.
Since the aperture corrections to go from fitted PSF mags to instrumental mags for each image were calculated independently for each chip, the zero-point in any chip can scatter about the mean for that image by a few hundredths of a mag (or in pathological cases by worse amounts).  To mitigate this problem for the template images (to which all photometry calibration is eventually referred), they were compared against other images of similar depth (deep to deep and shallow to shallow) obtained in photometric conditions.
Let $m_{0j}^{k}$ be the measured aperture corrected magnitude of star $j$ on image $0$ (template) and on chip $k$.   Let the same star as measured on image $i$ and also on chip $k$ be designated by $m_{ij}^{k}$, where image $i$ was also in photometric conditions. If we selected only those $j$ for which the reported measurement errors are small, and for which DoPHOT has reported that the object has an unambiguously stellar PSF,  we can construct:

\begin{equation}
\delta_{i}^{k}  =   \sum_{j} ( m_{ij}^{k} - m_{0j}^{k} )  / N
\end{equation}

and 

\begin{equation}
 \Delta_{i} = \sum_{k} \delta_{i}^{k} / M
\end{equation}

where $N$ is the total number of selected stars (over index $j$) in chip $k$ being compared, and $M$ is the number of chips (over index $k$).
$ \Delta_{i} $ is the overall offset between the instrumental mags of image $i$ relative to the template image: and since it is an average over $\sim 60$ chips, is essentially unaffected by small random errors in evaluating the zero-points on individual chips. The offset $\Delta_{i}$ can be caused by differences in atmospheric extinction (different airmass), and over longer durations by differences in system response and transmission.  Consider the chip-to-chip fluctuation about this mean difference:

\begin{equation}
\epsilon_{i}^{k} = \delta_{i}^{k} - \Delta_{i}
\end{equation}
which shows the aggregate result of individual chip-to-chip aperture correction errors for both images $i$ and $0$.

If we have $n$ images against which such a comparison can be made for the template, we can calculate the 
ensemble average like quantity $C^{k}$ from the $\epsilon_{i}^{k}$'s:

\begin{equation}
C^{k} = \sum_{i=1}^{n} \epsilon_{i}^{k} / (n+1)
\end{equation}

which is a robust estimate of the correction to be added to the instrumental  magnitudes for the template frame for each chip $k$.

All of the photometry lists in each passband were then matched one by one to the ``grand'' template for that passband, from which the offsets in the instrumental magnitudes relative to the ``grand template'' magnitudes were calculated on a chip-by-chip basis. All instrumental mags for the individual epochs were adjusted (single additive magnitude offset per chip) to put all the instrumental mags for all epochs on the scale of the ``grand'' template. The lists with the instrumental mags thus normalized were then matched individually to the master-template (same as the 
$r$ band ``grand'' template), and the object ID's from the master list were then attached to the matched objects in each object-list for each epoch and for each passband.  With this labeling, the measurements for any object can be extracted for any epoch and passband, along with all of the associated information discussed above. The instrumental photometry for every epoch is normalized to that in the respective ``grand'' template for the relevant passband. Henceforth, all variability analyses can be carried out using either these normalized instrumental mags or using the calibrated AB-magnitudes described below.  Calibration to any system of magnitudes requires only the determination of zero-points for the ``grand'' template of the respective passbands, details of which are provided in the following paragraph. The normalized instrumental and object-labeled photometry for each epoch of each passband were then stored in a MySQL database, providing convenient access for subsequent variability analyses. There are 9,623,873 distinct objects in the database, each with multiple measurements at different epochs and different passbands (not all objects have all epochs in all passbands). They are labelled by an object ID corresponding to the running ID of the object on the master-template. In all, the database  for B1 measurements contains over $1.6 \times 10^{9}$ individual photometric measurements.  The above procedure ensures that all photometric measurements are placed on the same uniform instrumental system, independently in each of the passbands.
Comparison of these instrumental magnitudes for high signal-to-noise
ratio (S/N) objects across different exposures show that the self-consistency in the instrumental magnitudes is better than $0.02$~mag rms.

On photometric nights, two of the newly calibrated DA white dwarf
standards from \citet{nar16} were also observed through a range of
air-masses. These stars have calibrated spectral energy distributions,
from which their true AB-magnitudes were calculated according to the
prescription of \citet{fuk96} for each of the 5 passbands.  We then
derived photometric solutions relating instrumental mags to AB-mags. In the $u,g,r,i$ bands, photometric solutions have residuals with $\sim 0.01$ mag rms scatter. In the 
$z$ band, which encompasses telluric water bands that can vary on time
scales of minutes, as well as with position, the scatter is $ \sim
0.02$ to $0.03$ mag.

When these solutions are applied to the photometry of the ``grand'' templates (one for each passband), we obtain calibrated AB-mags for the native DECam passbands.  This is the same system used in 
\citet{vivas17} where the luminosity and color relations for RR~Lyraes are derived for precisely this system of magnitudes, making their results directly applicable to the data for the B1 field.  Combining the scatter in the self-consistency in instrumental magnitudes discussed above, and the total calibration accuracy, we estimate that the 
systematic uncertainty in the calibration of any exposure is thus $\approx 0.02$~mag rms in $u,g,r,i$ and $\approx 0.035$~mag in $z$.  Measurement errors for any object on any exposure are additional, and are estimated by the measurement procedures, including by the DoPHOT program.

It should be pointed out that the analysis presented in this paper does not depend on what system of magnitudes we adopt, as long as the same system is used for all targets, including the globular cluster Messier~5 (NGC~5904), hereafter M5, where the color properties of the RR~Lyrae stars are derived.

\section{Variability Analysis} 
\label{sec:variability}

\subsection{An Independent Identification of Variable Sources}
\label{sec:allsources}

Each object was tested for variability independently in each passband.  However, a variability test is only meaningful if there are sufficient number of measurements of adequate quality. It is important to remove reported observations that have a high likelihood of being pathological. 
For each object and for a given passband, each measurement (by epoch) was subject to the following ``interrogation:''

\begin{enumerate}

\item Is the object's centroid located within 50 pixels of the edge of the respective CCD?

\item Are there any detected sources (including cosmic ray hits) within 1 FWHM (of the PSF) distance from the object's centroid?

\item Does the value of the fitted background, $s$, fall outside the range $s2 \leq s \leq 2\, s90 - s2$, where $s2$ and $s90$ are the 2nd and 90th percentile values respectively for the fitted value of the background for {\it all} stars on that image?

\item Does the reported measurement error exceed $err_{lim}$, as defined in \S~\ref{sec:photometry}?

\item Did DoPHOT assign the object a type other than 1 or 3 (which are
  for objects with an unambiguously stellar PSF, see
  \S~\ref{sec:photometry} for DoPHOT types)?

\end{enumerate}

If the answer to any of the above questions is positive, the measurement was excised from further consideration.
At least 15 measurements for a given object in a given passband must survive in order to proceed with variability assessment. Criterion 5 above is particularly severe in eliminating faint measurements. For our primary purpose 
of detecting and measuring the RR~Lyrae stars, this is not an obstacle (as will be demonstrated below), but it may well inhibit the detection of variables to the faint limits that the photometry would otherwise allow. However, it is clear that the ``purity'' of the variable candidate list declines rapidly if DoPHOT type 7 measurements (those for which the S/N is too low to unambiguously ascertain if they have stellar PSFs) are allowed. 

In the final analysis, only 450,344 objects in $u$, 1,082,121 in $g$, 1,950,425 in $r$, 2,509,906 in $i$ and 2,347,075 in $z$ passed the above ``interrogation'' and were examined for variability.  These numbers correspond to about 20\% of all objects detected on the best and deepest available images in the respective passbands.  The process can be easily re-run with changes in any and all of the parameters mentioned above.

\begin{figure}[htb!]
\centering
\epsscale{0.85}
\plotone{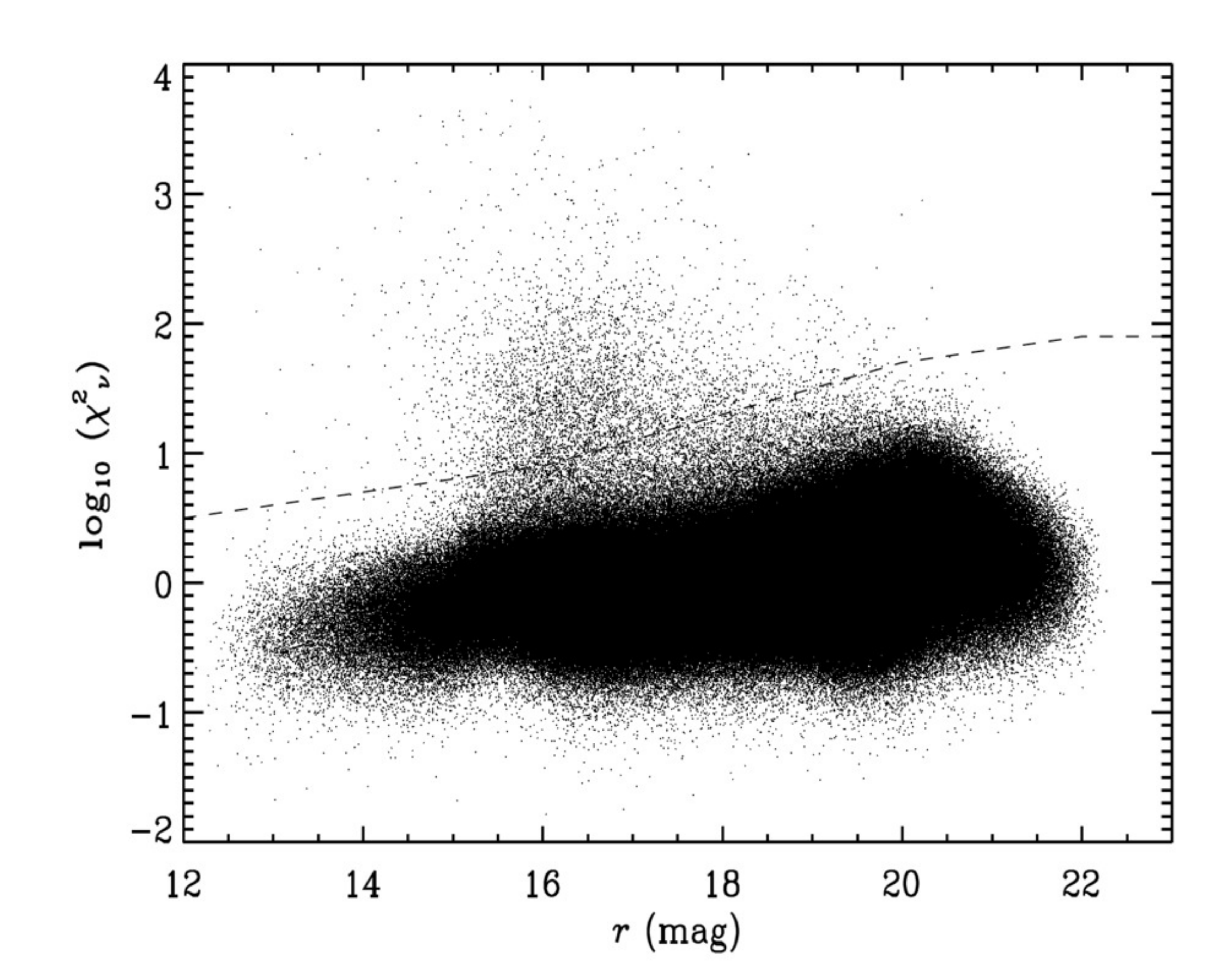}
\caption{Individual values on a log scale of the reduced chi-square
  $\chi^{2}_{\nu}$ (as described in \S~\ref{sec:variability}) as a
  function of the average brightness ($r$-band magnitude) for
  approximately 2 million stars in the $r$ band.  For accurate
  reported errors, and for a Gaussian error distribution for an
  unvarying object, the most probable value of $\chi^{2}_{\nu}$ is
  unity.  Accordingly, the figure suggests that for the brighter stars, the reported errors are overestimates, while for $r > 17$ the distribution peaks close to $\log \chi^{2}_{\nu} = 0$, indicating that error estimates reported by DoPHOT are reasonably correct.  Objects with significantly higher values of $\chi^{2}_{\nu}$ are candidate variables. The program allows the user to interactively set the threshold for flagging something as a putative variable, such as by the dashed line shown in the figure.  As described in the text, the $\chi^{2}_{\nu}$  values are computed for each object using a bootstrap procedure with 100 draws, which provides a more robust estimate by suppressing the contribution from one or two pathological extreme measurements of brightness.}
\label{fig:chisq}
\end{figure} 

The variability search was carried out using the method laid out in \citet{saha90}.  A reduced chi-square, $\chi_{\nu}^{2}$,  with respect to the mean value of available magnitude measurements is computed using the available magnitude measurements and associated reported errors.
We used a boot-strap sampling of the magnitude and error measurements for each of the available epochs to generate a robust estimate for the $\chi_{\nu}^{2}$, using 100 resamples.  Ideally the $\chi_{\nu}^{2}$ for a non-variable (given the reported noise) should hover around unity.  However, since in reality the distribution of noise 
is not fully expressed by a single Gaussian, and because reported errors are themselves subject to bias, we see that the mean of $\chi_{\nu}^{2}$ can change weakly with the brightness (see Figure~\ref{fig:chisq}).  Accordingly, the mode value of $\log \chi_{\nu}^{2}$ was calculated for 0.5 mag wide bins of mean magnitude, and an object was flagged as a variable if its $\log \chi_{\nu}^{2}$ is higher than the relevant mode value by 1.3 (i.e. 20 times or more higher than the mode).  The program also allows the user to interactively set the detection threshold with varying brightness. The variable lists from the above analysis in each passband were merged. A total of 
4877 putative variables were flagged with this procedure, where each candidate is flagged in at least one of the 5 passbands.

\subsection{Identification of ab-type RR Lyraes and derivation of colors at minimum light}
\label{sec:classification}

All of the 4877 candidate variables flagged above were run through the
period finding procedure described in \citet{saha17} (hereafter
{\small PSEARCH}) in ``batch'' mode using $\Delta\phi_{max} = 0.05$ (as defined
in their Equation 9). We identified the highest resultant peak of the $\Psi$ periodogram of {\small PSEARCH} for each object, and the resulting folded light curves in all available bands were plotted.  A visual examination of the plots very quickly reveal the objects that are possibly RRab (radially pulsating in the fundamental mode) variables.  The {\small  PSEARCH} code was rerun interactively with $\Delta\phi_{max} = 0.02$ \citep[for details see][]{saha17} for the objects that were flagged as possibly RRab's to confirm the classification and to select the most likely period from among any aliases. 

The preliminary light curves mentioned above for all the of the 4877
candidate objects reveal a wealth of different variables.  Binaries with short periods that are relatively well sampled with the observed cadence show very convincing light curves, as do short period pulsators like $\delta$-Scuti and/or SX~Phe stars.  Many RRc's can be discerned by their slightly skewed near-sinusoidal light curves, while those without perceptible skewness are likely hiding as indistinguishable from amongst contact binaries with sinusoidal light curve shapes.  There are also variables for which no believable folded light curves could be obtained. Their periodograms show peaks at much longer periods for which the data at hand cannot be used to derive believable periods. These are likely to be long period or semi-regular or irregular variables.  Since the OGLE project surveyed the same region of the sky, and with much more extensive timing coverage than the present one (which was optimized to get light curves of the RR Lyraes), many of our identified variables can be matched to OGLE identified variables, for which their variable classification is available.  While their data are primarily in one band, the panchromatic information from the 
data set of our study here can be used in novel ways to develop new classification methods, and is the subject of an ongoing study. 

\begin{figure}[htb!]
\epsscale{0.9}
%\plotone{t1448112.pdf}
\plotone{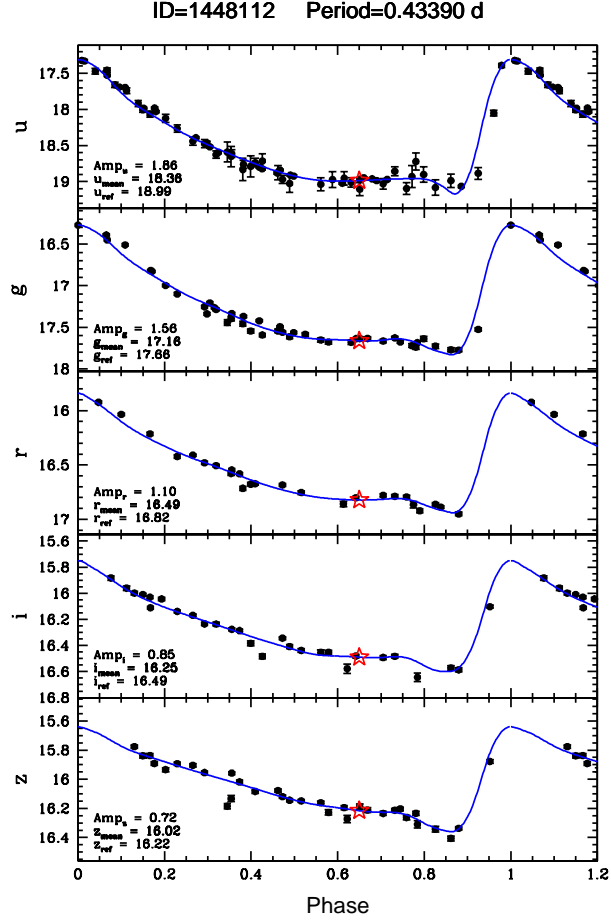}
\caption{Example of lightcurve and best fitted templates in $ugriz$ for one of the RR Lyrae stars in field B1. The red star indicates the magnitude at  minimum light ($m_{\rm ref}$), which was set to be measured at $\phi=0.65$. The complete figure set (474 images) is available in the online journal.}
\label{fig:lc}
\end{figure}

We then ran the list of 491 possible RRab stars through a template fitting program from which the properties of the light curves (mean magnitude, amplitude, initial phase) were derived. This process is particularly important to define the initial phase (phase at maximum light) of the light-curves. The use of templates is helpful when the observations do not sample well that part of the pulsation period. We used the library of lightcurve templates set up by \citet{sesar10} from RR Lyrae stars in SDSS Stripe~82. The library contains between 10 and 20 templates in each filter for type ab RR Lyrae stars, and only 1 or 2 for the types c. During the fitting process we allowed for variations around the period found by {\small PSEARCH} ($\pm 0.001$ d, in steps of $1 \times 10^{-6}$ d), the observed amplitude ($\pm 0.2$ mag in steps of 0.01 mag), the magnitude at maximum light ($\pm 0.2$ mag in steps of 0.01 mag), and the initial phase ($\pm 0.2$ in steps of 0.01) which was initially set by the time of the observation with the brightest magnitude in the lightcurve.  The best template is found from $\chi^2$ minimization. Initially the fit is done only in the filter with the largest number of epochs available, which sets up the period and initial phase for that star. Then, the template fitting procedure is repeated for the other 4 filters but allowing variations only in the amplitude and maximum magnitude. Light-curves and the fitted templates for all stars are available as Figure~\ref{fig:lc}. They are also available via a Github repository\footnote{\it https://github.com/akvivas/Baade-s-Window}.  The epoch-by-epoch photometry in all bands for the bona-fide RRab stars is presented in Table~\ref{tab:timeseries}.

During this process, we found 16 objects in the list to be poor fits
to RRab templates.  These objects were classified as other kinds of variables in the OGLE catalogs. In the final analysis, we have 474 surviving ab-type RR Lyraes, for which the periods, mean magnitudes,  magnitudes at minimum light, and amplitudes in the 5 passbands are listed in Table~\ref{tab:fitparams}.  The mean magnitudes were calculated by integrating the template light curve in each band after it had been converted to intensity units. The magnitudes at minimum light correspond to the magnitude of the fitted template at phase $\phi=0.65$. The same procedure was applied to the RR Lyrae stars in the globular cluster M5 presented by \citet{vivas17}, whose calibration will be used here to estimate the reddening. 

\begin{deluxetable}{ccccc}[htb!]
\tablecolumns{5}
\tablewidth{0pc}
\tablecaption{Photometry of ab-type RR Lyrae Stars in Field B1 \label{tab:timeseries} }
\tablehead{
\colhead{ID}  & \colhead{HJD-2400000.0} & \colhead{Filter} & \colhead{Mag} & \colhead{Error} \\
}
\startdata
 1448112 & 56423.665618 & $u$ & 17.528 & 0.012  \\
 1448112 & 56423.814222 & $u$ & 18.799 & 0.015   \\
 1448112 & 56424.718744 & $u$ & 18.923 & 0.014   \\
 1448112 & 56424.810501 & $u$ & 18.971 & 0.015   \\
 1448112 & 56424.888096 & $u$ & 19.067 & 0.016   \\
\enddata
\tablecomments{Table~\ref{tab:timeseries} is published in its entirety in the electronic edition of the journal. A portion is shown here for guidance regarding its form and content.}
\end{deluxetable}

\begin{deluxetable}{cc}
\tablecolumns{2}
\tablecaption{Description of Table of RRab stars and their Fitted Parameters \label{tab:fitparams} }
\tablehead{
\colhead{Column} & 
  \colhead{Description}  \\
}
\startdata
1 & Object Identifier \\
2 & RA (J2000) \\
3 & DEC (J2000) \\
%4 & Variable Type \\
4 & Fitted Period (days) \\
5 & HJD - 2400000. Reference Epoch of zero phase ($\phi = 0$) \\
6 & No. of available $u$ measurements \\
7 & $u$ band Amplitude \\
8 & $u_{mean}$  (mags) \\
9 & $u_{ref}$ (mags) [Fitted value at $\phi=0.65$]  \\
10 & No. of available $g$ measurements \\
11 & $g$ band Amplitude \\
12 & $g_{mean}$  (mags) \\
13 & $g_{ref}$ (mags) [Fitted value at $\phi=0.65$] \\
14 & No. of available $r$ measurements \\
15 & $r$ band Amplitude \\
16 & $r_{mean}$  (mags) \\
17 & $r_{ref}$ (mags) [Fitted value at $\phi=0.65$] \\
18 & No. of available $i$ measurements \\
19 & $i$ band Amplitude \\
20 & $i_{mean}$  (mags) \\
21 & $i_{ref}$ (mags) [Fitted value at $\phi=0.65$] \\
22 & No. of available $z$ measurements \\
23 & $z$ band Amplitude \\
24 & $z_{mean}$  (mags) \\
25 & $z_{ref}$ (mags) [Fitted value at $\phi=0.65$] \\
26 & Cross-matched OGLE Identifier \\
\enddata
\tablecomments{Table~\ref{tab:fitparams} is published in its entirety in the electronic 
edition of the {\it Astrophysical Journal}.  The description is shown here 
for guidance regarding its form and content.}
\end{deluxetable}

\subsection{Completeness Estimates}
\label{sec:completeness}

The correlation of our final list of RRab stars with the RRab's from the RR~Lyrae stars compilation of \citet{sosz14} from the OGLE survey (hereafter OGLE RRab's) allows us to make a quantitative estimate of the discovery completeness from {\emph {both}} surveys. 

The DECam pointings at the various epochs were intended to repeat exactly. Hence the excluded regions, such as from gaps between chips were also  repeated, and any variables that occupy that excluded region would not be detected. Recall also that a measurement of any object that at a given epoch fell within 50 pixels ($\sim $13\arcsec) of any chip's edge was discarded.  With such caveats in mind, we denote the total number of RRab stars 
present on the complex operational portion of the DECam footprint by $N$. Let $n_{O}$, be the number of these actually present in the \citet{sosz14} compilation and $n_{U}$ the number found by us in this work.  Let $p_{O}$ and $p_{U}$ denote the discovery completeness of OGLE RRab's and the present work respectively.   We estimate from counting the OGLE RRab's within the discoverable area of our B1 field that:
\begin{equation}
 n_{O}  = N\, p_{O} \approx 560 ~(\pm 15)
\end{equation}
where the uncertainty arises from the difficulty of counting stars in the avoidance zones within our footprint.

From our own data and procedures described above, we have independently identified 474 ab-type RR~Lyraes, 
so that:
\begin{equation}
n_{U} = N\, p_{U} = 474
\end{equation}

Matching the list of RRab's from \citet{sosz14} with ours using a 2 arcsec matching tolerance, we find that there are 472 RRab's in common, so that:

\begin{equation}
N\, p_{O}\, p_{U} = 472
\end{equation}

It follows from the above that: 
\begin{eqnarray}
p_{O} \approx 0.996  \\
p_{U} \approx 0.843
\end{eqnarray}

It should be emphasized that these numbers are valid for RRab stars only. Other types of variables suffer different selection effects. Specifically our images typically have better seeing, and greater depth than OGLE, but we have much fewer epochs and a shorter total time baseline. Consequently we optimized our observing cadence for detecting RR Lyrae stars (our primary goal) at the expense of other kinds of variables with different temporal characteristics.  OGLE has many more epochs and better coverage of the window function compared to the present study, albeit only in the $I$ band.

\section{Reddening from the RR~Lyrae colors at Minimum Light}
\label{sec:minlightreddening}

\citet{sturch66} showed that the $B-V$ colors of fundamental mode
RR~lyrae stars (i.e., the Bailey ab-type) are invariant in the phase
range $0.5 < \phi < 0.8$ (where the phase at maximum light is defined
as 0.0)  and  that, aside from small metallicity and period dependent de-trending, the instrinsic $(B-V)$ colors are the same from star to star to within a few percent. Thus these serve as standard color sources, and have been used to determine interstellar reddening, including calibrating reddening from HI maps and galaxy counts by \citet{burstein78}.  This paradigm has recently been re-examined using DECam filter passbands by \citet{vivas17}, who have presented expected colors for several passband combinations (their Equation 1, Table~6, and Figure~3) from a study of the RR~Lyraes in the globular cluster M5.   Their minimum light colors are derived from fitted light curve magnitudes at $\phi = 0.65$.   Specifically, for zero reddening, we have from their paper that:

\begin{equation}
\label{eqn:rmz0}
(r-z)^{0}_{min}  ~~~  =  ~~~    0.095 - 0.322\,(\log P)^2  ~~~~ (rms = 0.016)
\end{equation}

\begin{equation} 
\label{eqn:gmi0}
(g-i)^{0}_{min} ~~~ = ~~~    0.347  -  0.973\,(\log P)^2  ~~~~ (rms = 0.026)
\end{equation}

\begin{equation}
\label{eqn:umg0}
(u-g)^{0}_{min} ~~~ = ~~~ 0.665 - 0.669\,(\log P)^2  ~~~~ (rms = 0.029)
\end{equation}

where $P$ is the period in days, and the $0$ superscript implies intrinsic colors.
The information in Table~6 of \citet{vivas17} also enables us to
derive the intrinsic minimum light colors in any color combination as
a function of period. We do not list them all here explicitly.

\begin{figure*}[htb!]
\epsscale{1.1}
\plottwo{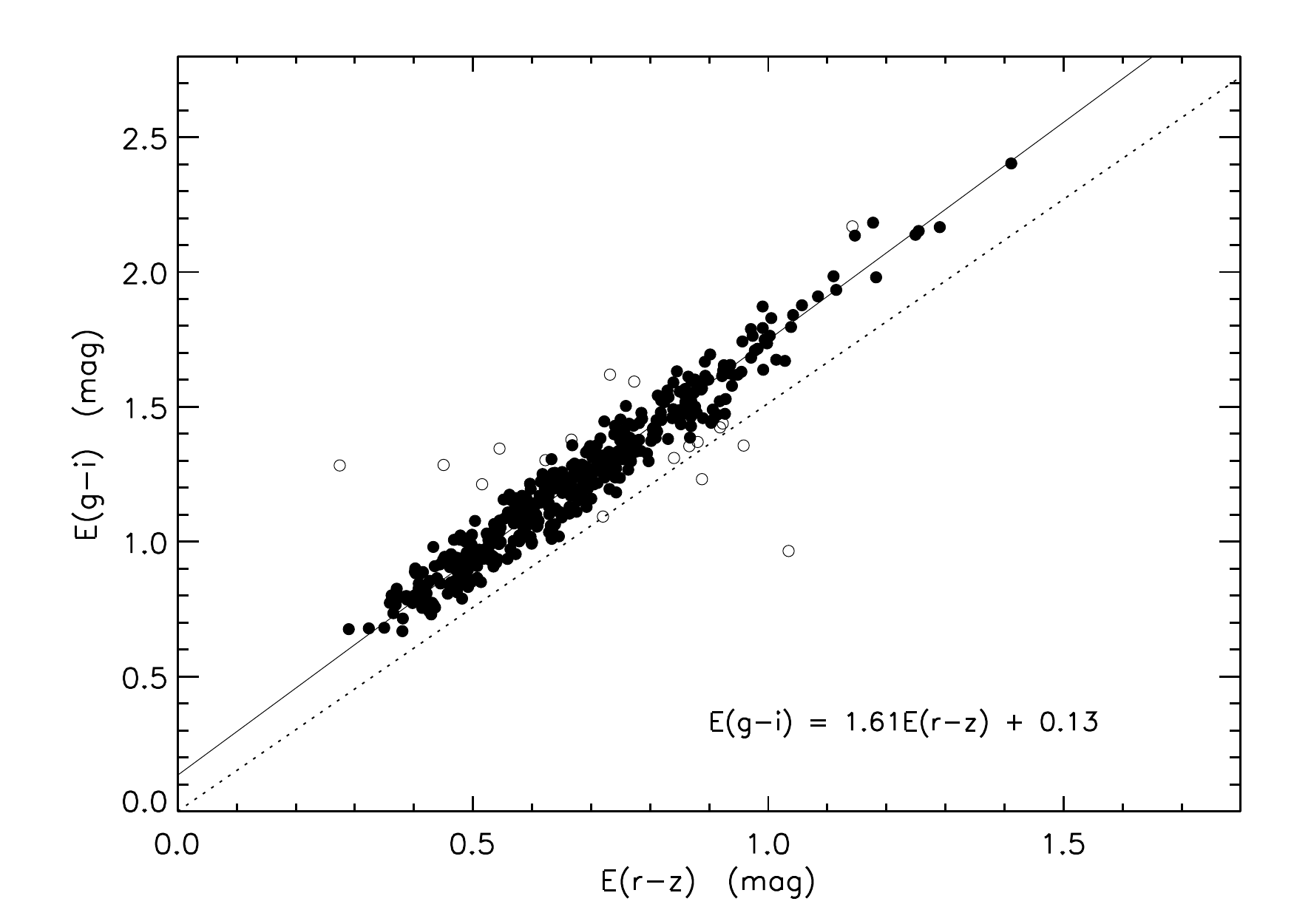}{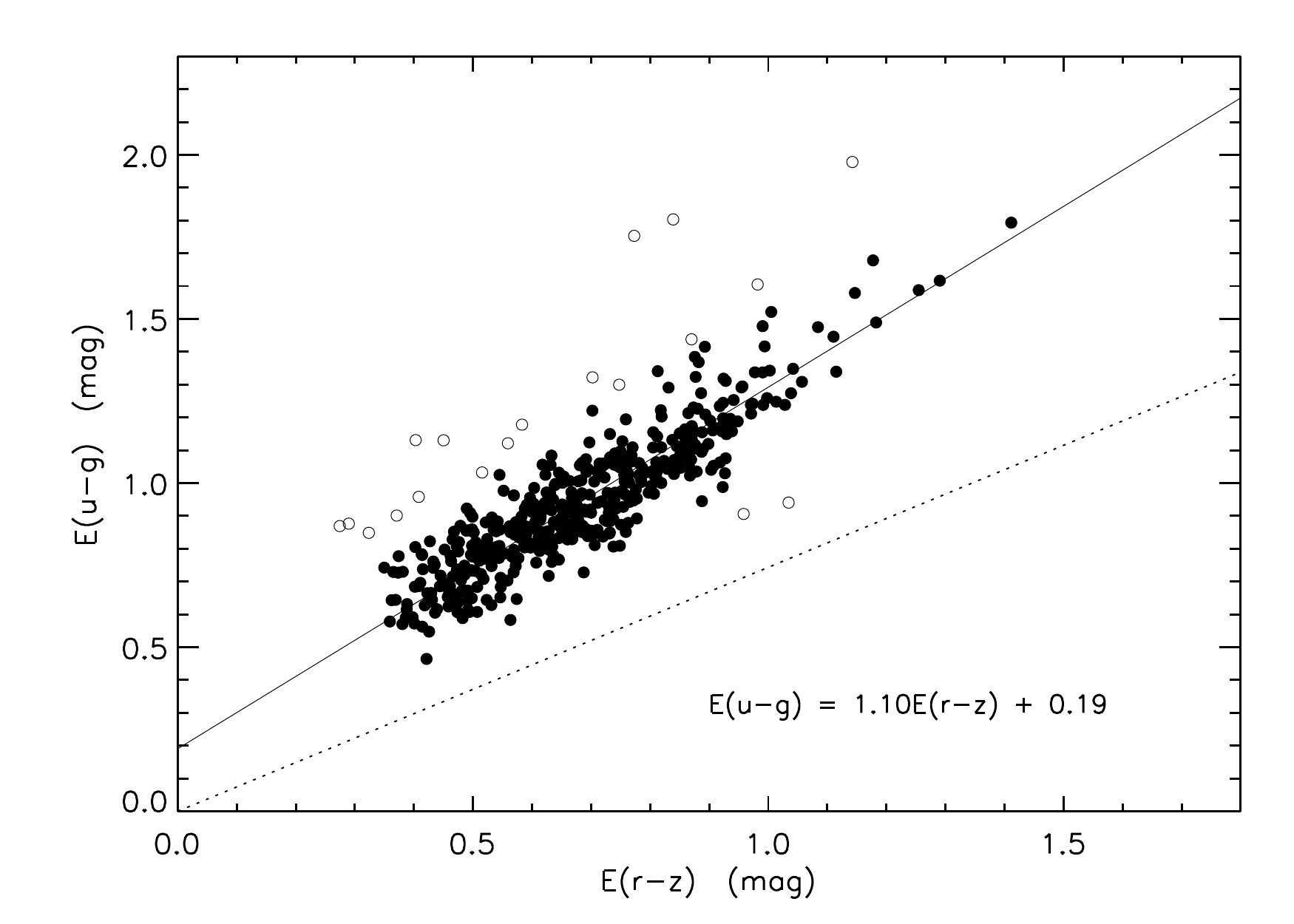}
\caption{Color excesses in $E(g-i)$ versus that in $E(r-z)$ as derived
  from the minimum light colors of individual RRab stars for which the
  brightness at $\phi = 0.65$ could be determined in {\it all} 5 bands
  (left panel). Due to the large differential extinction in the field, the RRab ensemble samples a large range of reddening.  A straight line fit to the data, allowing errors in both axes and 3.0$\sigma$ iterative rejection (rejected points shown as open circles),  is shown, whose slope $E(g-i)/E(r-z)$ is a reflection of the reddening law.  The 0.13 mag intercept is unexpected, and is discussed in the text. The dotted line shows what would be expected from the \cite{odonnell94} reddening law with $R_{V}=3.1$. The right hand panel shows the same for $E(u-g)$ versus $E(r-z)$: the intercept is slightly larger, as is the discrepancy of the slope wrt the O'Donnell law.}
\label{fig:EgmivsErmz} 
\end{figure*}

The fitted magnitudes at phase 0.65 (the middle of the phase range where Sturch demonstrated that colors are constant) of the individual RR~Lyraes in Table~\ref{tab:fitparams}  are listed in the same table.  From these we can construct the observed $(r-z)_{min}$ and $(g-i)_{min}$ values for the RRab's, in Table~\ref{tab:fitparams}  and obtain the color excess $E(r-z)$ and $E(g-i)$ from Equations~\ref{eqn:rmz0} and \ref{eqn:gmi0}.  The differential reddening across the field provides an opportunity to study the relationships of reddening across different color combinations.  Figure~\ref{fig:EgmivsErmz} shows the individual reddening values $E(g-i)$  and $E(u-g)$ as a function of the reddening in $E(r-z)$ for all the RRab's in Table~\ref{tab:fitparams}.  Both panels show a linear trend.  The $u$ filter has a small known red leak. For RR~Lyrae stars, which are relatively blue, this should not have a noticeable effect. Further, if the reddening to M5 and field B1 were the same, the effect of the red leak would affect both fields equally, and the effect would be nulled out.  However, since B1 has substantially more reddening it is prudent to ask if this is a potential problem.  We note that given the large range of individual reddening among the RRab's in B1, the effect of a red-leak in the $u$ band would express itself in the $E(u-g)$ vs $E(r-z)$ relation as a departure from linearity. We do not see such an effect in Figure~\ref{fig:EgmivsErmz}, demonstrating that any adverse contribution from a $u$ band red leak is below the accuracy imposed by the scatter seen in the figure.

The best fit (allowing for errors in both axes, and utilizing iterative $3.5\sigma$ outlier rejection) that relates the reddening in the two cases are given by:
\begin{equation}
\label{eqn:Egmi}
E(g-i) = (1.614 \pm 0.019)\,E(r-z) + (0.134 \pm 0.013)
\end{equation}
and 
\begin{equation}
\label{eqn:Eumg}
E(u-g) = (1.101 \pm 0.026)\,E(r-z) + (0.191 \pm 0.018)
\end{equation}
where the uncertainties are estimated from the scatter. 
The figure shows the fitted relations, as well as the expectation from
an O'Donnell law \citep{odonnell94} with $R_{V} = 3.1$.
%({ \bf the O'Donnell extinction in a given passband was evaluated for a black-body source at 6000 K})  
We see differences in fitted vs expected slope,  but what is problematic is the vertical intercept, since we expect the relation to pass through the origin. We consider and discuss the following possible explanations:

\begin{enumerate}
\item 
Calibration errors/uncertainties in either the M5 reference data or the data presented here, or both.  Since the data in both these figures come from observations taken on the same nights, this is unlikely.  We have gone over the procedures several times to ascertain that a careless error has not been made.
\item
Cumulative random errors in the photometric calibrations. Ascertaining zero-points for any one band for each of the M5 and current data-sets 
can suffer from random errors of up to 0.02 mag, so each color can have zero-point errors of 0.03 mag. Each axis of these figures is a difference of the same color in B1 and M5, so the total uncertainty in the zero-point of each axis can be .04 mag.  Given that the slope of $E(g-i)$ vs $E(r-z)$ is $\approx 1.6$, we can thus expect y-axis intercepts of $\approx 0.065$ mag at the $1\sigma$ level due to systematic errors in measuring the $E(r-z)$ alone (due to a shift in the x-axis zero).  If all errors add in quadrature, total rms uncertainty in the intercept is  $ (0.04^{2} + 0.065^{2})^{0.5}  ~=~ 0.075$ mag.  This makes the 
observed offset in $E(g-i)$  almost a $2  \sigma$ effect.  For $E(u-g)$, the offset is much larger, but metallicity differences between the globular cluster M5 and the RRab's in the B1 field can induce all or part of the observed discrepancy  \citep[see Figure~5 of][]{vivas17} in $E(u-g)$.
\item
The bulge RR~Lyraes have different properties from those in M5.  We are examining this possibility by studying several additional RR~Lyrae bearing clusters, that differ from M5 in metallicity and Oosterhoff type. Our sample includes clusters in the bulge that have much higher metallicities and unusual period distributions relative to their metallicity.
\item
There is some peculiar reddening that is shared equally by all objects (which means it must be relatively local before encountering a spiral arm where different lines of sight must produce differential extinction that cause the large spread in reddening) with a much steeper value of $E(g-i)/E(r-z)$.  Since the RR~Lyraes here are all piled up in the bulge, and we don't see any with $E(r-z) < 0.3$, there is some effect that is hidden from us. Some of our other fields, especially B5, which passes clear of the plane may illuminate whether this is a possibility.  The equivalent of Figure~\ref{fig:EgmivsErmz} for the other fields along different directions in the Galactic bulge may shed some light on whether this is plausible. 
\end{enumerate}

\begin{figure*}[htb!]
\epsscale{1.1}
\plottwo{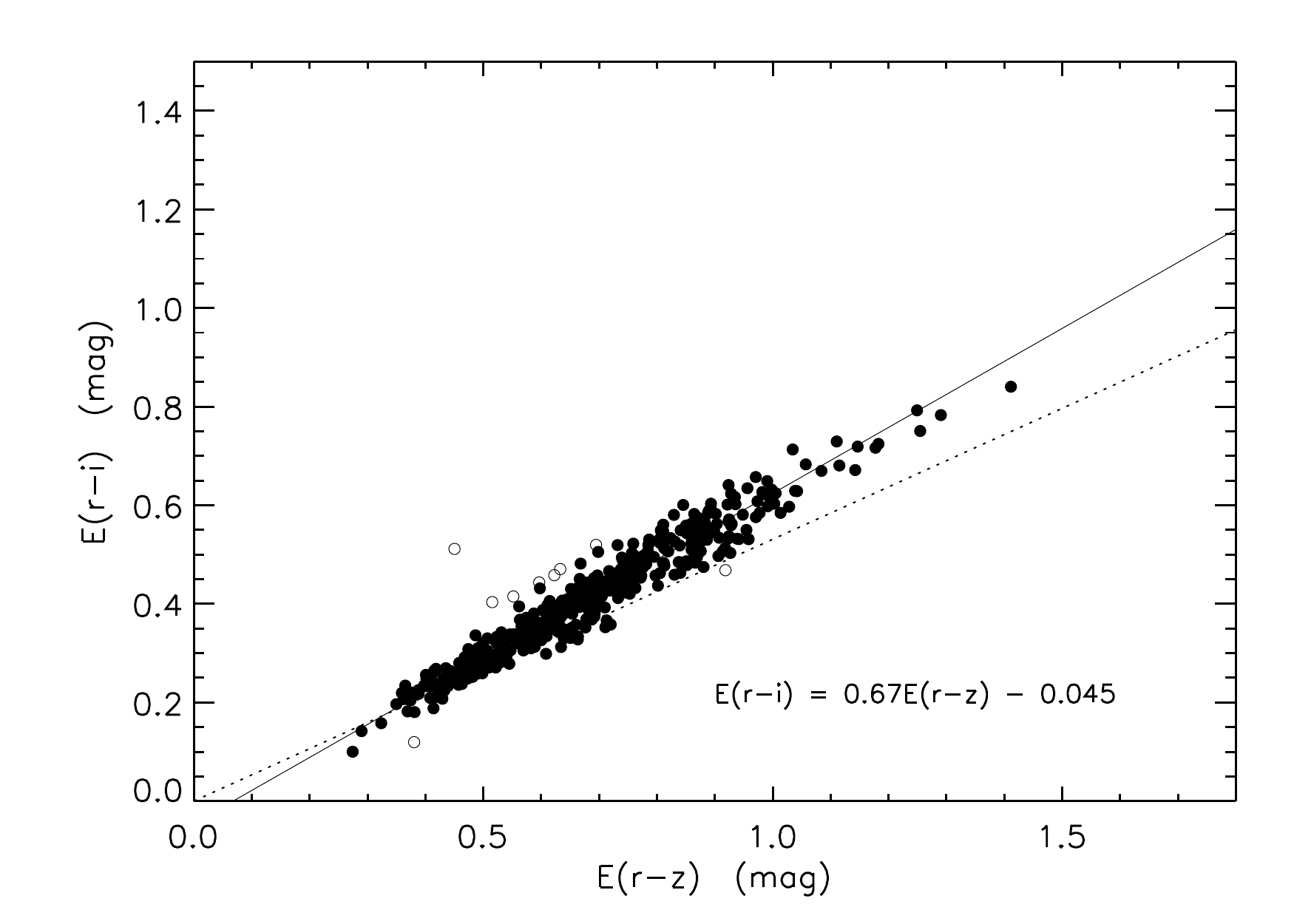}{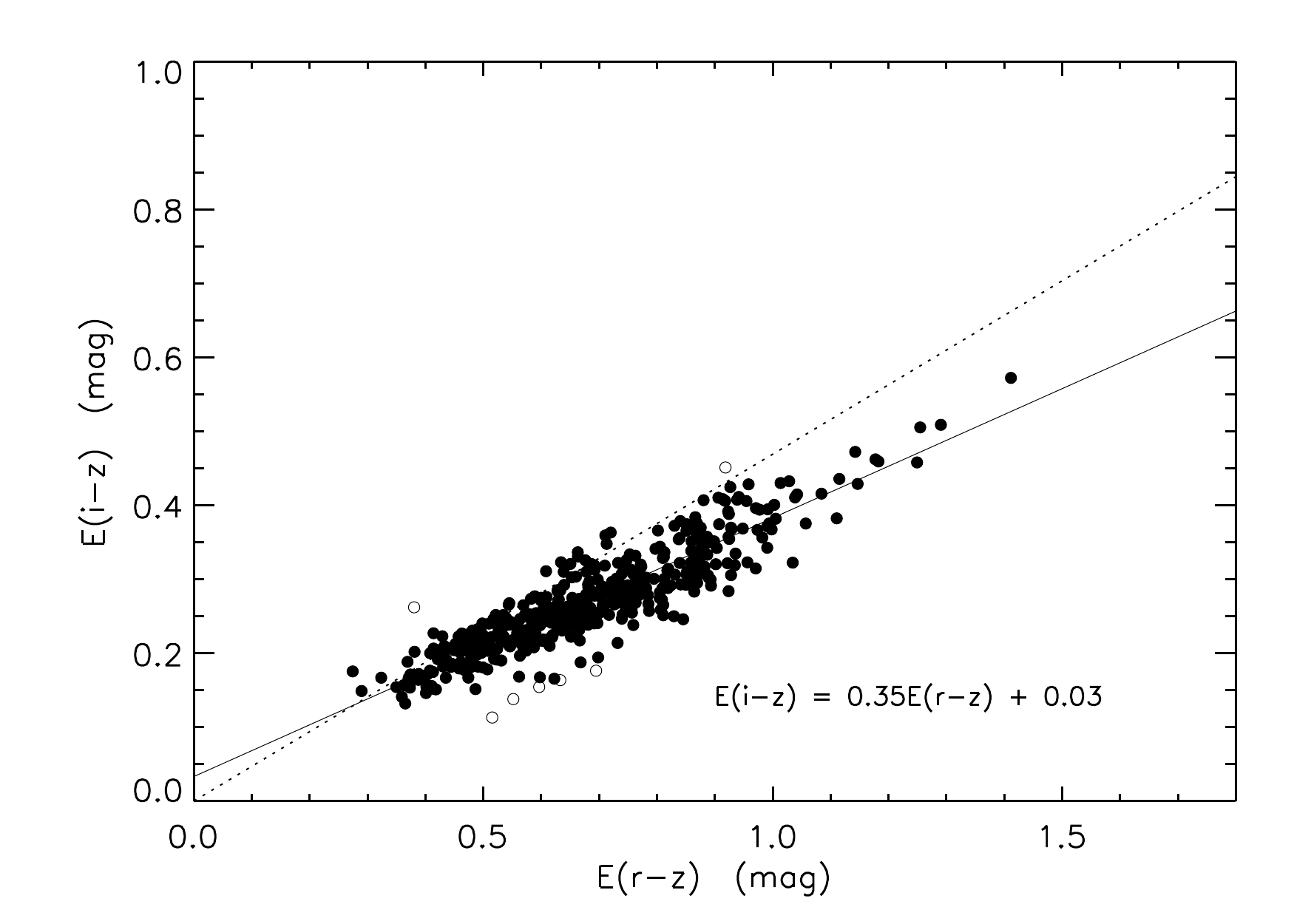}
\caption{Same as Figure~\ref{fig:EgmivsErmz}, but the left and right panels now have $E(r-i)$ and $E(r-z)$ as ordinate. Notice that unlike in Figure~\ref{fig:EgmivsErmz} the linear fits pass very close to the origin, well within expected errors. Note also that for $E(i-z)$, the fitted slope is shallower than for the \citep{odonnell94} law, whereas the slopes for $E(g-i)/E(r-z)$, $E(u-g)/E(r-z)$ and $E(r-i)/E(r-z)$  are steeper than predicted by the O'Donnell law with $R_{V} = 3.1$. }
\label{fig:ErmivsErmz} 
\end{figure*}

  The behavior in color combinations with the  $r, i$, and $z$ bands do not exhibit an anomalous intercept.  This is illustrated in Figure~\ref{fig:ErmivsErmz}.  The linear fits are shown below in Equations.~\ref{eqn:Ermi} and \ref{eqn:Eimz}, and have intercepts entirely consistent with expected calibration uncertainties. However, the fitted slopes are steeper than predicted by \citep{odonnell94} with $R_{V} = 3.1$ for all cases except for $E(i-z)/E(r-z)$, signaling that the slope differences are not resolvable by a simple scaling of $R_{V}$, and shows a departure in shape from the O'Donnell law. This result is independent of the issue of the unexpected intercept.
\begin{equation}
\label{eqn:Ermi}
E(r-i) = (0.669 \pm 0.009)\,E(r-z) - (0.045 \pm 0.006)
\end{equation}\
\begin{equation}
\label{eqn:Eimz}
E(i-z) = (0.350 \pm 0.008)\,E(r-z) - (0.033 \pm 0.006)
\end{equation}

Differences in slope are indicative of non-standard reddening, and we investigate the implications below.  The intercepts must be accounted for when investigating distances, but it takes some contortion (item 4 in the above enumeration of possible reasons) to argue that they arise from non-standard reddening.  We keep this in mind in the arguments we make in the remainder of this section. Specifically, the analysis that follows does not depend upon the zero-point anomaly, but only on the slopes derived from the differential extinction.

 We evaluate the effective wavelengths in each of the 5 passbands for an F5 spectrum
and read the extinction at those wavelengths from the standard reddening law with $R_{V} = 3.1$ according to \cite{odonnell94}.  The F5 spectrum is a good approximation to an RRab star near minimum light, and thus appropriately matched for color-excesses as measured from them.  The extinction values in the 5 passbands determined in this way, and scaled so that $E(r-z) = 1$ provides the total to selective extinction ratios as follows: 

\begin{eqnarray}
\label{eqn:stdreddening}
A_{u} = 3.92\, E(r-z)  \nonumber \\
A_{g} = 3.24\, E(r-z)  \nonumber \\
A_{r} = 2.26\, E(r-z)  \\
A_{i} = 1.73\, E(r-z)  \nonumber \\
A_{z} = 1.26\, E(r-z)  \nonumber 
\end{eqnarray}

We also quote from Equation~3 of  \citep{vivas17}, who calibrated the RRab absolute magnitudes in the globular cluster M5: 
\begin{eqnarray}
\label{eqn:absmags}
M_u \; (ab) &= (-0.10 \pm 0.24) \, \log P  + (1.10 \pm 0.13) \nonumber  \\
M_g \; (ab) &= (-0.57 \pm 0.17) \, \log P  + (0.43 \pm 0.12) \nonumber \\
M_r \; (ab) &= (-1.28 \pm 0.11) \, \log P  + (0.12 \pm 0.11)  \\
M_i \; (ab) &= (-1.59 \pm 0.09) \, \log P  + (0.07 \pm 0.11) \nonumber  \\
M_z \; (ab) &= (-1.68 \pm 0.08) \, \log P  + (0.03 \pm 0.11)  \nonumber 
\end{eqnarray}

\begin{figure*}[hbt!]
\epsscale{1.1}
\plotone{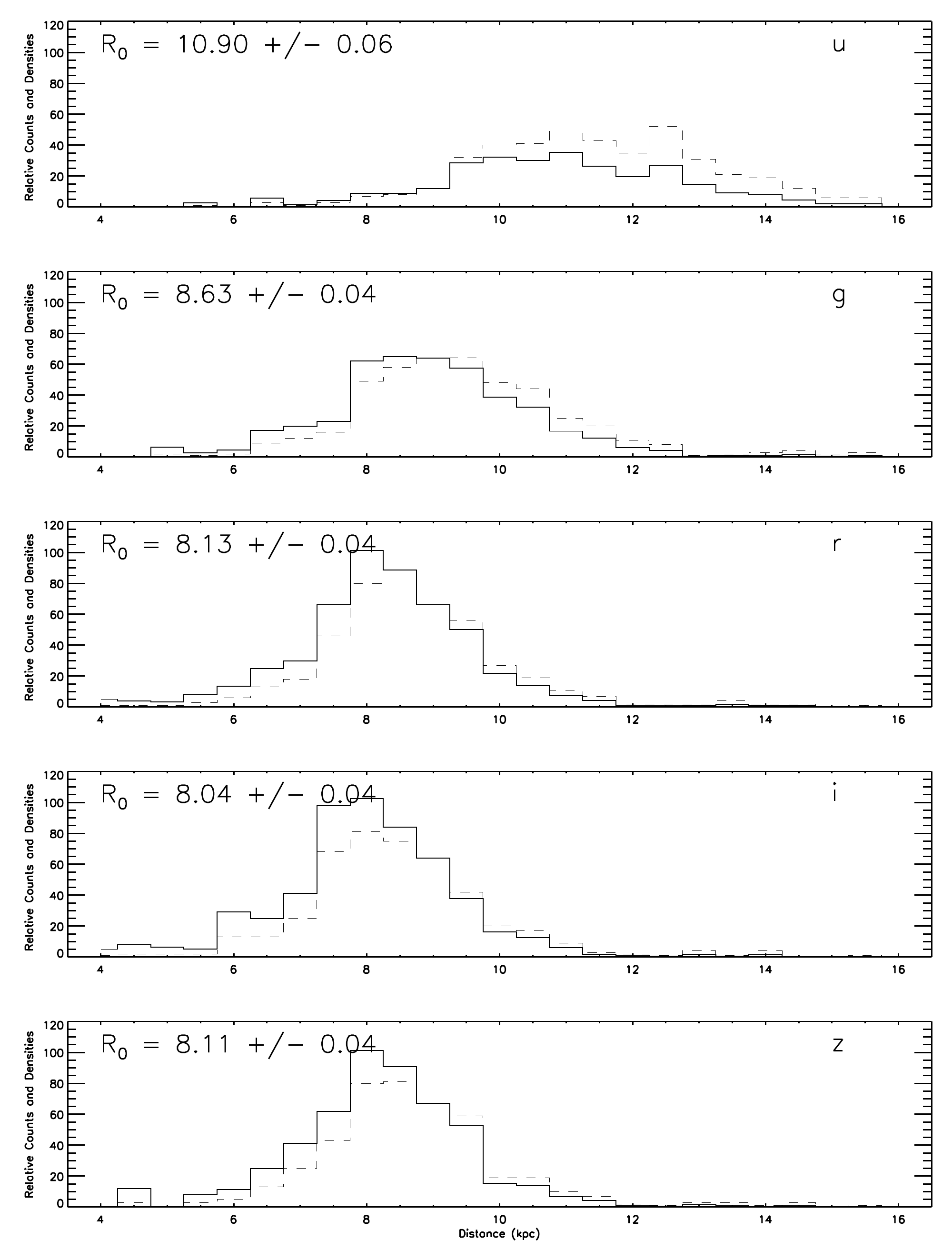}
\caption{Distribution in distance of the RRab stars from Table~\ref{tab:fitparams} calculated from data in each of the 5 passbands, using the standard extinction law with $R_{V}$ = 3.1, as enunciated in \cite{odonnell94}. The dashed lines show the histogram of star-counts, while the bold line shows {\emph relative} density by correcting for the larger sampled spatial volume at larger distance by scaling each bin with star-counts by $(9./d)^{2}$, where $d$ is the distance for that bin. This scales the relative density to equal the star count at 9 kpc. Note the disagreement across the various passbands in distance to the peak of the distribution, as well as in the shape of the inferred distance distribution. Since in reality these cannot depend on the passband, it suggests strongly that the standard reddening law used to produce this figure, does not apply for this line-of-sight.}
\label{fig:disthist_odonnell}
\end{figure*}

Given our observed mean magnitudes in each of these bands for each of
the RRab stars in the B1 field, we derive individual distances to
them.  Figure~\ref{fig:disthist_odonnell} shows the histogram of star
counts at the derived distances.  

Since the line-of-sight passes very close to the Galactic center, the peak of the relative density histogram should occur very nearly at the distance of the Galactic center $R_{0}$.  There are well defined peaks for the histograms in $r,i,z$ near a distance modulus of $8.1$ kpc, but shifts to $8.6$ kpc in $g$, with the histogram peak less sharply defined.
In $u$ the histogram essentially disintegrates to a flat skewed extension, and the highest density is nearly at $10$ kpc.  We submit that this inconsistency between the various bands is due to the use of the incorrect reddening law, as already surmised from the disagreement between the observed reddening vectors from those predicted by the standard reddening law.  Differences in the reddening law affects not only the central tendency of the density histogram, but, because of the large differential extinction, also redistributes the relative distances among the RRab stars, changing the structure and shape of the density histogram from one band to another.

We can seek to derive the correct reddening law that will yield not only the observed reddening vectors, but also bring into accord the distance histograms in all 5 passbands.
Since RR~Lyrae, and especially the RRab's are both standard candles, and standard ``crayons,'' they lend themselves naturally to such an analysis. We show (in \S~\ref{sec:geom}) that such an analysis yields a reddening law that can rectify disagreements of the distance distance distribution highlighted above, and makes them consistent across all 5 passbands.

%The derived values for $R_0$ are now much lower in $r,i,z$, and lower but less so in $g$. This is consistent with the relative extinction differences by passband shown in Figure~\ref{fig:redlaw} between our own derived values (Equation~\ref{eqn:total2selective}) and the values from the standard \citet{odonnell94} form with $R_{V} = 3.1$.  The $u$ now yields an increased distance which is consistent with the lower total to selective extinction in $u$ for the standard law than what we have derived. However the derived density  distrbution with distance is no longer sharply peaked, and very dissimilar to that from the other passbands.  All of the density distribution profiles appear skewed. The average $R_{0}$ from $g,r,i,z$ is now $8.3 \pm 0.2 ~{\rm kpc}$, which is consistent with the recent compilation of $R_{0}$ values from various methods and sources.by \citet{Vallee17}.  

\subsection{Derivation of the Extinction from the Reddening}

In light of the results and discussion above we must proceed with caution, clearly enunciating our assumptions so that the impact of this reddening ``offset'' (especially as seen in Figure~\ref{fig:EgmivsErmz})  can be tracked and examined at any point.  We will choose to use $E(r-z)$ as fiducial reddening, since the evidence from 
Equations~\ref{eqn:Egmi} through \ref{eqn:Eimz} indicate that the $r,i,z$ bands are better behaved, whatever be the source of difficulty with $g$ and $u$.  However, the discrepancy in slopes with respect to the standard reddening law, and the discordance of the space density histogram with distance along the line-of-sight  means that in order to calculate extinction from the reddening $E(r-z)$, we will be well served to derive $A_{X}/E(r-z)$ for any passband $X$ for our line-of-sight from our own data.  Since the distribution of the RR~Lyraes along the line-of-sight is very sharply peaked at the Galactic center, we can exploit the standard candle property of
RR~Lyraes.

The absolute magnitude vs period relationships in Equation~\ref{eqn:absmags} are known to depend weakly on metallicity, and are strictly valid for the metallicity of M5. However \citet{walker91} showed from spectroscopic analysis the metallicity distribution of RR~Lyrae in the bulge peaks at ${\rm [Fe/H]} \approx -1.0$, whereas for M5 ${\rm [Fe/H]} = -1.25 \pm .05$ from \citet{dias16}. The similarity in metallicity supports using the absolute magnitudes from Equation~\ref{eqn:absmags}.   For the purpose
of deriving extinction from reddening by the method described below, any net offsets in the absolute magnitudes are not important, but their dependencies on period, as gleaned from Equation~\ref{eqn:absmags} are useful to consider.

The distance modulus $\mu_{0}$ of any RRab star, its mean magnitude $m_{X}$, and its extinction $A_{X}$ in band $X$ are related by: 
\begin{equation}
A_{X} = m_{X} - M_{X} - \mu_{0}
\end{equation}

 $M_{X}$ is adopted for the appropriate band from Equation~\ref{eqn:absmags}.   $\mu_{X} = m_{X} - M_{X}$ is the apparent distance modulus in $X$.  $m_{X}, E(r-z)$ for the RRab's are known
from a combination of Table~\ref{tab:fitparams}, Equation~\ref{eqn:absmags} and \S~\ref{sec:minlightreddening}.  Since $M_{X}$ can have no explicit dependence on $E(r-z)$, we can write:

\begin{equation} 
\label{eqn:mux2ax}
 \partial  A_{X}  /  \partial(E(r-z)) =  \partial  \mu_{X}  /  \partial(E(r-z))  -  \partial  \mu_{0}  /  \partial(E(r-z))
\end{equation}

 If all the RRab's are at the same distance, i.e., $\mu_{0}$ is constant,  then measuring $ \partial  \mu_{X}  /  \partial(E(r-z)) $  would directly give us the value of  $A_{X}/E(r-z)$.   In the present case,  $\mu_0$ is not constant, but has a peaked distribution (we anticipate the result shown later in this paper that it peaks exponentially). If we can pick out the ones that are very near or within the sharp peak, and for which $ \partial  \mu_{0}  /  \partial(E(r-z)) $ is negligible (which is true if the bulge itself has insignificant reddening), Equation \ref{eqn:mux2ax} would again yield the desired value of $A_{X}/E(r-z)$. 
Thus, given a large enough ensemble of RRab's, a wide enough range of $E(r-z)$, and a sufficiently peaked distribution in $\mu_{0}$ it should be possible to derive $\partial A_{X} / \partial E(r-z)$ from the slope $\partial \mu_{X} / \partial E(r-z)$.  Note that if extinction/reddening within the bulge is \emph{not} negligible, then objects at farther distance will on average have higher reddening, so that $ \partial  \mu_{0}  /  \partial(E(r-z)) $ is positive. This implies that the true value of $ \partial  A_{X}  /  \partial(E(r-z)) $ is, if anything, {\it smaller } than the measured value of $ \partial  \mu_{X}  /  \partial(E(r-z)) $, or that the above procedure yields at worst an upper bound on the total to selective extinction ratio. A potential further complication is that an individual RRab can have a metallicity different from the mean, producing some departure in absolute magnitude. However,  this is not expected to exceed $\sim 0.2$ mag for any given such object, which is much smaller than the individual variations due to actual distance along the line-of-sight.

\begin{figure*}[htb!]
\epsscale{1.1}
\plottwo{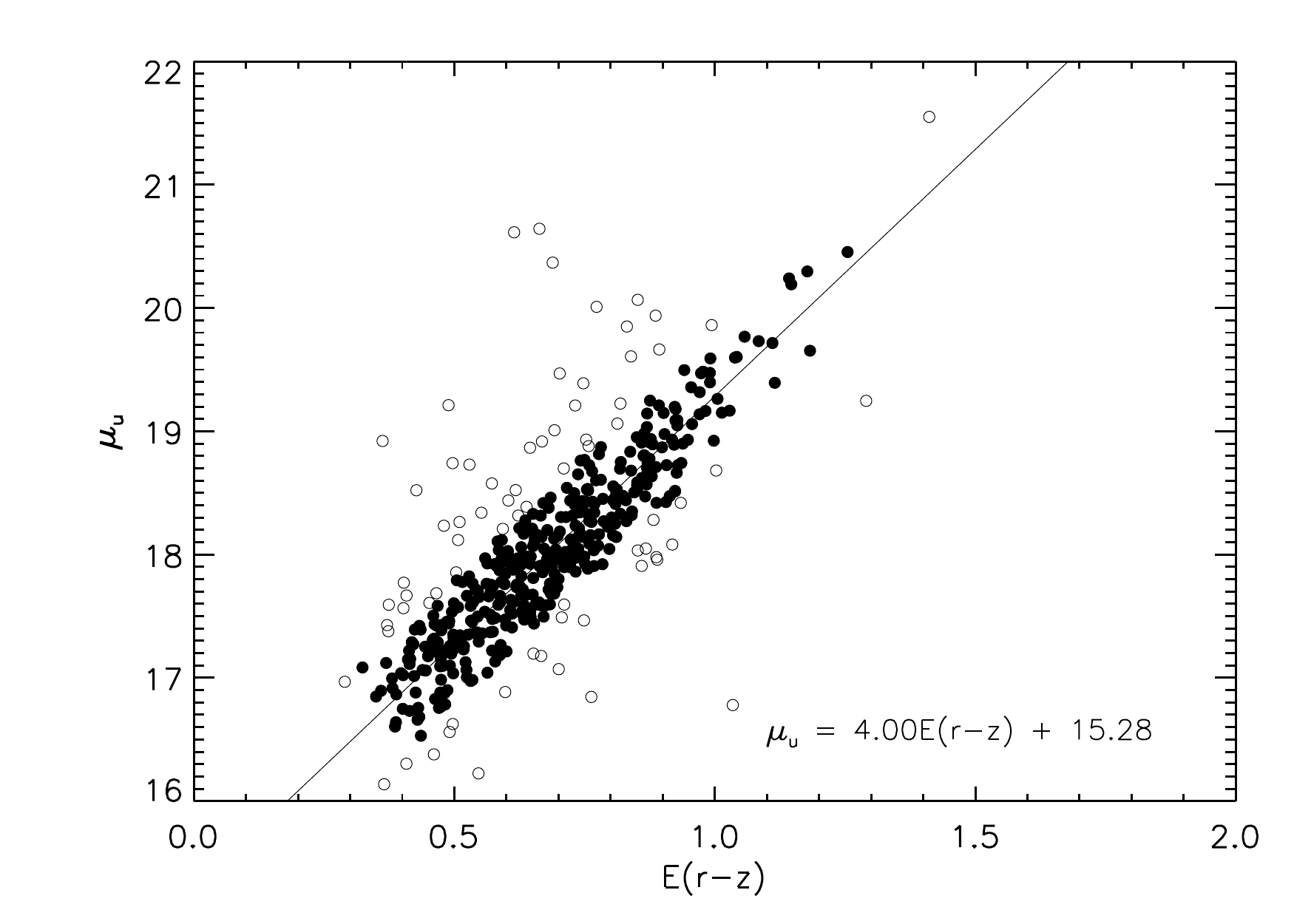}{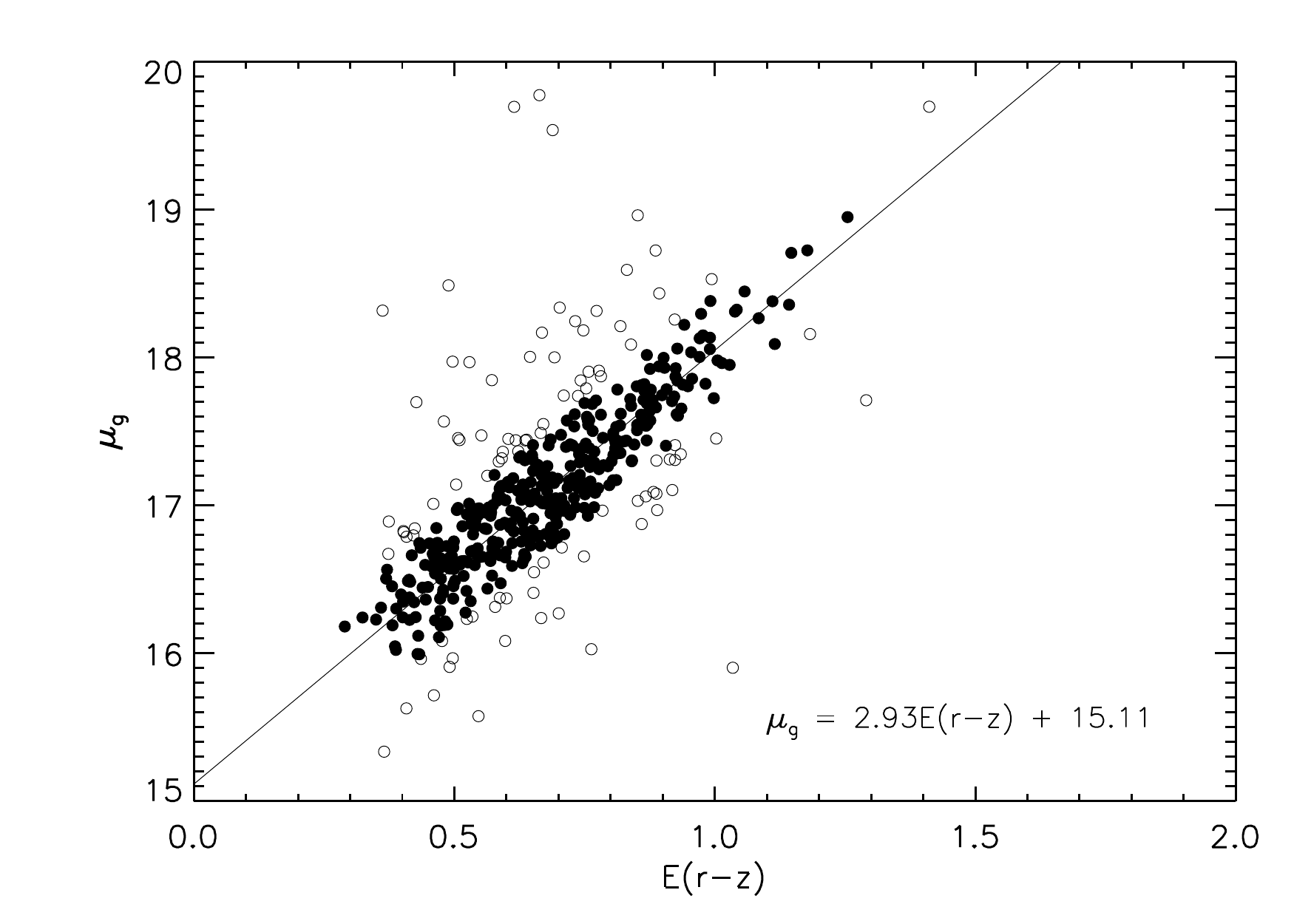}
\plottwo{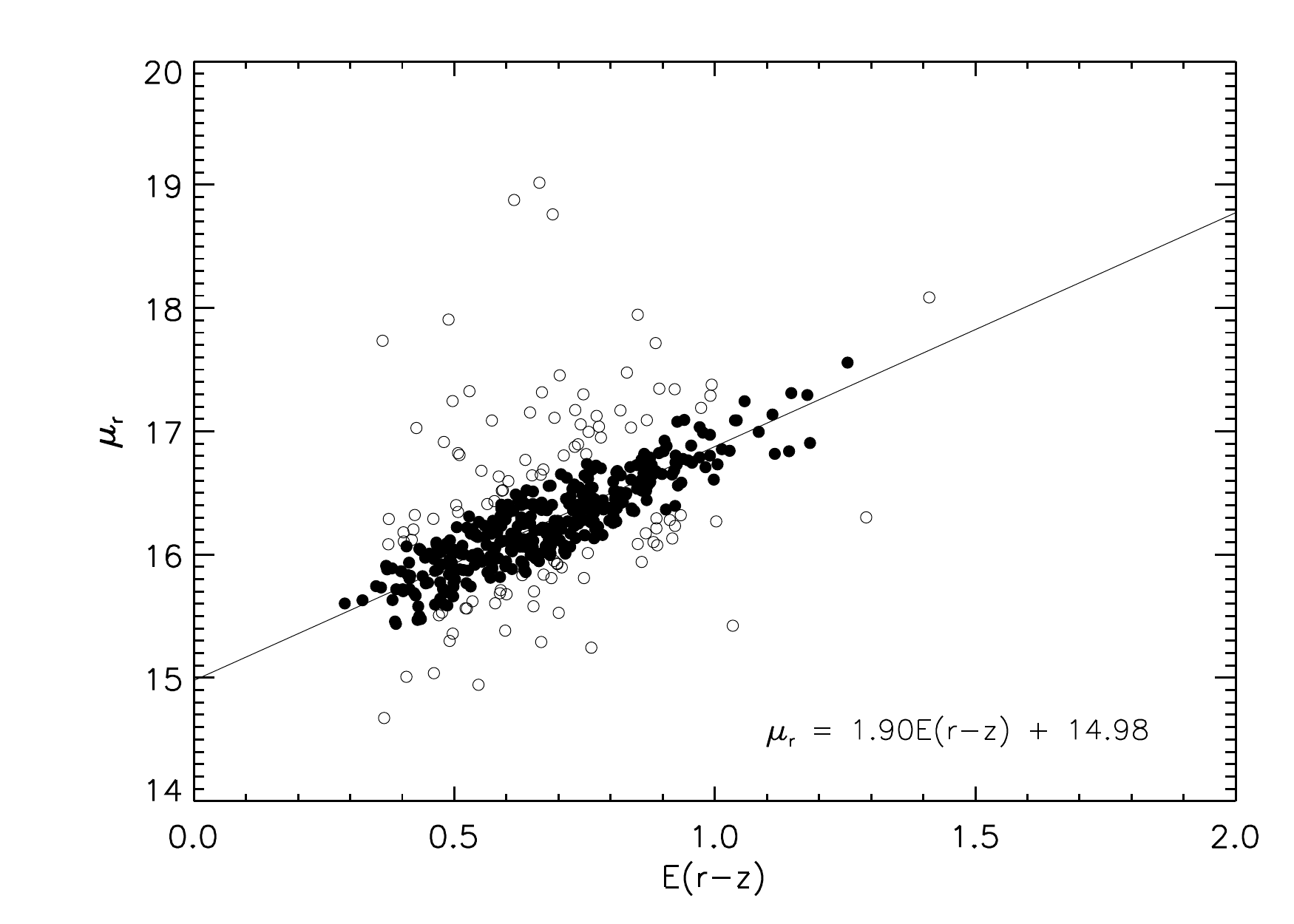}{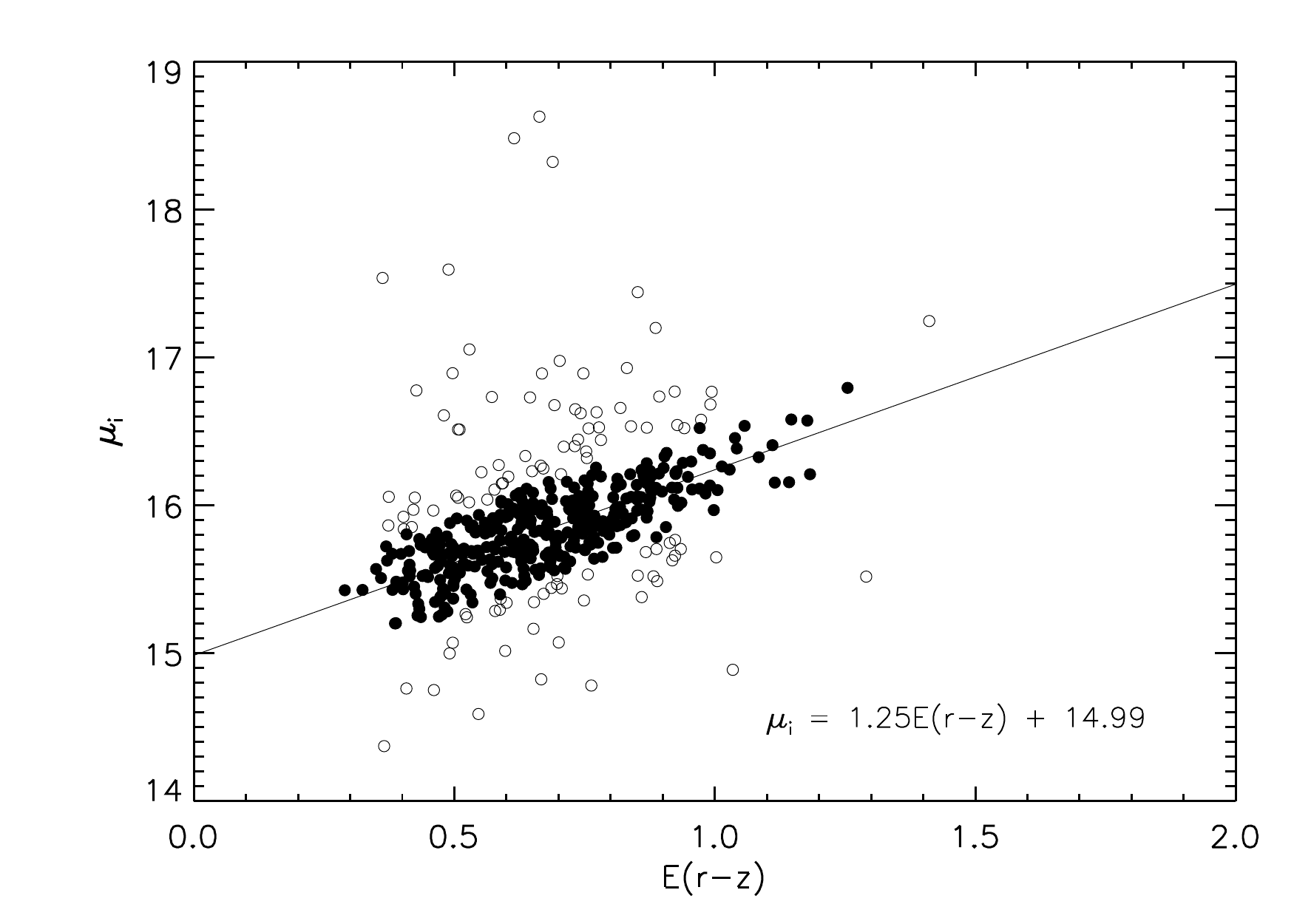}
\epsscale{0.55}
\plotone{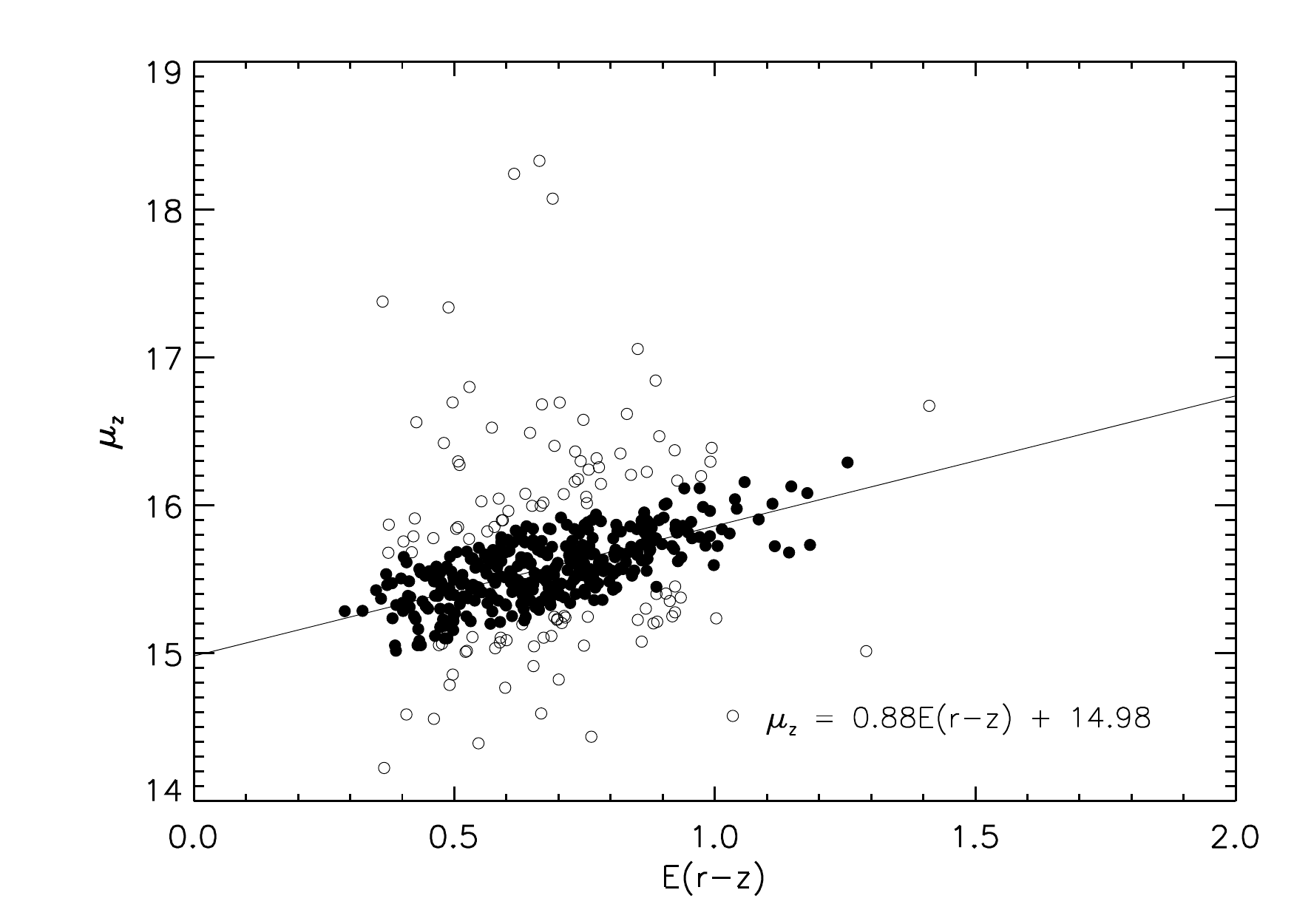}
\caption{Apparent distance moduli in the 5 passbands of all individual RRab's in Table~\ref{tab:fitparams} for which there are available mean magnitudes in all 5 bands against their color excess $E(r-z)$.  
The linear fits, obtained by regression with errors in both axes and iterative $2\sigma$ outlier rejection is shown in each case by the straight line. The stars used in the fit are indicated by filled circles.  The fitted slope for each band is a determination of total to selective line-of-sight extinction for that band (see text).}
\label{fig:Au} 
\end{figure*}

Figure~\ref{fig:Au} shows $\mu_{X}$ for all bands as a function of
$E(r-z)$ for the ensemble of all RRab's for which mean magnitudes are
available from Table~\ref{tab:fitparams} for all 5 bands. Again, a
linear fit with iterative outlier rejection was performed with
uncertainties on both axes.  A value of $\sigma = 0.05$ was used for
$E(r-z)$ and $\sigma = 0.20$ was used for $\mu_{X}$: the substantially larger uncertainty for $\mu_{X}$ allows for the back to front rms depth in the distances of individual RRab's about the distance at which the space density peaks. The iterative rejection threshold was set at $2\sigma$.  The intercepts in each case notionally provide the true mean distance modulus of the RRab in the sample. The few background and foreground RRab's are clearly seen as outliers in these figures.  

The parameters and their uncertainties for the fitted linear regression for each of the 5 passbands are listed below:

\begin{eqnarray}
\label{eqn:total2selective}
\mu_{u} = (4.003 \pm 0.087)\,E(r-z)  + (15.281 \pm 0.061) \nonumber  \\
\mu_{g} = (2.933 \pm 0.069)\,E(r-z)  + (15.115 \pm 0.048) \nonumber  \\
\mu_{r}  = (1.898 \pm 0.058)\,E(r-z)  + (14.978 \pm 0.041)  \\
\mu_{i}  = (1.254 \pm 0.055)\,E(r-z)  + (14.985 \pm 0.038) \nonumber  \\
\mu_{z} = (0.880 \pm 0.053)\,E(r-z)  + (14.981 \pm 0.037) \nonumber
\end{eqnarray}

 As argued above the slopes in Equation~\ref{eqn:total2selective} correspond to the values of $A_{X}/E(r-z)$.  We should note that 
the slopes (as well as intercepts) in the above derived relations are
consistent within the errors with the corresponding slopes (and
intercepts) in  Equations~\ref{eqn:Egmi} through \ref{eqn:Eimz}.
However, the larger uncertainties in
Equation~\ref{eqn:total2selective} are a result of the scatter
introduced by the spread in actual distances to the individual RRab's.
However, Equation~\ref{eqn:total2selective} is necessary to infer the
extinction.  The color to color reddening relations (Equations.~\ref{eqn:Egmi} and \ref{eqn:Eumg}) alone do not allow us to do that. 
In the absence of independent determinations of the total to selective absorption, the default practice is to use a standard reddening law such as \cite{odonnell94}, but Equation~\ref{eqn:total2selective} makes it possible to check whether that is appropriate for the line-of-sight to our field B1.

\begin{figure*}[hbt]
\epsscale{0.70}
\plotone{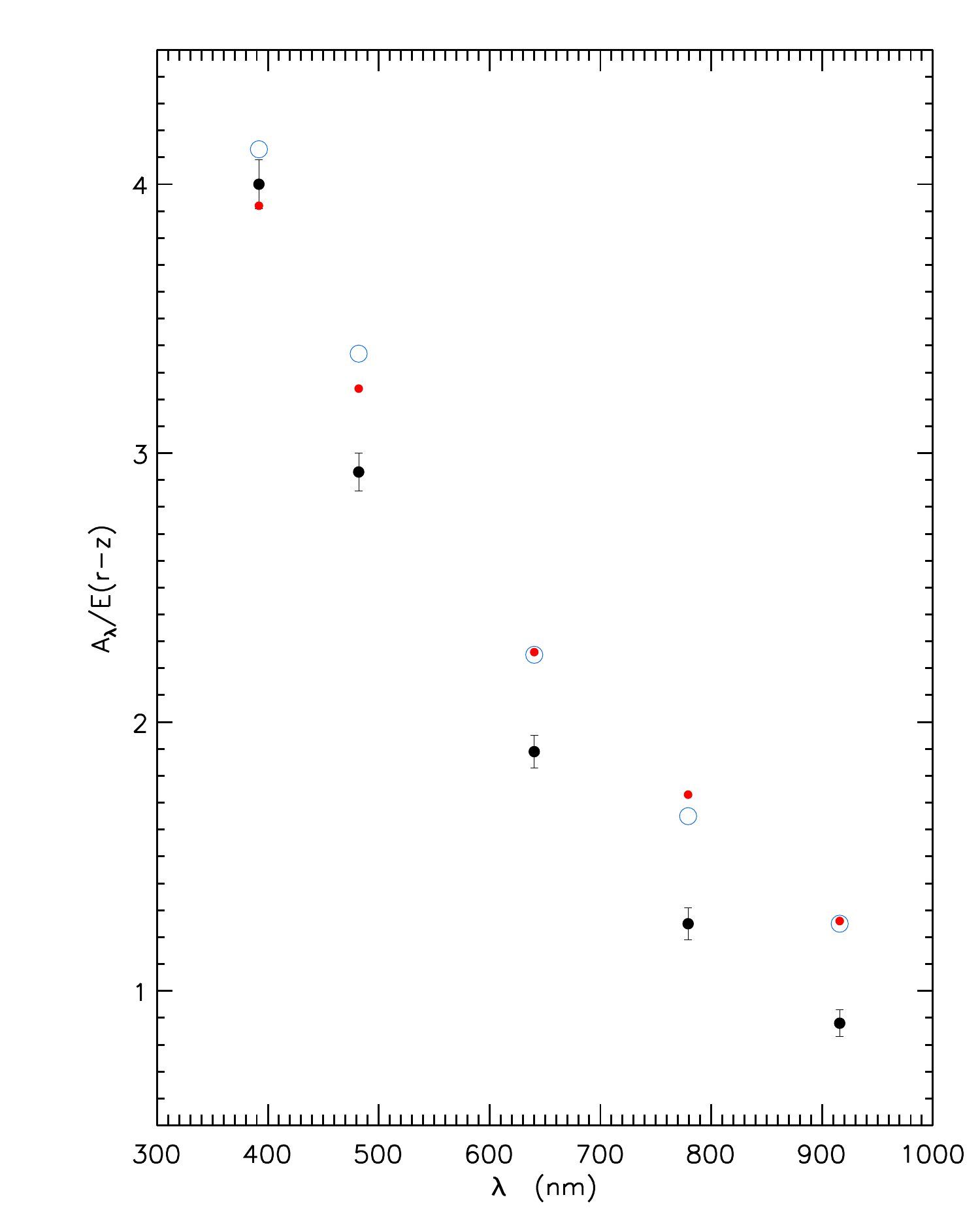}
\caption{Comparison of the derived total to selective absorption for the five DECam passbands used here (by the black dots, along with their uncertainties) to the values implied by the $R_{V} = 3.1$ reddening law from \citet{odonnell94} (shown by the red dots). The x-axis shows the effective wavelength corresponding to the passbands.  The 
\citet{fitz99} reddening law with $R_{V} = 3.1$, which is also commonly used in the literature as a default, is also shown for comparison (blue circles).  Note that except in the $u$ band, the differences between O'Donnell and Fitzpatrick laws with $R_{V} = 3.1$ differ by less than the difference of the law derived here with respect to either of them.
As implied by the y-axis label, all 3 sets of points are normalized so that $E(r-z)$  is unity.}
\label{fig:redlaw}
\end{figure*}

It is worth pondering whether the presence of RR~Lyraes in the  two globular clusters, NGC~6522 and NGC~6528, which fall within our field B1, bias our results. 
NGC~6522, which was placed in the gap between two chips, has 11 RR~Lyrae stars listed in the Clement catalog \citep{clement01}\footnote{Updated version at http://www.astro.utoronto.ca/~cclement/cat} within 2 half-light radii ($r_{h}$). All but two of these are masked by our pointing, and the ones that remain are first harmonic oscillators that are thus not in our list of RRab's.  Similarly, 
NGC~6528 has only 2 known RR~Lyrae associated with it, both within $2\, r_{h}$ \citep{skot15}, one of which may be an RRab, but is not in our list.  Looking at it another way, there are  3 RRab's in our list within $5\, r_{h}$ of NGC~6522 (and none within $5\, r_{h}$ of NGC~6528).  Even in the remote event that these 3 are bona-fide members of NGC~6522, removing them from the analysis does not change the derived coefficients in Equation~\ref{eqn:total2selective} by more than a small fraction of the stated uncertainties.

The intercept values in Equation~\ref{eqn:total2selective} strongly anti-correlate with the corresponding slope value.  The numbers correspond to the effective true distance modulus of where along the line-of-sight the RR~Lyraes pile up the most, but at farther distances the field-of-view samples a bigger volume, so this requires tempering before it indicates the distance at which the RRab density peaks. Also note that the value of $E(r-z)$ that follows from Equation~\ref{eqn:total2selective} by subtracting $A_{z}$ from $A_{r}$, while not identical to the input $E(r-z)$,  is self-consistent within the errors.  This is because we have fitted allowing uncertainties in both axes.  Reading from Equation~\ref{eqn:total2selective}, we therefore adopt:

\begin{equation}
\label{eqn:t2s0}
A_{i} = 1.254\,E(r-z)
\end{equation}
\begin{equation}
\label{eqn:t2s1a}
A_{r} = 1.898\,E(r-z)
\end{equation}
\begin{equation}
\label{eqn:t2s1b}
A_{z} = 0.880\,E(r-z)
\end{equation}

Recognizing that the accuracy of $E(u-g)$ and $E(g-i)$ from Equations~\ref{eqn:Egmi} and \ref{eqn:Eumg} is far superior to the uncertainties presented in Equation~\ref{eqn:total2selective}, and that the determined value of $A_{i}$ is both better constrained and less volatile with respective to errors in measuring $E(r-z)$ (because it has a multiplier of only 1.25, compared to 2.93 for $A_{g}$ and 4.00 for $A_{u}$),  it is prudent to determine $A_{u}$ and $A_{g}$ as follows (since they are better anchored to the data):
\begin{equation}
\label{eqn:t2s2}
A_{g}  = E(g-i) + A_{i}  = E(g-i) + 1.254\,E(r-z)
\end{equation}
\begin{equation}
\label{eqn:t2s3}
A_{u}  =    E(u-g) + A{g} =  E(u-g) + E(g-i) + 1.254\,E(r-z)  
\end{equation}

For the RRab's themselves, the individual $E(g-i)$ and $E(u-g)$ are directly derivable using Equations~\ref{eqn:gmi0} and \ref{eqn:umg0}.
In what follows, we will only require Equations~\ref{eqn:t2s0} to \ref{eqn:t2s3} to correct magnitudes for extinction, while we will correct (deredden) colors using Equations~\ref{eqn:Egmi} to \ref{eqn:Eimz}.

For the 5 DECam passbands used, using the extinction law of
\citet{odonnell94} with $R_{V} = 3.1$ gives $A_{X}/E(r-z)$ to be
$4.00, 3.26, 2.28, 1.73$, and $1.27$ for $X = u, g, r, i,$ and $z$ respectively.  Figure~\ref{fig:redlaw} shows both sets of values for $A_{X}/E(r-z)$. Since they are sufficiently different from our derived values we adopt the latter as given in Equations~\ref{eqn:t2s0} to \ref{eqn:t2s3}.

\section{Color-Magnitude diagrams}
\label{sec:CMDs}

\subsection{Differential Extinction and their Effect on the Observed CMDs}
\label{sec:rawCMDs}

\begin{figure*}[h]
\epsscale{1.1}
\plottwo{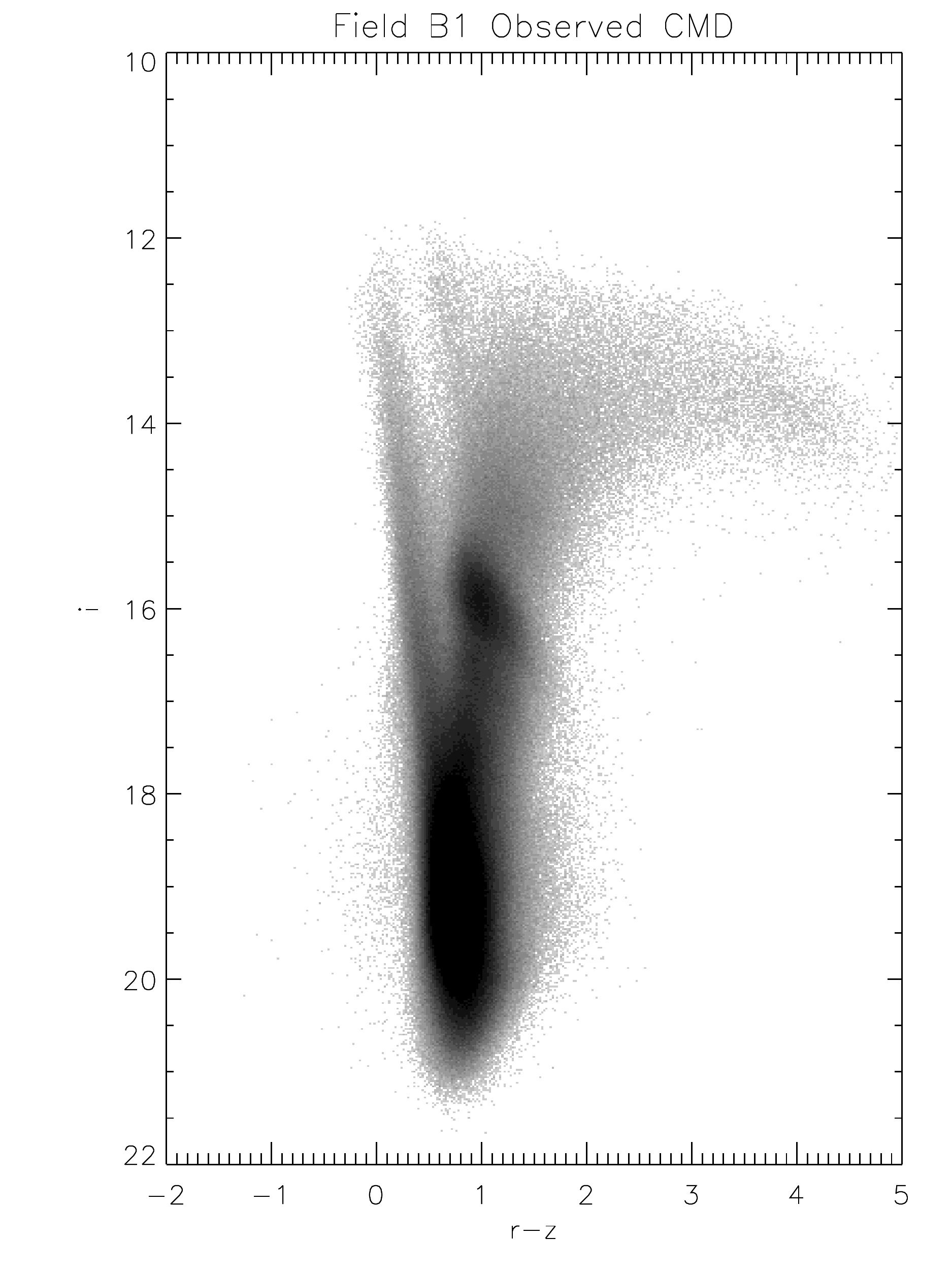}{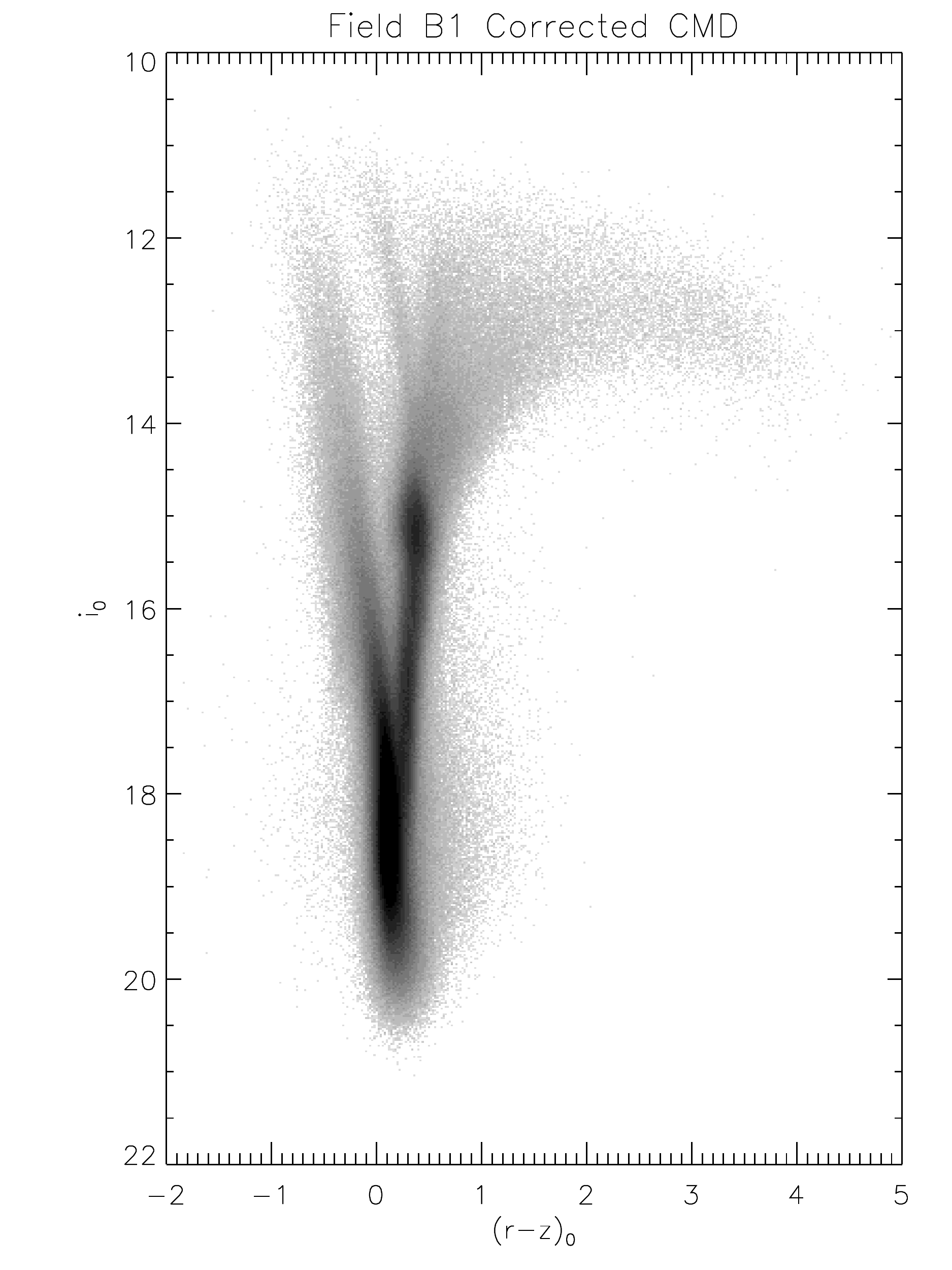}
\caption{Observed CMD as $i$ vs $(r-z)$ (left panel).  The right hand panel shows the same CMD corrected for reddening and extinction using the procedure described in 
\S~\ref{sec:justRRLs} and \S~\ref{sec:useCCH}. Note how in the left panel the red clump is extended along the reddening vector due to differential extinction,  whereas in the reddening/extinction corrected right hand panel the red clump fall in a very narrow color range, but show vertical extension corresponding to the distance distribution. The various features are discussed in \S~\ref{sec:CMDfeatures} }. 
\label{fig:cmd_rmz}
\end{figure*}

\begin{figure*}[h]
\epsscale{1.1}
\plottwo{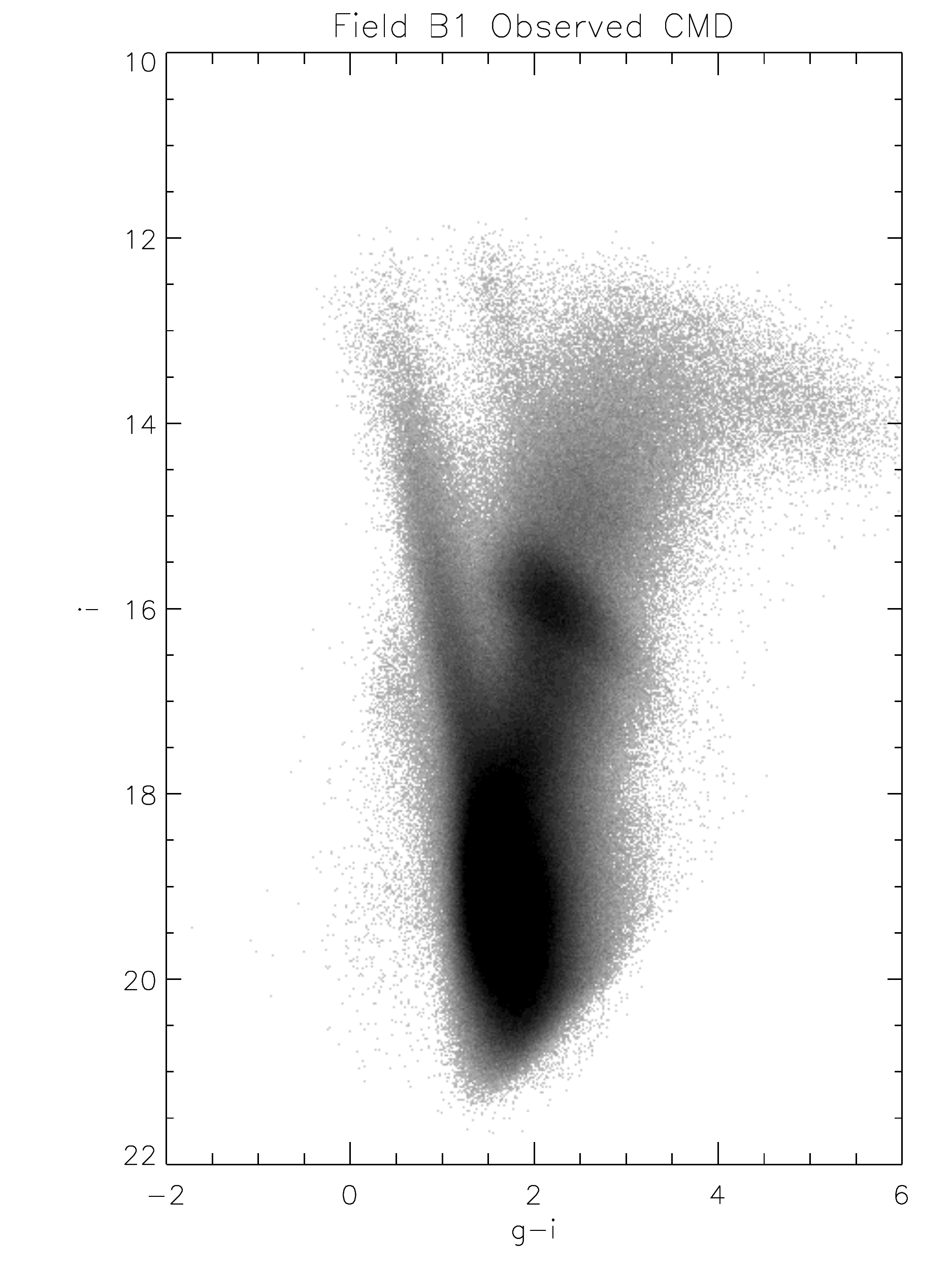}{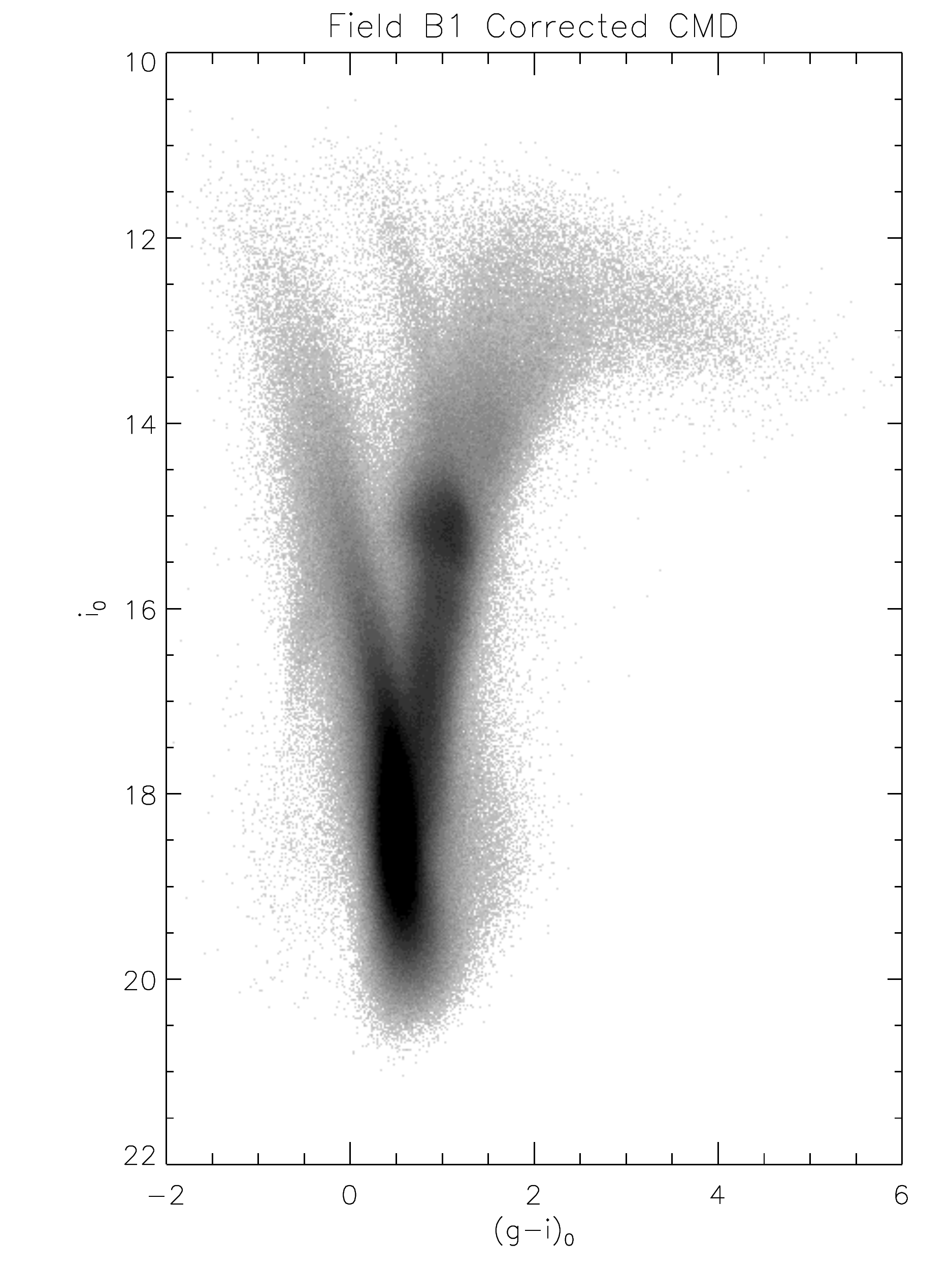}
\caption{Same as Figure~\ref{fig:cmd_rmz}, but for $i$ vs $(g-i)$.  See \S~\ref{sec:CMDfeatures} for discussion of the features}.
\label{fig:cmd_gmi}
\end{figure*}

\begin{figure*}[hbt]
\epsscale{1.1}
\plottwo{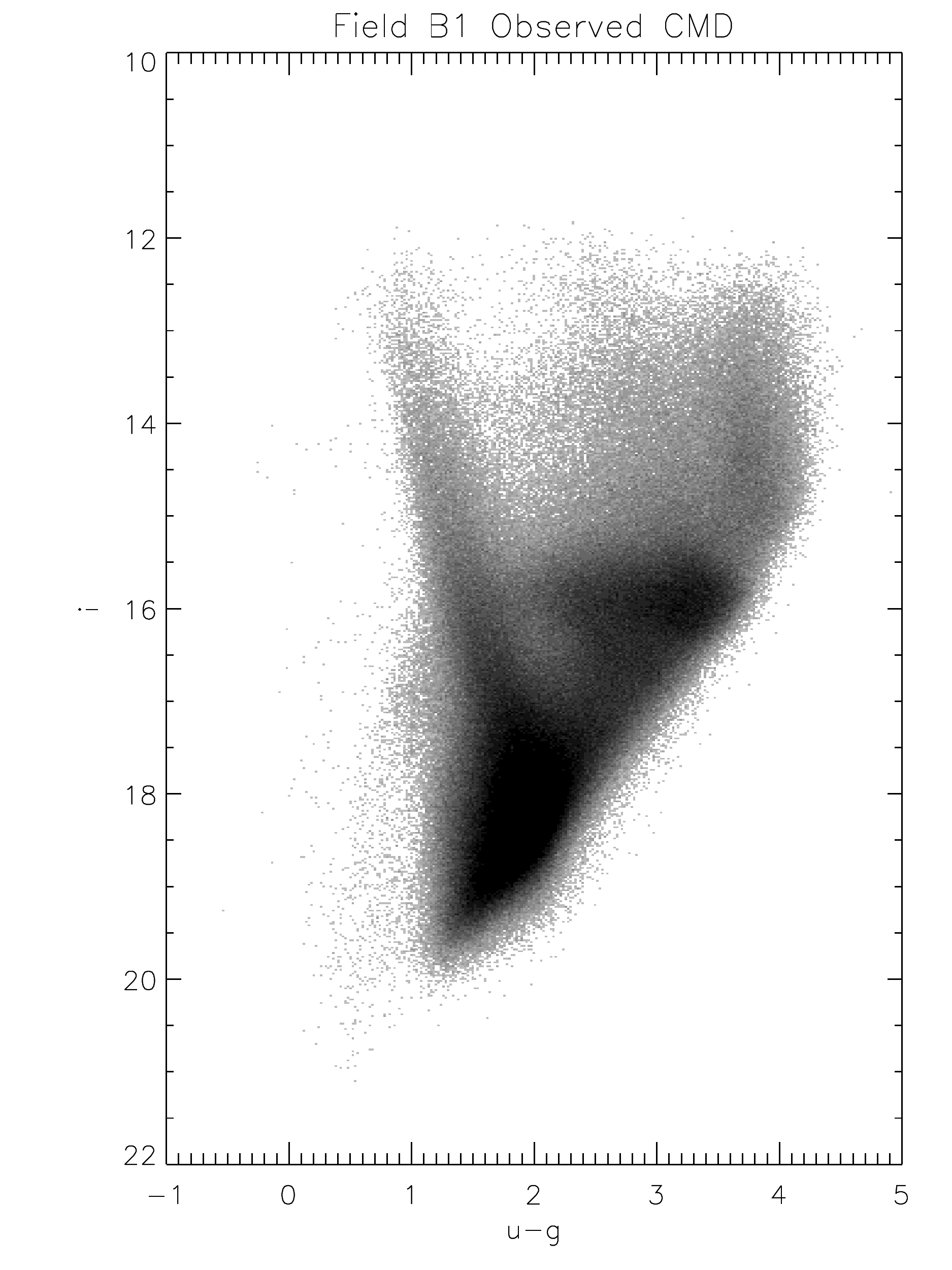}{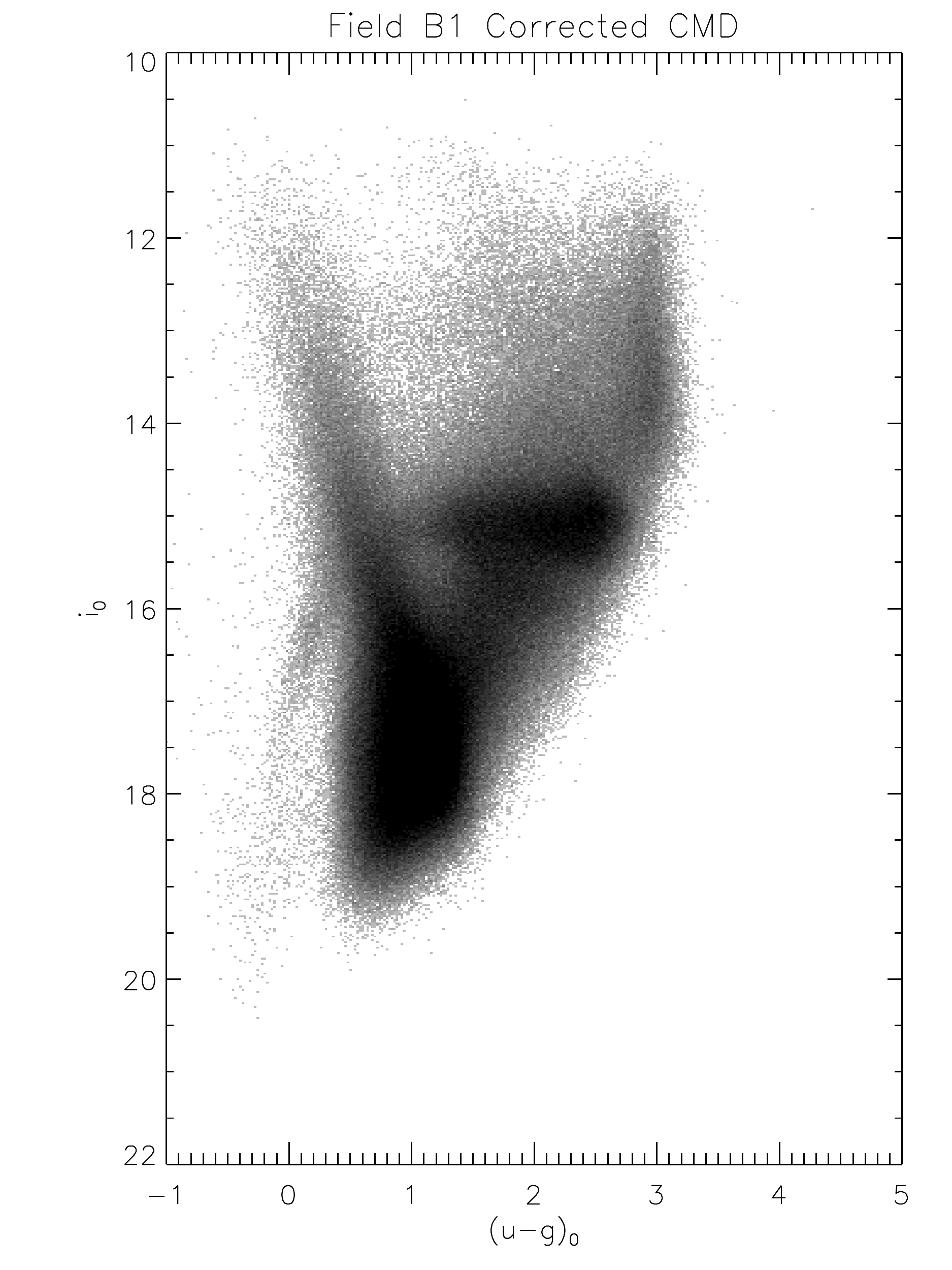}
\caption{Same as Figure~\ref{fig:cmd_rmz}, but for $i$ vs $(u-g)$. The faint cut-off in the $u$ band, which is much brighter than for the redder bands, explains the the sharp oblique truncation on the lower right side of the CMD. Note how the clump stars spread to the blue due to intrinsic properties 
manifest in these passbands. See \S~\ref{sec:CMDfeatures} for further discussion}
\label{fig:cmd_umg}
\end{figure*}

There are in all $9,623,873$ distinct possible stellar objects in the
master-list of all objects in the field B1, as described in
\S~\ref{sec:photometry}.  All available measurements in all epochs in
each passband were evaluated for the rejection criteria enumerated in
\S~\ref{sec:variability}, and if 3 or more such measurements in each
band survived the cut, they were averaged. Because of the extreme
crowding in the fields, there is a rather severe elimination of faint
objects, which are measured cleanly only in the best seeing and
deepest images. In all there are a little over 2.5 million stars for
which we have average magnitudes with this preselection in all of
$g,r,i,z$, and 906,449 where average mags in all 5 bands are
available.  The observed color-magnitude diagrams, with different
colors in the abscissa but using $i$ mags in the ordinate for all
cases, involving various (but not exhaustive) combinations of the
passbands are shown in the left panels of Figures~\ref{fig:cmd_rmz},
\ref{fig:cmd_gmi}, and \ref{fig:cmd_umg}. Differential reddening and
extinction contribute to a washed out appearance: most notably the red
clump giants are smeared out along the reddening line for the redder
abscissae, while for $u-g$ vs $i$, where the clump's extension into
the blue is an intrinsic feature, the reddening vector is no longer
recognizable by the structure of the clump.

%\begin{figure}[h]
%\epsscale{0.45}
%\plotone{cmd_umg_i.pdf}
%\caption{Color magnitude diagram $i$ vs $(u-g)$ is shown. the $u$ band effective faint limit is much brighter than for the other redder bands, 
%for which reason there is a sharp truncation in the lower right side of the diagram. The extension of the clump into the horizontal branch is discernible in this CMD, since $u-g$ stretches out the bluer features.}
%\label{fig:cmd_umg}
%\end{figure}

\subsection{Correcting for reddening and extinction using the RR~Lyrae stars}
\label{sec:justRRLs}

As the reddening and extinction to the individual RRab's are established as described in \S~\ref{sec:minlightreddening}, we can apply these derived values to other stars close to them along the line-of-sight.  However, we can see from the patchiness in the star counts on the images (especially in the $u$ band) that the extinction varies on angular scales of an arc-minute.  The image of the full field was subdivided into rectangular bins, 30\arcsec\ on a side (reason for the choice of bin size is explained below).  If a bin contains one or more RRab's from Table~\ref{tab:fitparams}, we assign the reddening and extinction from the RRab (averaging if there is more than one in a bin) to all the stars in that bin.  For the first pass, stars in bins without an RRab are ignored.  The resulting corrected CMD is shown in the left panel of Figure\ref{fig:rrdered}.  Most notably, the clump stars no longer show the signature of differential extinction as they do in the uncorrected corresponding CMD in the left panel of Figure~\ref{fig:cmd_gmi}, showing that the method works as it should. However, there are only 31,804 stars in this CMD, compared to over $ 2.5 \times 10^{6} $ stars that define the one in Figure~\ref{fig:cmd_gmi}, which is only 1.2\% of all stars with adequate photometry.  The corrected CMD is also over-represented by RR~Lyrae stars because of how it was constructed (only bins containing an RRab were used). The clump of stars near $(g-i)_{0} \approx 0.0$ and $i_{0} \approx 15.5$ are thus the over-represented RRab's.  Making the bins bigger increases the number of surrounding stars, but the angular structure of the differential reddening prevents us from using bins larger than than 60 arcsec, before deterioration from the differential effects becomes apparent.  Clearly, there are too few RR~Lyraes to directly de-redden all the stars in this way: we would need 50 times or more of them to do so.

 We resort to a secondary method, anchored to the RRab's.  The right panel of Figure~\ref{fig:rrdered} shows the dereddened color-color diagram 
of the same stars that are in the left hand panel.  The primary shape of the distribution of stars in this color-color plane is an extension along a direction that is almost degenerate with the reddening vector (shown in the figure with a dashed line). However, the star counts at various points on the color-color diagram locus provide a third dimension, and there is a lot of structure in the relative counts of stars, so that it is in effect a  de-reddened color-color histogram (CCH) of stars.  We assert that, however complex the stellar population components may be along the line-of-sight, this CCH is self similar across the entire B1 field.  Thus, the distribution shown in Figure~\ref{fig:rrdered} is the measured intrinsic CCH, made up of multiple sub-samples taken from over 460 locations randomly scattered across the DECam field-of-view.  However, it is affected by the faint cut-off, which varies across the 4 passbands used, and, because the faint cut-off for the {\emph {dereddened}} colors, varies from place to place depending on the reddening and extinction.  The former affects all field areas equally, so should not adversely affect what we are about to do, but the latter could affect us if there are features in the CMD near the faint cut-off.  Since the faint cut-off is on the the main-sequence of bulge stars, and below the turn-off, variation in the cut-off of intrinsic magnitudes affects the CCH by changing the histogram value at the cut-off colors. 
We show later, from a diagnostic from the de-reddening procedure described below, that this is fortunately not a problem in the present case.\\

\subsection{Correcting for reddening and extinction using color-color histograms}
\label{sec:useCCH}

\begin{figure*}[htb!]
\epsscale{0.9}
\plottwo{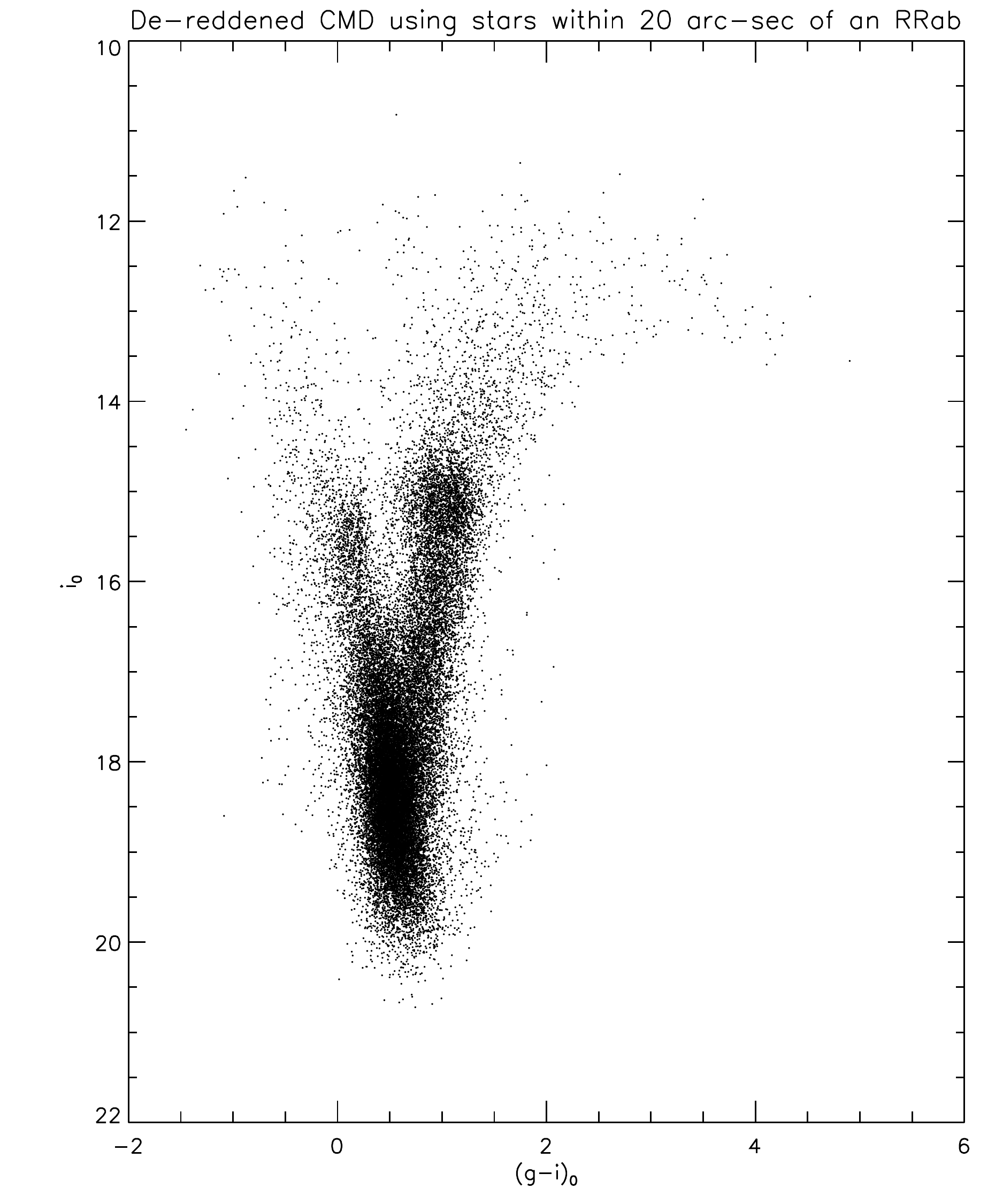}{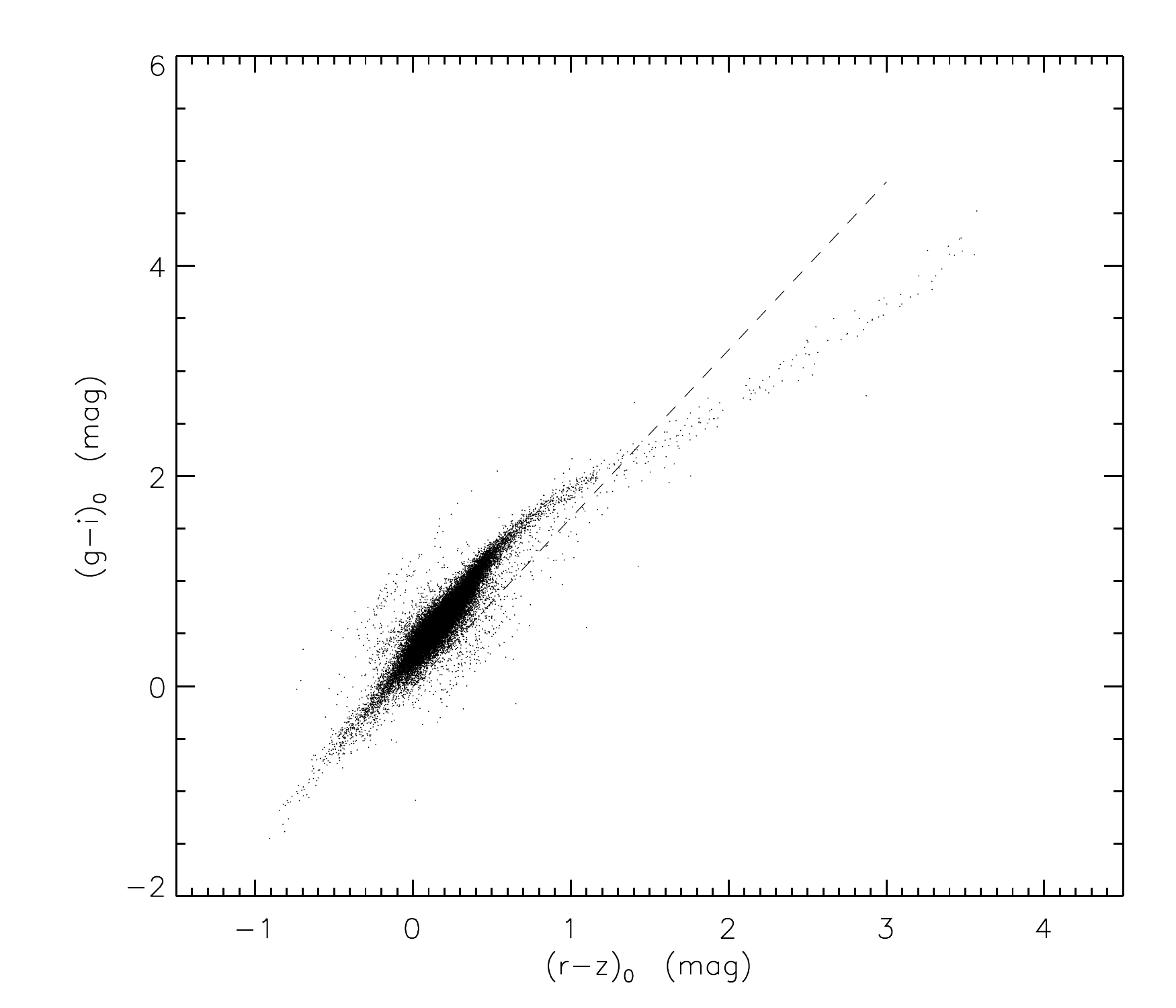}
\caption{Extinction corrected CMD in $i$ vs $g-i$ for stars around
  RRab's in Table~\ref{tab:fitparams} (left panel), using the extinction correction derived from the RRab. Note how the clump stars have lost their extension, and form into a well defined blob. The right panel shows the reddening corrected color-color-diagram using the same stars corrected for reddening and extinction in the same way. The dashed line is the reddening vector with the slope from Equation~\ref{eqn:Egmi}. See text for further details.}
\label{fig:rrdered}
\end{figure*}

We constructed a CCH using the stars shown in Figure~\ref{fig:rrdered} in the $r-z$ and $g-i$ color-color plane, with bin sizes of 0.02 mag along both axes.  This represents the dereddened CCH that we have asserted applies to all sub-regions within the $\sim 3$ square degree DECam field. We denote this as the reference CCH.  Consider the CCH constructed in the same way, but with uncorrected magnitudes and colors from any line-of-sight bin in the field. This should differ from the reference only to the extent of a translation in colors corresponding to the reddening of that field in $E(r-z)$ and $E(g-i)$.  These shifts can be evaluated by a cross-correlation in the two color axes, i.e., by determining the values of $E(r-z)$ and $E(g-i)$ that provide the best match to the reference CCH.  It is possible to force a one-axis cross-correlation by demanding that the reddening obeys Equation~\ref{eqn:Egmi}, but allowing both color-excesses to be derived simultaneously provides an important cross-check.

\begin{figure*}[htb!]
\epsscale{1.1}
\plottwo{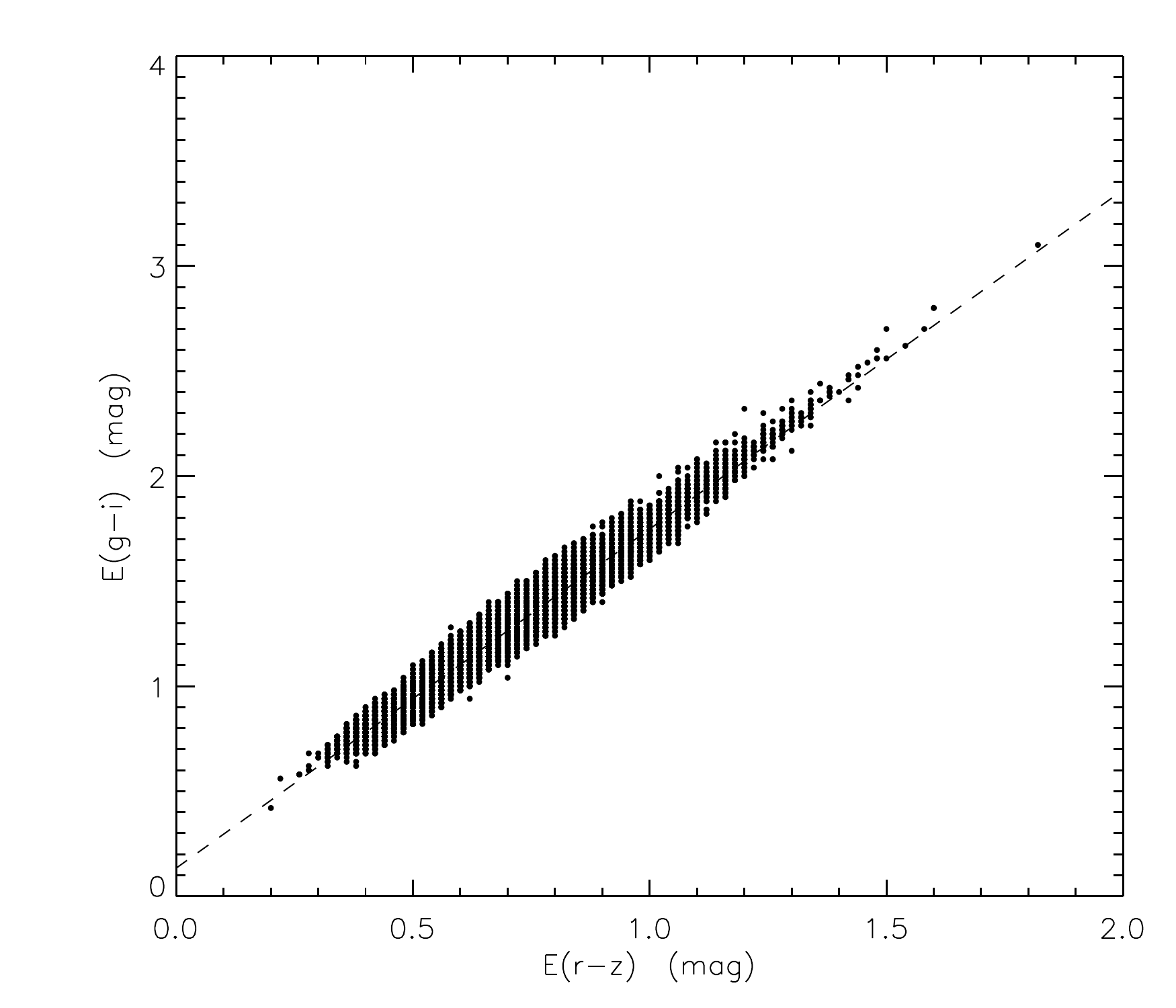}{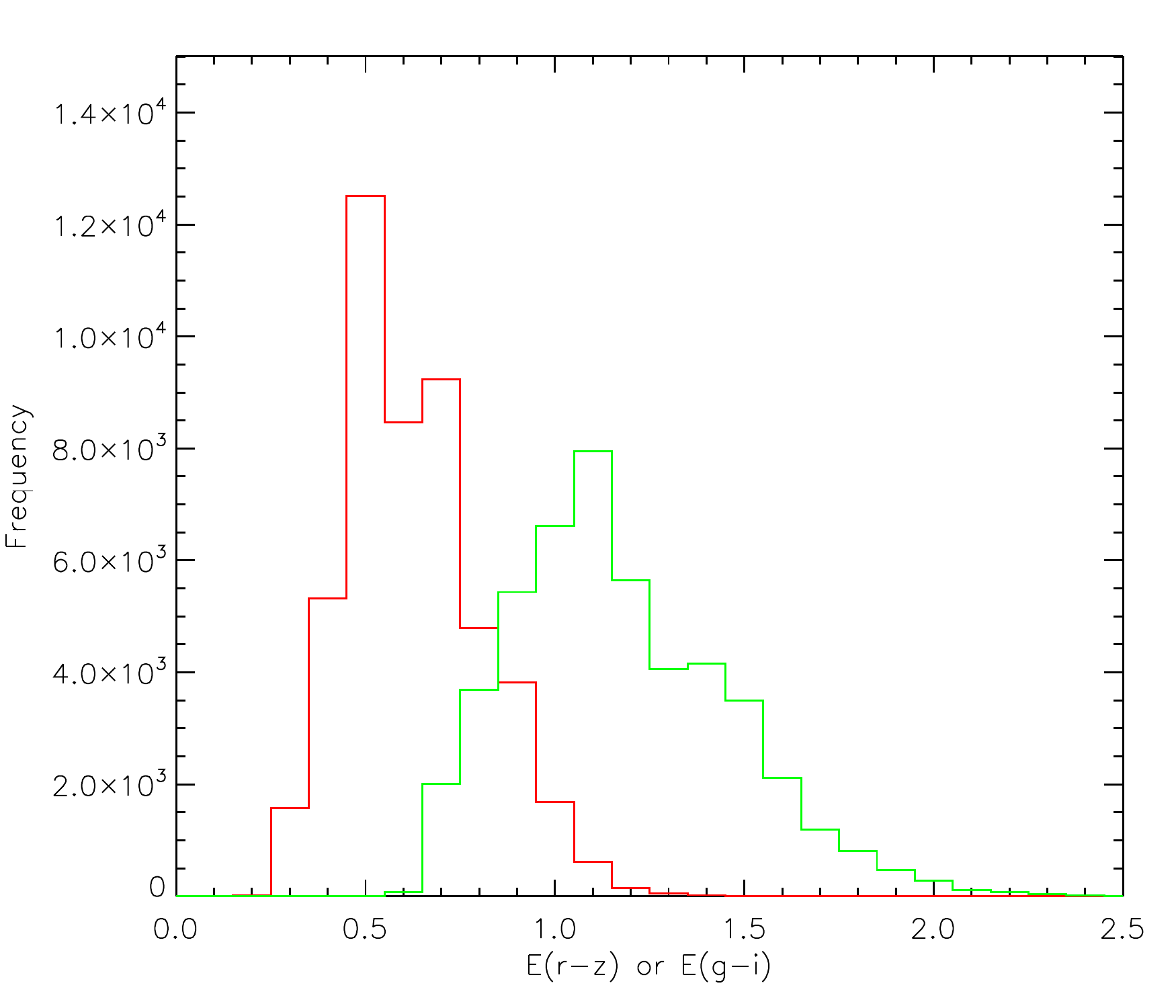}
\caption {In the left panel each point shows the derived $E(r-z)$ and $E(g-i)$ along a $30\arcsec \times 30\arcsec$ sight line within the B1 field, computed using the procedure described in \S~\ref{sec:useCCH}. The dashed line is {\emph not} a fit to the points, but represents Equation~\ref{eqn:Egmi}, which is derived solely from the RRab's.  In the right panel, the histograms of reddening values over the field area are shown: red for $E(r-z)$ and green for $E(g-i)$.  }
\label{fig:ccresult}
\end{figure*}

\begin{figure}[htb!]
\epsscale{0.9}
\plotone{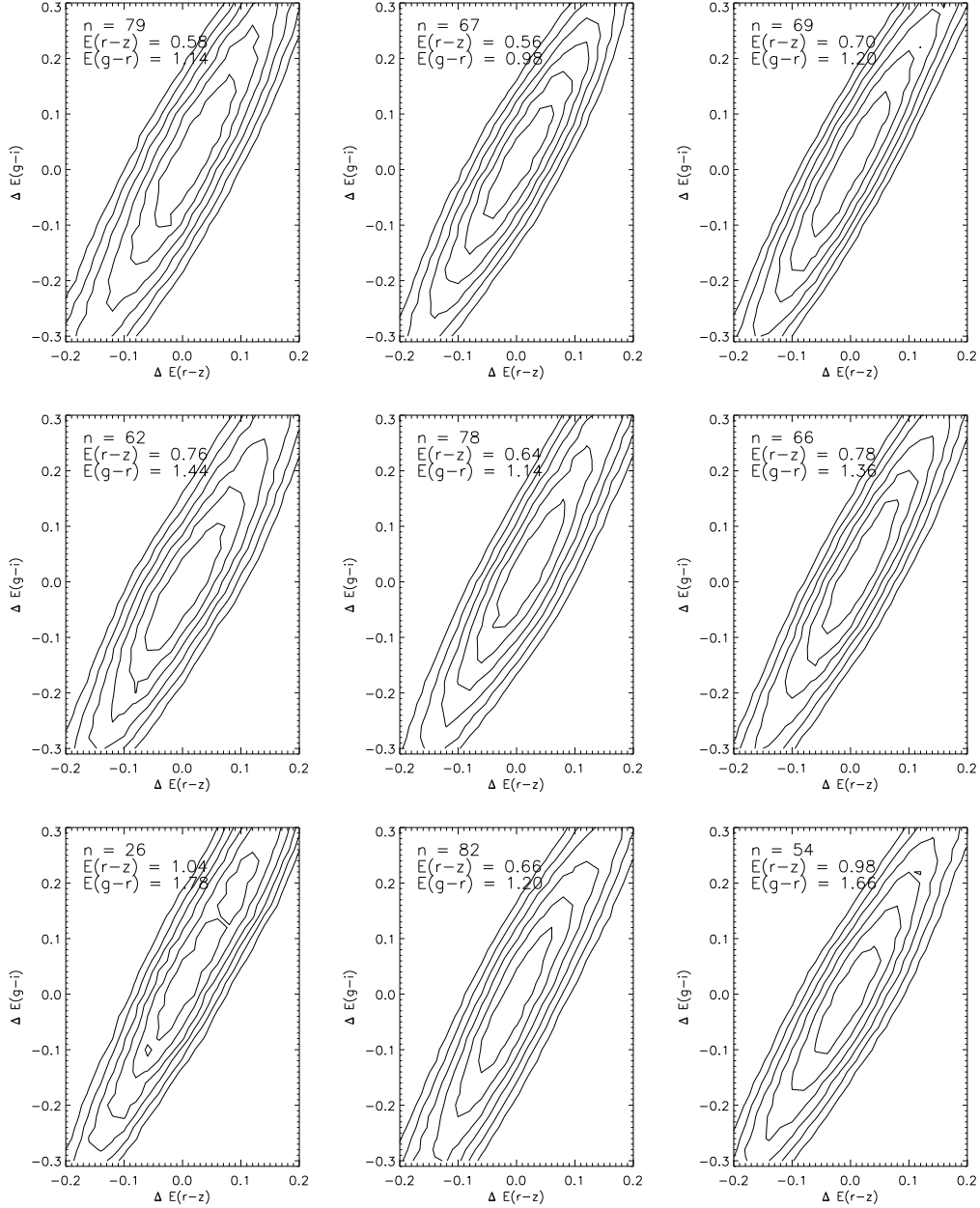}  
\caption{Contours of peaks in the cross-correlation matrix in the $E(r-z)$ - $E(g-i)$ plane for 9 (out of over 37,000) randomly selected 30 arcsec square angular bins, used in the color-color histogram matching procedure described in \S~\ref{sec:useCCH}.  The contour innermost contour is at 90\% peak value, proceeding outward at 10\% intervals. Each example is annotated with the derived values of $E(r-z))$, $E(g-i)$, with the contours showing the correlation matrix centered about those values. The number of available stars, $n$, in the bin is also shown.  Except when $n$ is very small, the contour levels have a common center, and are evenly shaped. The elongation in the contour ellipses is due to the degeneracy of the reddening vector with the shape of locus of stars in the color-color histogram. Uncertainties in centering on the peak appear to be consistent with the scatter seen in Figure~\ref{fig:ccresult}. }
\label{fig:ccmat}
\end{figure}

The result of this exercise is illustrated in Figure~\ref{fig:ccresult}. Each point represents the derived value of $E(r-z)$ and $E(g-i)$ for one of nearly 40,000 line-of-sight bins. As mentioned above, no external constraint was placed on the interdependence of the two axes. It is therefore very satisfying to see that the outcome is in accordance with Equation~\ref{eqn:Egmi}, which is represented by the dashed line. This is of course expected, because it is an essential ingredient of the reference CCH.  If it were not recovered, it would signal that the procedure for matching the observed CCH of each line-of-sight bin to the reference CCH is not working correctly. Rather, the fact that the slope and spread closely follow that of the left panel of Figure~\ref{fig:EgmivsErmz} assures us that the caveat raised at the end of \S~\ref{sec:justRRLs} is not a manifest problem.
  It is also a diagnostic for ascertaining the optimal line-of-sight
  binning size.  With the 30\arcsec\ bins, there are about 70 stars
  per bin that make up the observed CCH. There are places where there
  are fewer.  For example, when a bin straddles an inter-chip gap. To
  take such situations in stride, a condition was imposed to not use
  any bins where there are fewer than 10 stars with available averaged
  photometry in all of the $u,g,r,i,z$ bands (instead for such bins we
  interpolate the results from neighboring bins).  Smaller bins with
  fewer stars suffer from Poisson noise issues and the equivalent of
  Figure~\ref{fig:ccresult} steadily deteriorates for bins smaller
  than 30\arcsec\ on a side, and show greater scatter. Bins that are
  much larger allow more variation in the reddening within their
  extent: with resulting ambiguity in the cross-correlation. To see
  this we need to examine the two-dimensional structure of the
  correlation function peak, which we have found is often distended
  (or double peaked) along the reddening vector for bin sizes larger
  than 60\arcsec\ on the side.  Our choice of 30\arcsec is guided by
  the desire to maximize the spatial resolution while minimizing the
  effects of Poisson noise from too few stars in a bin. This choice is
  customized for the B1 field.  For other fields with different stars densities and differential reddening structure, the optimal bin-size is expected to be different.  Figure~\ref{fig:ccmat} shows contours of the peaks in the cross-correlation matrix for 9 randomly selected 30\arcsec\ bins.  The peaks are highly elongated along the common direction of the reddening vector and the shape of the color-color locus, but the contour levels point to a common center. The 9 examples sample a range of reddening, as well as number of available stars in the respective CCH.  
  
The individual $E(r-z)$ and $E(g-i)$ values thus determined for each 30 arcsec square line-of-sight can then be used to calculate the reddening in other colors using Equations~\ref{eqn:Eumg} through \ref{eqn:Eimz}.  The extinction values in all 5 bands can be computed using 
the coefficients on $E(r-z)$ in the array of Equations~\ref{eqn:total2selective}, but ignoring the offsets therein.  For each bin, dereddened colors, and extinction corrected magnitudes in all 5 bands for all stars in that bin can be obtained in this way, and the accumulated results for the entire field can be derived.  The resulting CMDs are shown in the right hand panels of Figures~\ref{fig:cmd_rmz}, \ref{fig:cmd_gmi} and \ref{fig:cmd_umg}.
%Figure~\ref{fig:corrCMD}.

%\begin{figure}[h]
%\epsscale{0.9}
%\plottwo{cmd_rmz0_i0.pdf}{cmd_gmi0_i0.pdf} 
%\plottwo{cmd_umg0_i0.pdf}{cmd_umg0_u0.pdf}
%\caption{Same CMDs as in Figure~\ref{fig:cmd_rmz}, but corrected for reddening and extinction. The features are notably sharpened, and are discussed in the text. Note that in the bottom right panel, the RRab's are now arranged in a narrow color strip: the vertical extent represents the line-of-sight distance differences to individual RRab's, with a strong clumping at the imapct radius of the line-of-sight to the Galactic center.}
%\label{fig:corrCMD}
%\end{figure}

\subsection{Salient Features in the Corrected CMDs}
\label{sec:CMDfeatures}
The efficacy of our procedure is immediately clear upon comparing the left and right panels of Figures~\ref{fig:cmd_rmz} to \ref{fig:cmd_umg}. The difference is most striking 
for the $(r-z)$ vs $i$ CMD, where in the corrected version the red
clump stars are gathered into a narrow color range, but with a
vertical extent exceeding 0.5 mag from a combination of distance
spread and possibly from stars of different ages.  Both the lower main
sequence and the sub-giant branch are much narrower in the de-reddened
CMD, as we would expect.  The corrected CMD shows that the red giant
branch (RGB) and any asymptotic giant branch (AGB) stars fan out over
a considerable color range, indicating a wide range of metallicities,
as is already known from spectroscopy of bulge giants \citep[e.g.,][]{schult17}.  The bright plumes of the bluest stars however appear to be {\emph {more}} washed out in the {\emph corrected} CMDs. This is because the reddening estimates are anchored by the RR~Lyrae stars, which are clumped in the bulge, whereas the bright blue stars are 
foreground disk stars, for which the reddening has been overestimated: thus the ```corrected'' colors and magnitudes for these stars are incorrect.  

\begin{figure*}[htb!]
\epsscale{0.8}
\plotone{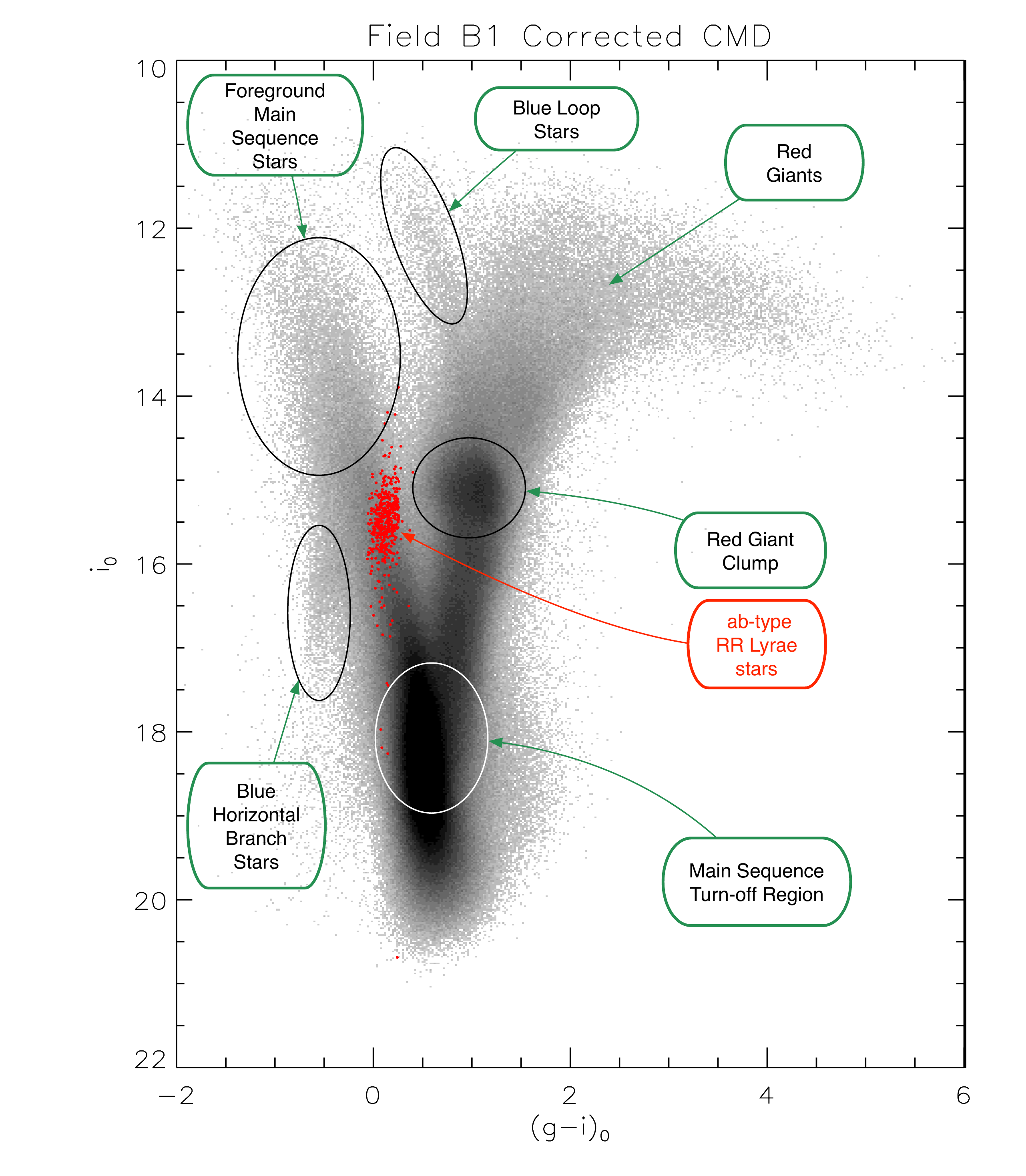}
\caption{Same as the right panel of Figure~\ref{fig:cmd_gmi}, but with labels marking the salient features discussed in \S~\ref{sec:CMDfeatures}.}
\label{fig:CMDlabels}
\end{figure*}

There are two additional curious features in the corrected $i$ vs $(r-z)$ CMD:  a plume of stars extending from the top of the clump star locus, arcing to the blue with increasing brightness ( $ 0 < (r-z)_{0} < 0.5$ and $i_{0} < 14.7$ ) and a sharpish blue edge for the RGB like stars.  These features occupy the expected  location for evolved stars where helium ignition in the core occurs before the core becomes degenerate, and the star ends up either as a red super-giant (that appears here as a pile up of RGB stars along a blue edge) or on the blue extremity of the helium burning ``blue loop.'' However, such locations in the CMD are populated by stars that are more massive than $\sim 2.5$ solar masses, implying that they are relatively young with ages of about 1~Gyr or even less.  The ``blue loop'' track is clearly visible also in the $i_{0}$ vs $(g-i)_{0}$  CMD, but the putative red super giants are indistinguishable from the rest of the RGB/AGB stars. Note that unlike the foreground main sequence stars, the ``blue loop'' plume appears sharper and more tightly bound in color after it is de-reddened, signaling that they are located beyond the distances where most of the reddening takes place. The fact that the structure of the plume continues to stay well bounded in color at all brightness levels, and that it arcs to the blue as it gets more luminous, are arguments that 
they are very unlikely to be foreground red clump stars.  

\begin{deluxetable*}{rrrrrrrrrrrrc} 
\tabletypesize{\scriptsize}
\tablewidth{0pt}
\tablecolumns{13}
\tablecaption{Mean magnitudes of all stars used in the CMD's \label{tab_allstars} }
\tablehead{
  \colhead{RA (J2000)} &
  \colhead{DEC (J2000)} &
  \colhead{$u$ } &
  \colhead{$\sigma(u)$} &
  \colhead{$g$ } &
  \colhead{$\sigma(g)$} &
  \colhead{$r$ } &
  \colhead{$\sigma(r)$} &
  \colhead{$i$ } &
  \colhead{$\sigma(i)$} &
  \colhead{$z$ } &
  \colhead{$\sigma(z)$} &
  \colhead{$E(r-z)$ }     \\ 
  \colhead{(degrees)} &
  \colhead{(degrees)} &
  \colhead{(mag)} &
  \colhead{(mag)} &
  \colhead{(mag)} &
  \colhead{(mag)} &
  \colhead{(mag)} &
  \colhead{(mag)} &
  \colhead{(mag)} &
  \colhead{(mag)} &
  \colhead{(mag)} &
  \colhead{(mag)} &
  \colhead{(mag)} \\   
}
\startdata
    270.554370  &  -31.002110  &  ~~---~~  & ~~---~~  &  20.134  &   0.008  &  18.977  &   0.007 &   18.451  &   0.005  &  18.186  &   0.005   &    0.88 \\
    270.562700  &  -31.001770  &  ~~---~~  & ~~---~~  &  21.355  &   0.026  &  19.687  &   0.032 &   18.845  &   0.018  &  18.457  &   0.020   &    0.88 \\
    270.554960  &  -31.002100  &  ~~---~~  & ~~---~~  &  20.623  &   0.012  &  19.445  &   0.008 &   18.810  &   0.007  &  18.481  &   0.008   &    0.88 \\
    270.561610  &  -31.002010  &  ~~---~~  & ~~---~~  &  21.116  &   0.018  &  19.770  &   0.012 &   19.141  &   0.010  &  18.799  &   0.010   &    0.88 \\
    270.563130  &  -31.002290  &  ~~---~~  & ~~---~~  &  21.582  &   0.025  &  20.244  &   0.013 &   19.733  &   0.017  &  19.428  &   0.011   &    0.88 \\
    270.558400  &  -31.004000  &   18.508  &   0.009  &  16.790  &   0.005  &  16.068  &   0.006 &   15.825  &   0.004  &  15.747  &   0.004   &    0.88 \\
    270.559340  &  -31.005540  &   21.608  &   0.032  &  18.413  &   0.006  &  16.751  &   0.007 &   16.004  &   0.004  &  15.585  &   0.004   &    0.88 \\
\enddata
\tablecomments{Table~\ref{tab_allstars} is published in its entirety in the electronic edition of the journal. A portion is shown here for guidance regarding its form and content.}
\end{deluxetable*}

\begin{figure*}[hbt!]
\epsscale{1.1}
\plottwo{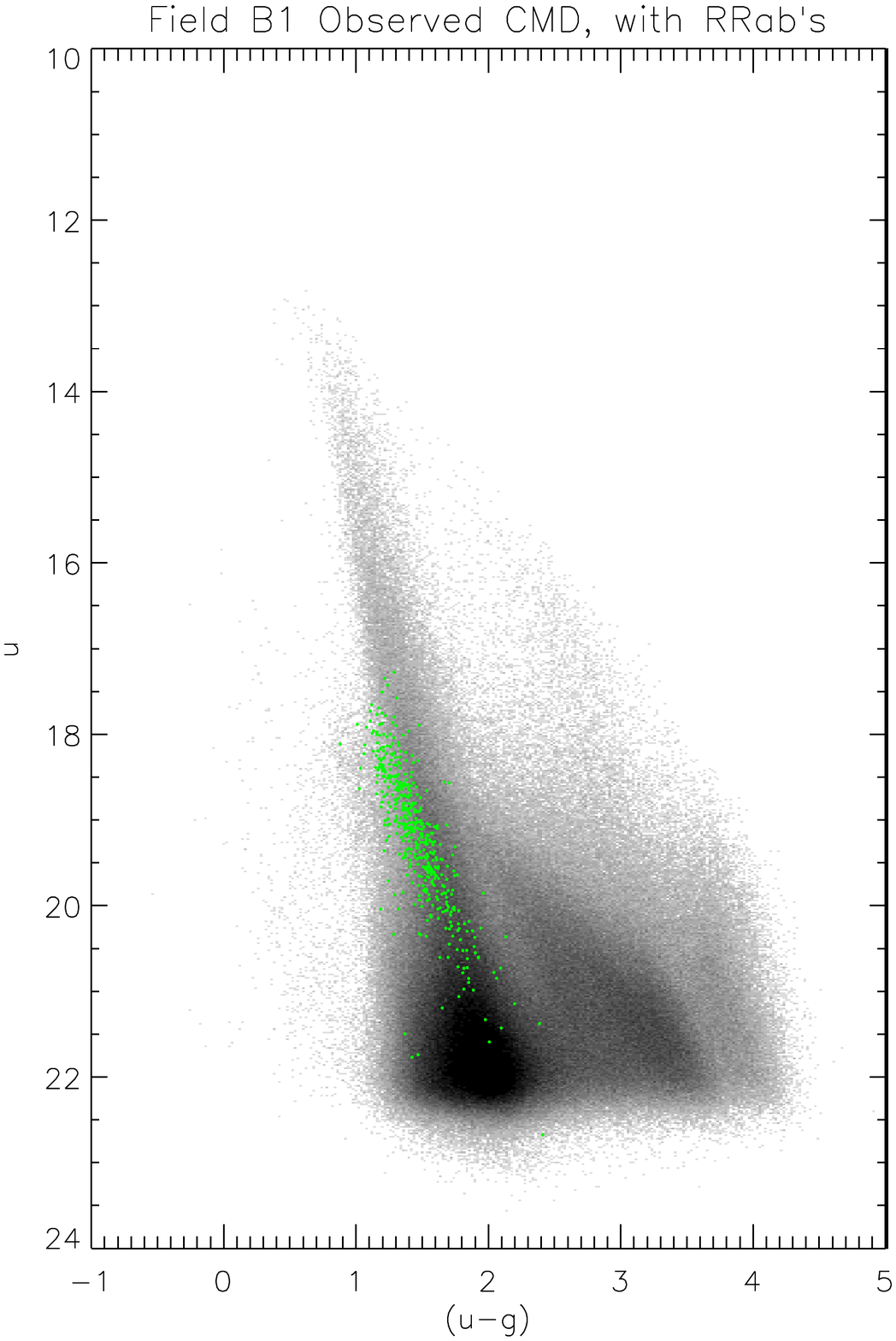}{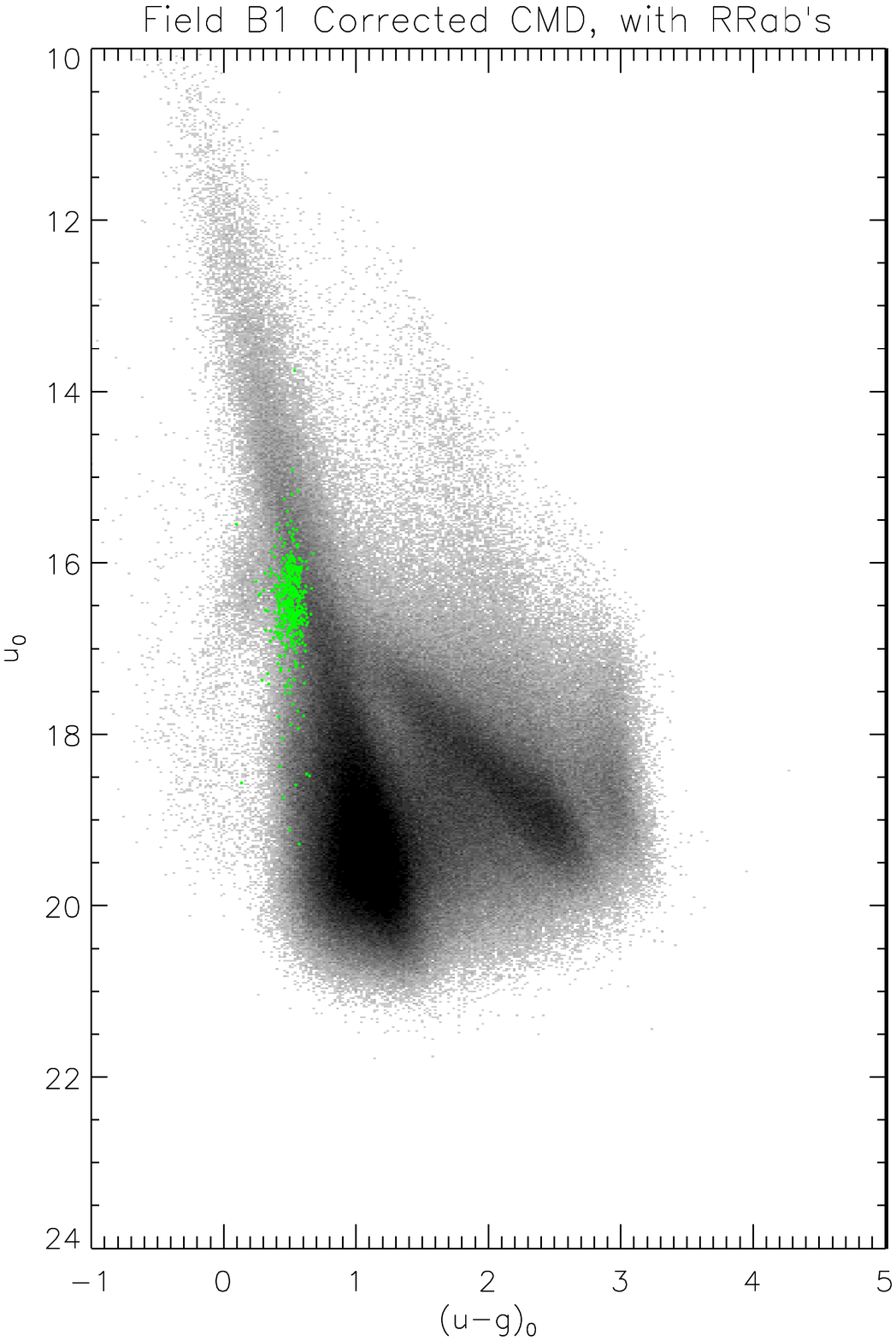}
\caption{The left panel shows the observed CMD of field B1 with $i$ vs $(u-g)$.  The green points mark the mean values for the ab-type RR~Lyraes. Their locus is spread out along the reddening vector, plus some vertical spread from the spread in distances. The right panel shows the reddening and extinction corrected CMD in the same bands. Note how the RRab stars line up within a narrow vertical range, lending credence to the correction procedure employed.}
\label{fig:rrverify} 
\end{figure*}

Figure~\ref{fig:CMDlabels} is a re-display of the right hand panel of Fig~\ref{fig:cmd_gmi}, but with labels pointing out the features mentioned above as they appear on the 
$i_{0}$ vs $E(g-i)_{0}$ CMD. There is a general broadening of features in this plane relative to the $i_{0}$ vs $(r-z)_{0}$  case, consistent with the fact that age and metallicity effects exhibit larger differences as we move to bluer colors.  Nevertheless, the near vertical feature near $(g-i)_{0} = -0.5$  and $i_{0} > 15.8$ is more clearly expressed in the 
$i_{0}$ vs $(g-i)_{0}$ CMD, which is undoubtedly the extension  of the horizontal branch as it ``droops'' in the blue.  Note also that the red clump stars are not as tightly confined in color in $(g-i)_{0}$ as they are 
in $(r-z)_{0}$, very likely because of the metallicity spread among
the stars.  Past attempts to de-redden using the red clump stars
\citep[e.g.,][]{kiraga97}  using colors like $V-I$ would have suffered
from the uncertainty and spread of intrinsic colors among the clump
stars in the bulge.   

The CMD in $i_{0}$ vs $(u-g)_{0}$  is severely
cut-off in the red because the $u$ band sensitivity of DECam as well
as the more severe attenuation due to dust in $u$ imposes a much
brighter faint limit. The pile up of bright red stars against a red
limit near $(u-g)_{0} \approx 3.0$ is likely the result of a red-leak
in the $u$ filter.  The highlight of this version of the CMD is that
it stretches out the track of stars in their post-helium flash phase,
emphasizes the color extension of the red clump, prominently traces
the entire extension of the horizontal branch, and sharply delineates
the ``droop'' in the far blue range of the horizontal branch.

Figure~\ref{fig:rrverify}  shows the observed (left panel) and corrected (right panel) CMD with $u$ vs $u-g$, which is the bluest possible CMD rendition of our data where reddening and extinction express themselves maximally.  The mean colors and magnitudes of the ab-type RR Lyrae stars are shown by the green points. On the uncorrected CMD, the RRab distribution is extended along the reddening vector, whereas in the 
corrected CMD they are distributed vertically in correspondence to their individual 
distances along the line-of-sight. Notice how the RRab distribution peaks where it intersects the horizontal branch (which in this 
color-magnitude configuration gets brighter in the blue relative to the clump).  This particular representation of the CMD uses the color and 
magnitude most affected by reddening, so this consistency in the outcome of our de-reddening is gratifying. 

Table~\ref{tab_allstars} lists the positions, observed magnitudes, and derived values of $E(r-z)$ for all stars used to produce the CMD's.

\subsection{A Spectroscopic Preview of the ``Blue Loop'' Stars}
\label{sec:blueloopspec}
There are over 1200 stars in the ``blue loop'' feature.  If this is confirmed as such, then this is just the high-mass end of the IMF for stars 
with ages  of order a few hundred Myrs.  Ten of the objects
falling within this blue loop from this sample of stars were
also observed as part of the Apache Point Observatory Galactic Evolution
Experiment (APOGEE), which is one of the experiments from SDSS III/IV
\citep{abol18}.  APOGEE is a high-resolution spectroscopic 
(R=22,400) survey in the near-IR ($\lambda$=1.51-1.70$\mu$m) which targets,
primarily, red giants from all Galactic populations; it is planned to
have observed $\sim$500,000 stars by 2020 \citep{maj17, holtz18, jonss18}.  
Survey results from APOGEE include
stellar parameters (effective temperature, $T_{\rm eff}$, surface
gravity (as log g), and microturbulent velocity), precise radial velocities,
and detailed chemical abundance distributions from, typically, 15 elements.
These results are derived from an automated analysis package called the
APOGEE Stellar Parameter and Chemical Abundance Pipeline \citep[ASPCAP; ][]{gper16}.

The 10 red giants observed by APOGEE that are included in this study have
ASPCAP calibrated parameters derived in the latest SDSS public Data Release 14 (DR14\footnote{https://www.sdss.org/dr14/irspec/}).  A separate paper (Smith et al.,
in preparation) will present a detailed analysis of these 10 stars, while
DR14 ASPCAP results will be discussed here.  The effective temperatures and
surface gravities of these red giants are consistent with their being
core-He burning giants, with a small range in $T_{\rm eff}$ and log g: the
mean values and their standard deviations are $T_{\rm eff}$=4765$\pm$110K
and log g=2.6$\pm$0.2.  The metallicities of this sample
of red giants are interesting, as all are quite metal-rich, with values of
[Fe/H] from $\sim$+0.1 to +0.4, which places them as likely members of
the bulge, based on the observed distribution of metallicities of APOGEE bulge stars \citep[e.g.,][]{gper18, zas19}.  The
mean value for the 10 giants is [Fe/H]=+0.25 $\pm$ 0.10.

Of interest to this study are values of the carbon-to-nitrogen ratios, C/N,
in these core-He burning stars.  Early stellar evolution models \citep[e.g.,][]{iben64} predicted that the C/N ratio in red giants, after the completion
of the first dredge-up, will depend upon the stellar mass.  The relation
between C/N and red giant mass is due to the deep convective envelope of a
red giant, which mixes material to the stellar surface that has undergone
H-burning via the CN-cycle, where $^{12}$C has been partially processed
into $^{14}$N, leading to lower values of C/N relative to the main-sequence
values.  More massive red giants have both deeper convective envelopes, as well
as higher internal temperatures, so the increase in the surface $^{14}$N 
and decrease in the surface $^{12}$C abundances are larger, resulting in
lower values of C/N with increasing red giant mass.  \citet{mart16}
have recently calibrated the relation between C/N and red giant mass, using
a combination of APOGEE spectra and Kepler asteroseismology \citep{pins14}, resulting in
mass estimates with rms errors of $\sim$0.2M$_{\odot}$.  The change in
C/N is largest between about 1-2M$_{\odot}$, making it useful for
estimating ages over a range of about 1-10Gyr.  The values of the C/N ratio
in these 10 red giants display only a small scatter, with a mean value and
standard deviation of $ \langle {\rm [C/N]} \rangle =-0.55 \pm 0.12$; based upon \citet{mart16}
this indicates a mass of $\sim$1.5-1.7M$_{\odot}$ for this sample of bulge
core-He burning giants.  As these particular stars targeted by APOGEE are
not the most luminous of the stars covered in the DECam sample, there are
more luminous, and thus more massive, and even younger members of these clump giants.

\section{The Reddening Map}
\label{sec:redmap}

\begin{figure*}[hbt!]
\epsscale{1.25}
%\plotone{B130egmi.pdf}
\plotone{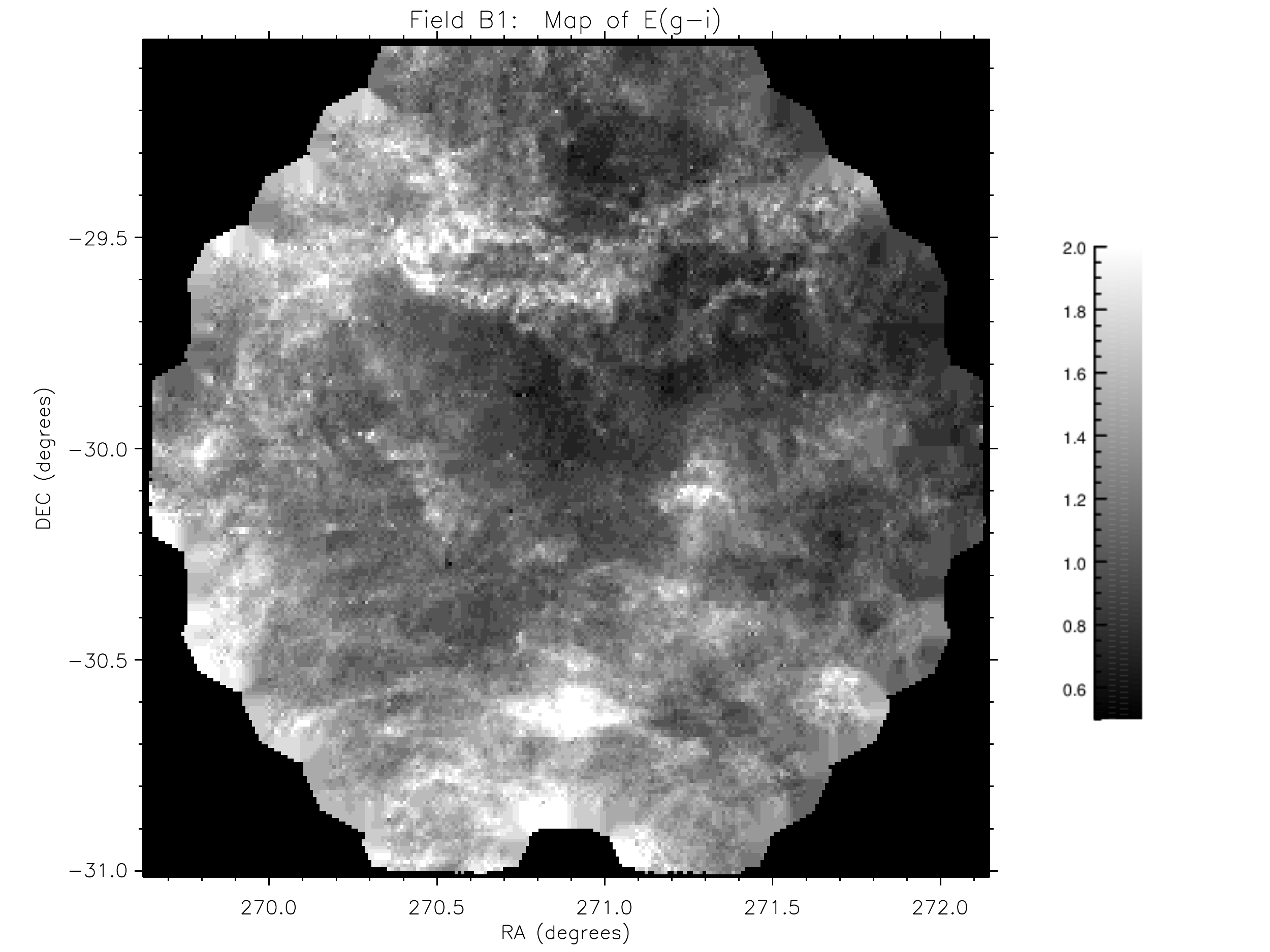}
\caption{Map of reddening values in $E(g-i)$ to the Galactic Bulge derived in this paper from the color-color diagram correlation and the minimum light colors of RRab stars. Dark areas correspond to lower reddening and higher transparency, while lighter areas indicate higher reddening. This map is very similar to the dust map from \citet{sfd98} of the same region, but has at least 10 times better linear resolution, and, while it differs quantitatively, is a good structural match to the 30 arcsec resolution $A_{V}$ map of \cite{stanek96} derived from Red Clump stars. The $E(g-i)$ and $E(r-z)$ maps are provided in FITS format as Data behind the figure.}
\label{fig:map_egmi}
\end{figure*}

The procedure described in \S~\ref{sec:useCCH} produces reddening
values in $E(r-z)$ and $E(g-i)$ for each 30 arcsec square cell over
the field of view of DECam, except where there are too few stars with
reported mean magnitudes in $g,r,i,z$.  These maps can be interpolated
to bridge gaps (where there are too few stars).  FITS images of these
maps (with WCS  encoding of RA and DEC) are available as Data behind
the Figure as well as in a Github
repository\footnote{https://github.com/akvivas/Baade-s-Window}.  The map of $E(g-i)$ is shown in Figure~\ref{fig:map_egmi}.

The central part of the reddening map in Figure~\ref{fig:map_egmi} shows relatively higher transparency without too much spatial variation in the reddening, and corresponds to the area chosen by \citet{baade46} to peer close to the Galactic center. There is considerable patchiness outside this central window, with blobs and filamentary structures from arc-minute scales on up to the better part of a degree.  Also visible are ring shaped structures with relatively low contrast scattered over the entire field that appear to be shells of dust. They range in size from about 5 to 15 arc-minutes in diameter.  These are likely to be ejecta from massive stars driven out by winds.  
Given the crowding in the field, and irregularities in the shapes of the rings, we are unable to find any unambiguous
 visual correspondence of the ring centers with bright stars. We wonder whether having very short lives, such stars have long disappeared, but we are not in a position to know how long the bubbles would last before they dissipate.

\section{Tracing the Bulge Geometry with the fundamental mode RR Lyraes}
\label{sec:geom}

\subsection{The Distance to the Galactic Center}
\label{sec:distance}

Consider a heliocentric Cartesian coordinate system, where the $Z$ axis points to the Galactic center, the $X$ axis is in the direction of the 
Galactic longitude $l$, and the $Y$ axis points towards the North Galactic Pole (NGP).  The projection on the $Z$ axis of a point in space at a distance $d$ from the Sun with Galactic coordinates $l$ and $b$ is then:
\begin{equation}
z = d \cos l \cos b
\label{eqn:zproject}
\end{equation}
The volume density in space of the RRab's of Table~\ref{tab:fitparams} in $z$ is expected to peak at the distance $R_{0}$ to the Galactic center provided the spatial distribution of the RRab's is spherical. However, for very small values of $l$ and $b$, the effects from non-sphericity in the distribution are small. For the stars in field B1, with direction centered at $l < 2.05^{\circ}$ and $|b| < 5.0^{\circ}$, the product of the cosine terms in Equation~\ref{eqn:zproject} differ from unity by less than 0.5\%, which mitigates any effects from moderate azimuthal and polar asymmetries in the density distribution.  In fact, for B1, the peak in the distribution of $d$ is by itself  a measure of $R_{0}$ to within a percent  if no azimuthal or polar asymmetries are present.

\begin{figure*}[hbt!]
\epsscale{1.1}
\plotone{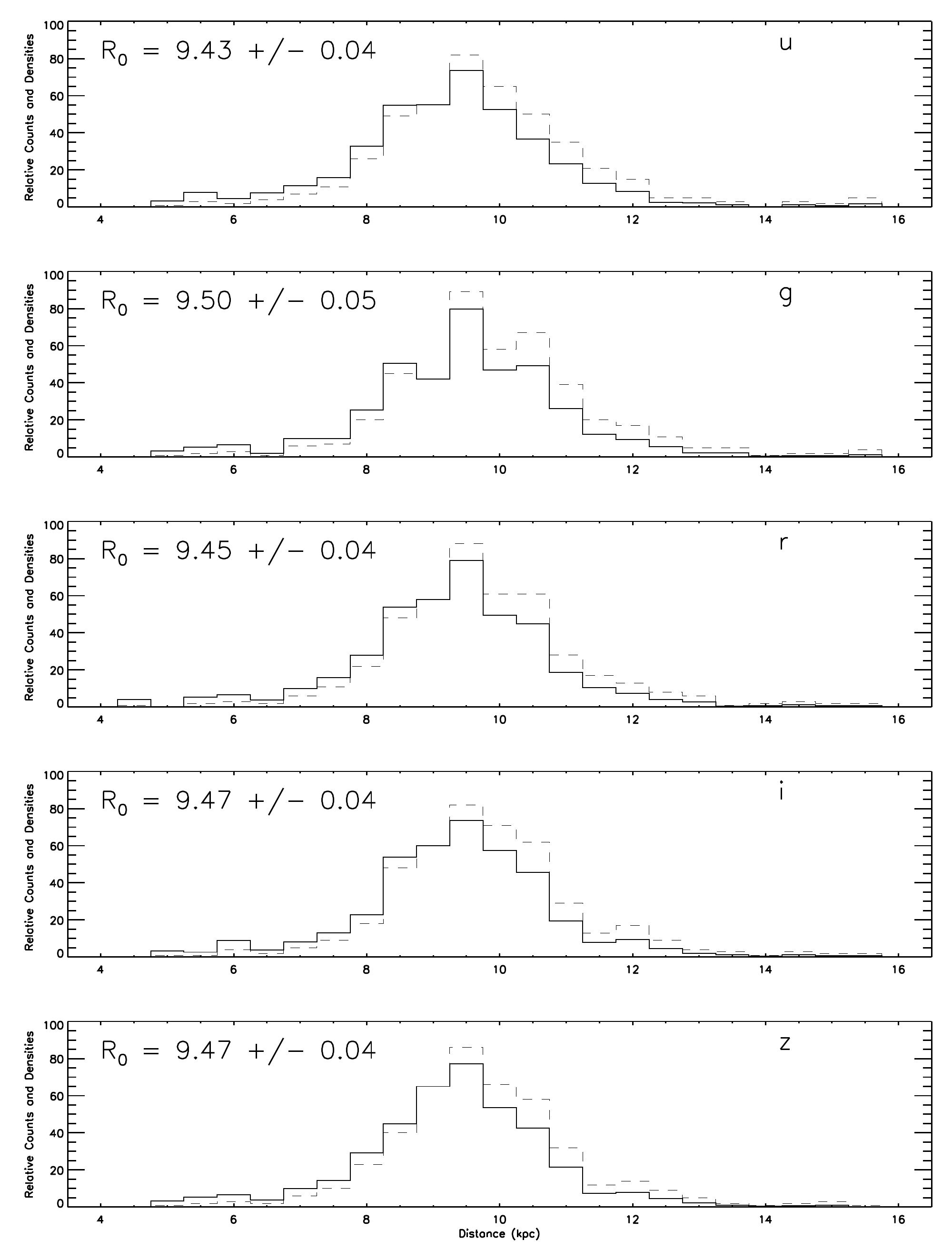}
\caption{The distribution in distance of the RRab stars from Table~\ref{tab:fitparams} calculated from data in each of the 5 passbands, and using the reddening law derived in this paper. The dashed lines show the histogram of star counts, while the bold line shows {\emph relative} density by correcting for the larger sampled spatial volume at larger distance as was done for Figure~\ref{fig:disthist_odonnell}.  
Contrary to what was seen in Figure~\ref{fig:disthist_odonnell}, the results for the 5 passbands (labeled in the figure) are remarkably concordant, though surprisingly large compared with extant values of $R_{0}$ in the literature. Moreover, the density distribution with distance is sharply peaked, symmetrical, and very similar in all bands. This reinforces the need for the custom derivation of the reddening law for this line-of-sight provided in this work.}
\label{fig:disthist}
\end{figure*}

Since we have the individual reddenings (Equation~\ref{eqn:rmz0} combined with observed minimum light colors from Table~\ref{tab:fitparams}) and extinctions (using  Equations~\ref{eqn:t2s0} to \ref{eqn:t2s3}) as well as mean observed magnitudes $m_{X}$ in any band $X$ for the individual RRab's in Table~\ref{tab:fitparams}, we can calculate their extinction corrected mean magnitudes $m^{0}_{X}$. We get their absolute magnitudes $M_{X}$ from Equation~\ref{eqn:absmags} using the period for the individual star from Table~\ref{tab:fitparams}, which yields the distance modulus $ DM_{X} = (M_{X} - m^{0}_{X})$ for stars for each of the 5 passbands.  The distance $d_{X}$ to an individual star calculated from data in the $X$ band is then given by: 
\begin{equation}
\label{eqn:distkpc}
d_{X} ~{\rm (kpc)} ~=~ 10^{0.2(DM_{X} - 10)}
\end{equation}
The distribution of $d_{X}$ for the RRab's in Table~\ref{tab:fitparams} is shown in Figure~\ref{fig:disthist} for each of the 5 bands using the dashed lines. 
The solid line shows the relative number density, in $0.5$ kpc bins, by accounting for the change in the sampled volume with distance.  
For each band, the value of $R_{0}$ is determined by finding the location of maximum density, calculated as the 
``center of mass'' from the 5 bins centered on the bin with the peak density.  The results for the $u,g,r,i,z$ bands (labeled in the figure) are remarkably concordant, though surprisingly large compared with extant values of $R_{0}$ in the literature. Moreover, the density distribution with distance is sharply peaked, symmetrical, and very similar in all bands.
The quoted uncertainties reflect \emph{only} the errors in finding the centroids in the histograms: systematic errors are discussed below separately.  The mean of the $u,g,r,i,z$ based results yields:
\begin{equation}
R_{0} = 9.47 \pm 0.04 ~{\rm kpc}
\label{eqn:rnought}
\end{equation}
where we do not reduce the uncertainties from the individual passband measurements because the departures from the respective centroids  are highly correlated.  Again the quoted uncertainty is only the random error estimated from the widths of the histogram peaks in Figure~\ref{fig:disthist}.

Our derived distance is at odds with the literature.  A good compendium of determinations of $R_{0}$ up to 2015 is available from \citet{degrijs16}, including different kinds of tracers, statistical parallax methods, and analysis of the kinematics of stars near the Galactic nucleus.  There are multiple reported values of $R_{0}$ from 7 to 9  kpc.    After homogenization of the various determinations, they arrived at a statistical determination of $R_{0} = 8.3 \pm 0.2 ({\rm statistical}) \pm 0.4 ({\rm systematic})~{\rm kpc}$.  By any account, our result presented here is about 10 percent higher than the norm.   
%We have shown above that this difference can be explained entirely by our use of the extinction law presented in this paper over the standard extinction % law.

The derived distances in each band depend on the $A_{X}/E(r-z)$
(slope) values in Equations.~\ref{eqn:t2s0}~to~\ref{eqn:t2s3}. Note
that while the derivation of these equations assumes a strong clumping
of distances of the RRab's, it does not place any external constraint
that the clump distance has to be identical across bands. That is, the intercepts in Equations.~\ref{eqn:total2selective} are determined independently from one band to another.  There are several possible reasons why the derived value of $R_{0}$ here is larger than similar determinations from RR~Lyrae stars in the past, but the three most pressing ones are the following:
\begin{enumerate}
\item
Our derived reddening is different from the standard reddening law, and as seen in Figure~\ref{fig:redlaw}, predicts lower extinction in $g,r,i,z$ than the standard reddening curve, making the corrected magnitudes fainter than what the standard law would give, and thus resulting in a larger distance.  Our result deserves some further scrutiny.
\item
We have adopted the absolute magnitudes derived for the globular cluster M5 in \citet{vivas17} to apply also to the RR~Lyrae in the Galactic bulge. If the RR Lyrae are different, for instance different Oosterhoff types, a distance discrepancy could result.  We discuss this issue below.
\item
The distance determination to M5,  and hence the inferred absolute magnitudes of the RR~Lyrae may be incorrect.
\end{enumerate}

\begin{figure*}[hbt!]
\epsscale{0.9}
\plotone{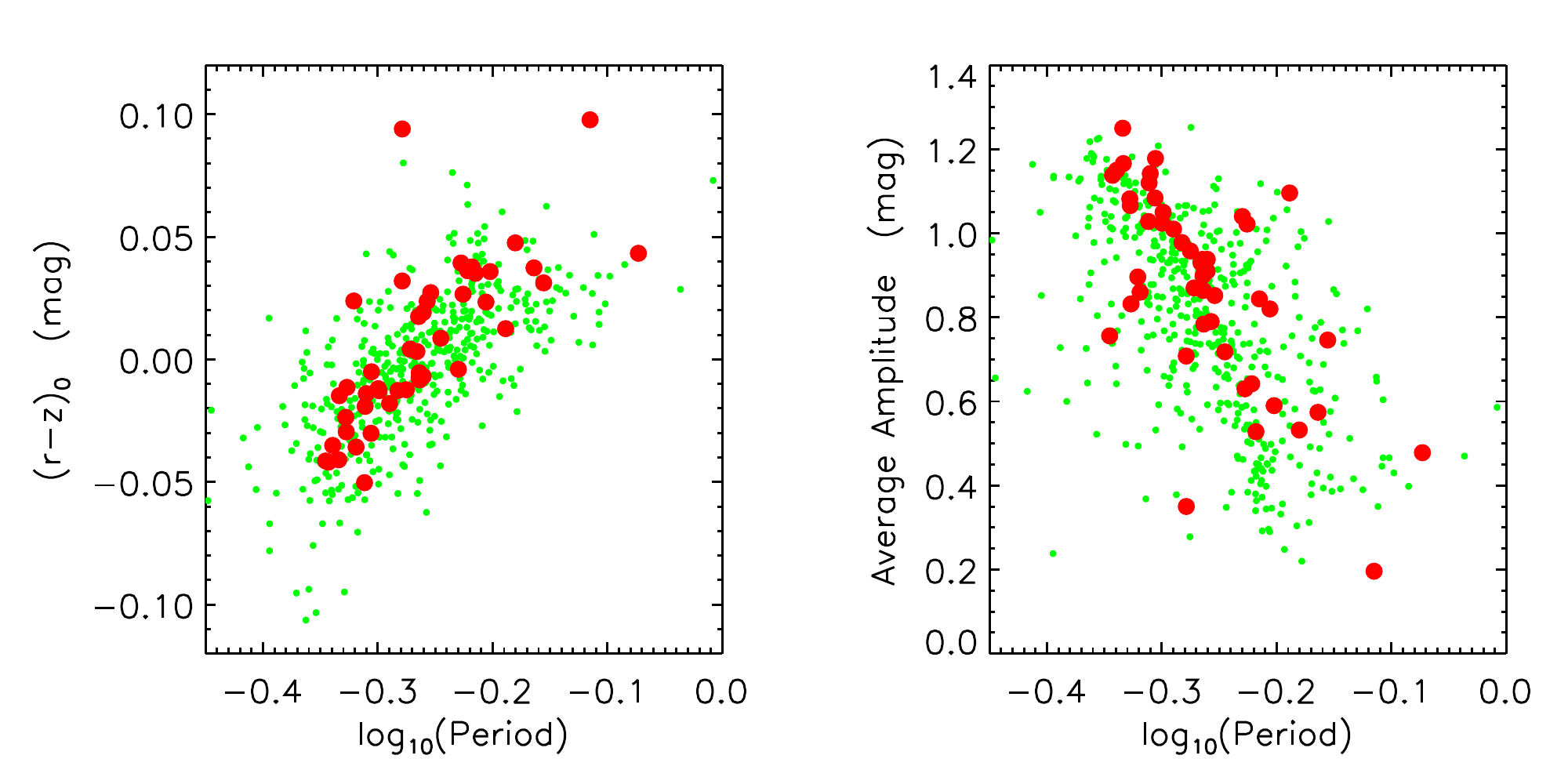}
\caption{Comparison of the Period-Color and Period Amplitude Relations for the RRab's in Field B1 (green dots) and the globular cluster M5 (red filled circles).  The agreement in the P-C and P-A distributions of both sets bolsters the assertion that the luminosity distributions of the RRab's for both instances are also the same, as argued in the text}
\label{fig:percolamp}
\end{figure*}

%\begin{figure}[h]
%\epsscale{0.45}
%\plotone{RRabdilemna.pdf}
%\caption{The histogram of the dereddened $g-r$ mean colors of the RRab stars:  a) the solid line shows the ditribution obtained by using the minimum light color method in this paper and governed by  Equation~\ref{eqn:gmi0}  and b) the dash-dot line using the standard reddening law.  In the latter case, the colors are too far red by 0.3 mag, and the distribution has a larger spread. Both are indications that the standard law is incorrect for this field.}
%\label{fig:RRabdil}
% \end{figure}
We examine these possibilities in turn in some more detail.  As mentioned above, our finding in this paper is that the standard reddening law is violated in the direction of our field. For a given reddening, be it $E(r-z)$, $E(g-i)$ or even $E(V-I)$, the extinction for all bands other than $450\, {\rm nm}$ is smaller with the reddening law derived here, compared to the standard formulation, as seen in Figure~\ref{fig:redlaw}.  Thus, for a given adopted absolute magnitude (in this case based on the adopted distance to M5), our reddening law yields larger distances compared to the standard law.  This is clearly seen in the comparison of Figures~\ref{fig:disthist_odonnell} vs. \ref{fig:disthist}. Note specifically that in the $i$ band, the standard law yields $R_{0} = 8.04~{\rm kpc}$ from Figure~\ref{fig:disthist_odonnell}, whereas Figure~\ref{fig:disthist} using the reddening law derived here gives $R_{0} = 9.47~{\rm kpc}$.  This 18\% difference in distance is exactly explained by the difference in total to selective extinction for the $i$ band given by equation~\ref{eqn:stdreddening} versus equation~\ref{eqn:t2s0}, given that the mean $E(r-z)$ in Field B1 is $\sim 0.68$.
The distances in all passbands when our derived reddening law is used are in agreement, whereas use of the standard law produces disparate distances across the passbands. This validates the derivation of our reddening law through Equation~\ref{eqn:total2selective}. 
%First, we can confidently rule out the first possibility, as we have shown in two different ways (in \S~\ref{sec:minlightreddening}) that 
%the standard reddening law is unacceptable in light of our multi-band RRab photometry. 
%This is illustrated in Figure~\ref{fig:RRabdil}.
%The solid histogram shows the distribution od mean de-reddened $g-r$ colors when the condition in Equation~\ref{eqn:gmi0} is imposed on the minimum light colors.  The dash-dot histogram shows the distribution when the standard extinction law is used, and produces mean RRab colors that are too red, and a distribution with a larger intrinsic mean color spread. In effect, in the second case we are using $E(r-z)$ to predict 
%the intrinsix $g-r$ colors using the standard extinction law for the RRab stars, and we find that it results in incorrect values. To our knowledge, this is the first time such a test has been applied, since photometry in more than two bands of RR Lyraes in the bulge has not previously been available. 
The departure from the
standard reddening law in fields in the Galactic bulge has previously been reported by \citet{Nataf16} and \citet{Nataf13}, who did a similar analysis with OGLE $V,I$ and VISTA $J,K_{S}$ photometry of red clump giants.  
%However Nataf et al. did not derive the total to selective extinctions independently for their bands, and so did not present an equivalent re-derivation of distance.
%The color of RRab's at minimum liight is much better understood (see discussion in \citet{vivas17}) than those of red clump stars: the latter have
%dependencies on metallicities and possibly ages. 

The luminosity of RR~Lyraes is less universal than the minimum light
colors, and can depend on metallicity, post zero-age horizontal branch
evolution, and on helium abundance. As remarked earlier, the average
metallicity distribution of  RR~Lyrae stars in Baade's window peaks at
${\rm [Fe/H]} \approx -1.0$ as shown by \citet{walker91} (see their
Figure~7).   The metallicity of the globular cluster M5 is ${\rm [Fe/H]} =
-1.25 \pm 0.05$ according to \citet{dias16}.  This similarity
notwithwithstanding, small differences in helium abundance can drive
much larger differences in luminosity.  A more robust test for
differences can be had through the Period-Luminosity-Temperature
($PLT$) relation.  Eddington's  pulsation equation $P \langle \rho \rangle^{1/2} = Q$
and its refinements \citep[e.g.,][]{vanalbada71}, in combination with
a mass-luminosity relation for any class of stars that share a common
evolutionary state, implies the existence of a $PLT$ relation for that
stellar class.  For RR~Lyraes on the zero-age horizontal branch
(ZAHB), it means that stars with the same $P$ and $T$ should have the
same luminosity $L$.   This precept has been used to examine the cause
of the Oosterhoff dichotomy \citep[e.g.,][and references therein]{sandage90}.  Using the dereddened mean color $(r-z)_{0}$ as a proxy for the effective temperature $T$, we compare the mean Period-Color
relations for RRab in the globular cluster M5 and in our Field~B1.  The left hand panel of Figure~\ref{fig:percolamp} shows that there is no net period shift at the same intrinsic color between M5 and the B1 field RRab's, thus implying that their luminosities are also at par.  The average amplitude (the mean of amplitudes from all 5 bands, which is most robust against measurement errors) as a function of $\log P$ is shown on the right hand panel: amplitudes have also been used in the literature as a proxy for temperature, but have been deprecated \citep[and references therein]{sandage90}.  The left hand plot is predicated on our determination of reddenings, while the right hand one is reddening independent but its quality as a proxy for temperature is less secure.  In both plots we see no indication of a difference between the RRab's in M5 vs. those in the B1 field, which supports the contention that the luminosities are the same for the RRab's in both locations.  Thus the second of the above possibilities is not a strong contender either.

This leaves us with the question of whether the calibration of absolute magnitudes through M5 could be in error.
The adopted distance modulus to M5, which is inherited from
\citet{vivas17} through Equations~\ref{eqn:absmags}, is based on main
sequence fitting. \citet{layden05} derived
 \begin{equation} 
\label{eqn:m5dist}
\mu_{0} [M5] = 14.45 \pm 0.11
\end{equation}
 \citet{layden05} also fitted the white dwarf sequence in M5 to the local white dwarfs with parallax distances, and obtained $\mu_{0} = 14.67 \pm 0.18$ which is both, more distant (which would make the RRab's brighter, and $R_{0}$ even larger), and more uncertain.  They report the average apparent $V$ magnitude of ab-type RR~Lyraes in M5 to be $15.025 \pm 0.011$ (from their Table~4). Correcting for extinction ($E(B-V)=0.035$ and standard extinction law), 
\begin{equation}
\label{eqn:Vmagm5}
\langle m^0_{V} \rangle [M5RRab]  = 14.92 \pm 0.01 
\end{equation}
which implies that the intrinsic $V$ band average absolute magnitude for RRab stars in M5 is:
\begin{equation}
\label{eqn:absVM5}
M^0_{V} [M5RRab] = 0.47  \pm 0.11
\end{equation}

A very recent determination of the distance to M5 by \citet{gont19} gives $(m-M)_{0} = 14.34 \pm 0.09$ by multi-band isochrone and main-sequence fitting, which, if adopted, would decrease $R_{0}$ to 9.04 kpc.  in the \emph{Gaia} era, we should look to astrometric measurements for a more definitive distance determination.
From \emph{HST} parallax measurements of field RR~Lyrae stars, \citet{benedict11} obtain 
\begin{equation}
\label{eqn:absparallaxM5}
M_{V}  = 0.50 \pm 0.05 
\end{equation}
for the M5 globular cluster metallicity of ${\rm [Fe/H]} = -1.25$.  This has a smaller formal uncertainty than Equation~\ref{eqn:absVM5}, while the $0.03$ mag smaller distance modulus to M5 implied by the  parallax based RR~Lyrae  absolute magnitudes is within the uncertainties, and yields 
%This result is in agreement with the \citet{mura18} \emph{Gaia} DR2 parallax analysis of nearby RR~Lyraes.
%Adopting this slightly smaller distance modulus, and incorporating the uncertainty from Equation~\ref{eqn:absparallaxM5} for the absolute magnitudes of %RRab's and allowing for 0.04 mag in the determination of colors from our observations, which propagate a $3\%$ uncertainty in the distance estimate, %we can modify Equation~\ref{eqn:rnought} to read:
\begin{equation}
\label{eqn:benedictr0}
R_{0} = 9.45 \pm 0.30  ~{\rm kpc}
\end{equation}
%This is our formal result for the distance determination to M5, based now on the RR~Lyrae absolute magnitude scale of \citet{benedict11}.     

It is known that the \emph{Gaia} DR2 results suffer from systematic errors in the measured parallax that vary with position in the sky as discussed in \S~4.2 and 4.3 of  \citet{arenou18} and by \citet{lindegren18}.  The reported parallax of 0.1135 mas for M5 \citep{helmi18} is thus likely to be an underestimate by a few tens of micro-arcsecs. It is expected that future \emph{Gaia} data-releases will ascertain and correct for this systematic error, but at the moment a direct distance to M5 based on published \emph{Gaia} parallaxes is not reliable. 

We can try instead to go through the \emph{Gaia} DR2 parallax distances to calibrate the absolute magnitudes of RR~Lyrae stars that are much closer to us than M5. 
\citet{mura18} present such an analysis for 401 RR~Lyraes, which include objects that still are at distances of several kpc, and so suffer from systematic uncertainties mentioned above.  They also present a restricted sample 
of 23 RR~Lyrae which have particularly well determined metallicities, but this sample too is not made of solely the nearest objects, and thus is not free of the parallax systematics. From the results in their Table~4 which gives a linear correlation of $M_{V}$ with ${\rm [Fe/H]}$ that is consistent with an LMC distance modulus of 18.5, we read:
\begin{equation}
\label{eqn:mura_absmag}
M_{V} = (0.26^{+0.05}_{-0.05}){\rm [Fe/H]} + 1.04^{+0.07}_{-0.07}
\end{equation}
Using  ${\rm [Fe/H]}  =  -1.25$ for the metallicity of M5 \citep{dias16}, we get 

\begin{equation}
\label{eqn:DR2absmagM5}
M_{V} = 0.72 \pm 0.07
\end{equation}

Combining Equations \ref{eqn:Vmagm5} and \ref{eqn:DR2absmagM5} we get a revised distance modulus to M5 of 
\begin{equation}
\label{eqn:mu0M5new}
\mu_{0} [M5] = 14.20 \pm 0.07
\end{equation}

This corresponds to a reduction of all distances by a factor of $ 0.89 \pm  0.03 $, implying that Equation~\ref{eqn:rnought} is changed to 

\begin{equation}
\label{eqn:finalr0}
R_{0} = 8.44 \pm 0.28  ~{\rm kpc}
\end{equation}

which is consistent at the 1-$\sigma$ level with the recent determination of distance to the central black-hole of $7.93 \pm 0.13$ kpc \citep{chu18} and $8.13 \pm 0.03$ kpc \citep{abuter18} from the analysis of the orbit of the star S2 around the central black hole.

The \citet{mura18} result includes data for the type-c RR~Lyraes, which we have avoided in our analysis in the bulge. For this reason we  
have independently analyzed the data for the 41  type-ab RR~Lyraes from the \citet{mura18} sample that have \emph{Gaia} DR2 parallaxes greater than 1 milli-arcsec, and so are least affected by the systematic uncertainties in the \emph{Gaia} parallax zero-point. Using the reported magnitudes and extinction estimates from their Table~1, we derive the equivalent of Equation~\ref{eqn:mura_absmag} for this sample (rejecting the 3-sigma outlier AT And) to be:
\begin{equation}
\label{mura_absmag_resample}
M_{V} \approx 0.35{\rm [Fe/H]} + 1.06     ~~~~~~ (rms = 0.13 ~{\rm mag})
\end{equation}
which yields $M_{V} \approx 0.62$ for the M5 RR~Lyraes.  This is brighter than the \citet{mura18} result, but significantly and definitely fainter than from Equation~\ref{eqn:absVM5}.   As \citet{mura18} have pointed out, there are many selection effects to worry about from such {\it ad hoc}  sample selection. The point of the exercise is to establish that there is enough uncertainty in the calibration of the RR~Lyrae absolute magnitudes that 15\% errors in distance determination are easily possible, and our derived value of $R_{0} = 9.47~{\rm kpc}$ based on the \citet{layden05} main sequence fitting distance to M5 
awaits modification at a future time when the \emph{Gaia} mission gives us a parallax distance to M5, or better yet, directly to the RR~Lyrae (tracers of the oldest stars) in the bulge.   For the present, we continue the discussion with our derived value in Equation~\ref{eqn:rnought} for the purpose of the remaining analyses of the structure of the bulge in this paper, noting that all quantitative distances will scale linearly with any change from $R_{0} = 9.47~ {\rm kpc}$.

\subsection{Dereddening and Distances to the OGLE RRab's}
\begin{figure}[ht!]
\epsscale{0.9}
%\plotone{DECAM2OGLE.pdf}
\plotone{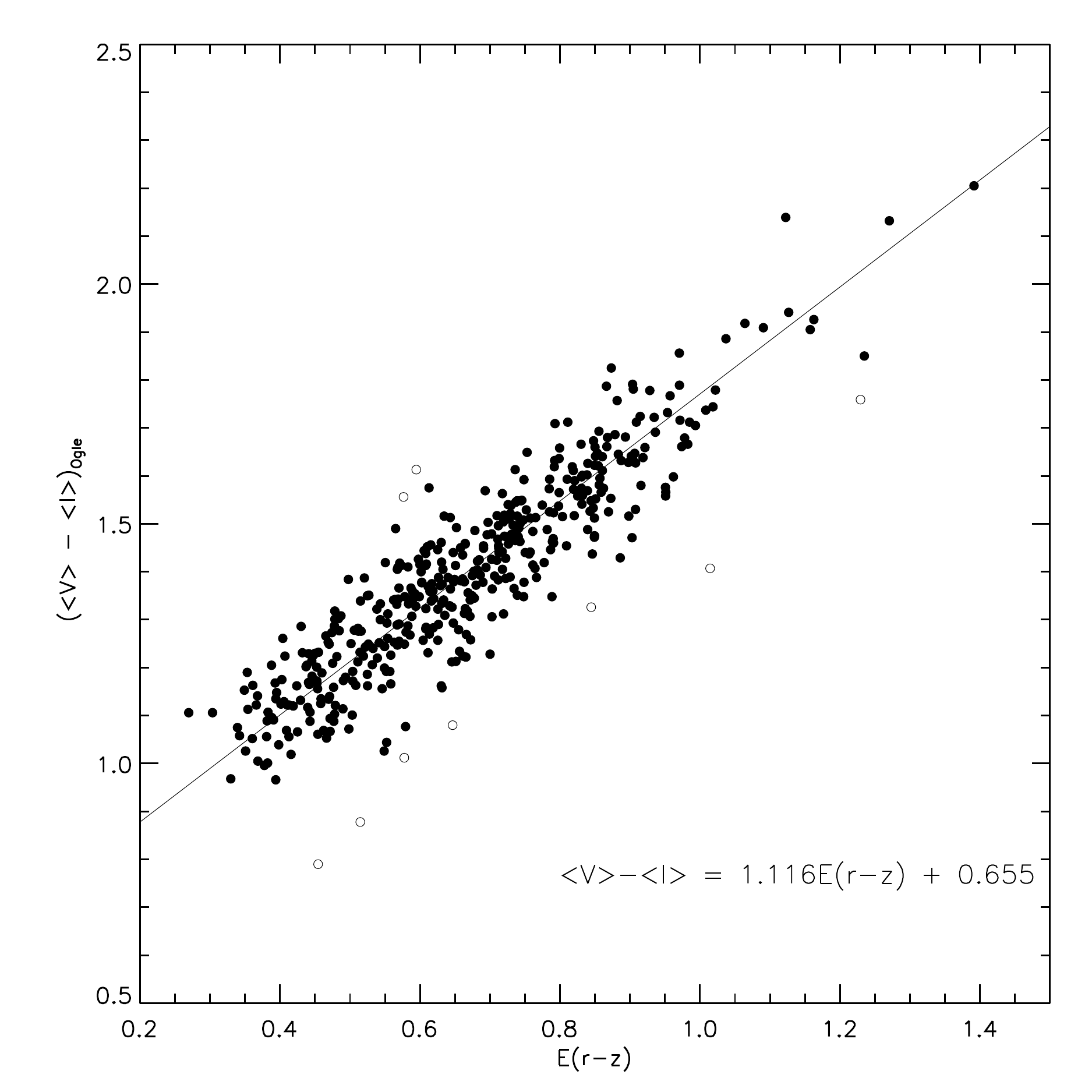}
\caption{OGLE colors of individual RRab's cross-identified in Table~\ref{tab:fitparams} (from their mean magnitudes) are compared to their
corresponding  $E(r-z)$ values determined in this paper. A good correlation is seen. The scatter ( 0.08 mag in $ \langle V \rangle - \langle I \rangle $) is much larger than the uncertainties in determining $E(r-z)$, which is a reflection of the fact that colors at minimum light are less variant from object to object than 
colors from mean magnitudes. The open circles indicate points that were rejected from the fit to a straight line}.
\label{fig:DEC2OGLE}
\end{figure}
While determining minimum light colors for the bulge RRab's in the OGLE catalog is made difficult because of the paucity of $V$ band measurements, we examine the possibility of using their mean $\langle V \rangle $ and $\langle I \rangle $  magnitudes to estimate reddening to individual RRab's. We utilize the 472 RRab's in Table~\ref{tab:fitparams} for which we have cross-matches to the OGLE-III catalog, from which we obtain their $\langle V \rangle$ and $\langle I \rangle$ values, and relate them to our values for $E(r-z)$. This is shown in Figure~\ref{fig:DEC2OGLE}.  We derive the following relation
\begin{equation}
\label{eqn:DEC2OGLE}
{   \langle V \rangle  -  \langle I \rangle  ~~=~~ 1.116 (\pm 0.027) E(r-z) + 0.655 (\pm 0.018) ~~~~  [ \sigma = 0.08 {\rm mag }]  }
\end{equation}
where $\sigma$ indicates the rms scatter in $ \langle V \rangle - \langle I \rangle $ for an individual RRab star.  Compare the derived $\sigma \approx 0.08 $ mag to the accuracy of better than $0.03$ mag with which we can predict $E(g-i)$ from $E(r-z)$ using Equation~\ref{eqn:Egmi}: this is a consequence of using mean mags instead of colors at minimum light.  A bi-variate correlation of the extinction corrected mean $\langle i \rangle$ magnitudes (DECam system) derived in this paper to the observed $\langle V \rangle $ and $\langle I \rangle $  OGLE magnitudes of the corresponding RRab's yields the following relation:
\begin{equation}
\label{eqn:OGLEtoI0}
\langle i \rangle _{0}  ~=~  1.986\, \langle I \rangle_{OGLE}  ~-~ 1.031\, \langle V \rangle_{OGLE} ~+~ 1.765   ~~~~ [\sigma = 0.115~{\rm mag}]
\end{equation} 
which predicts the extinction corrected mean $i$-band magnitude (DECam system used in this paper) using the {\emph {observed}} mean $V$ and $I$ OGLE magnitudes for any RRab star.  The scatter indicates that the prediction is uncertain with an rms of 0.115 mag.  Using this relation we can get distances to individual RRab in the OGLE catalog with a 6\% rms scatter (and additional systematic uncertainties from how well we know the absolute magnitudes of the RRab's).  

\begin{figure}[ht!]
\epsscale{0.9}
\plotone{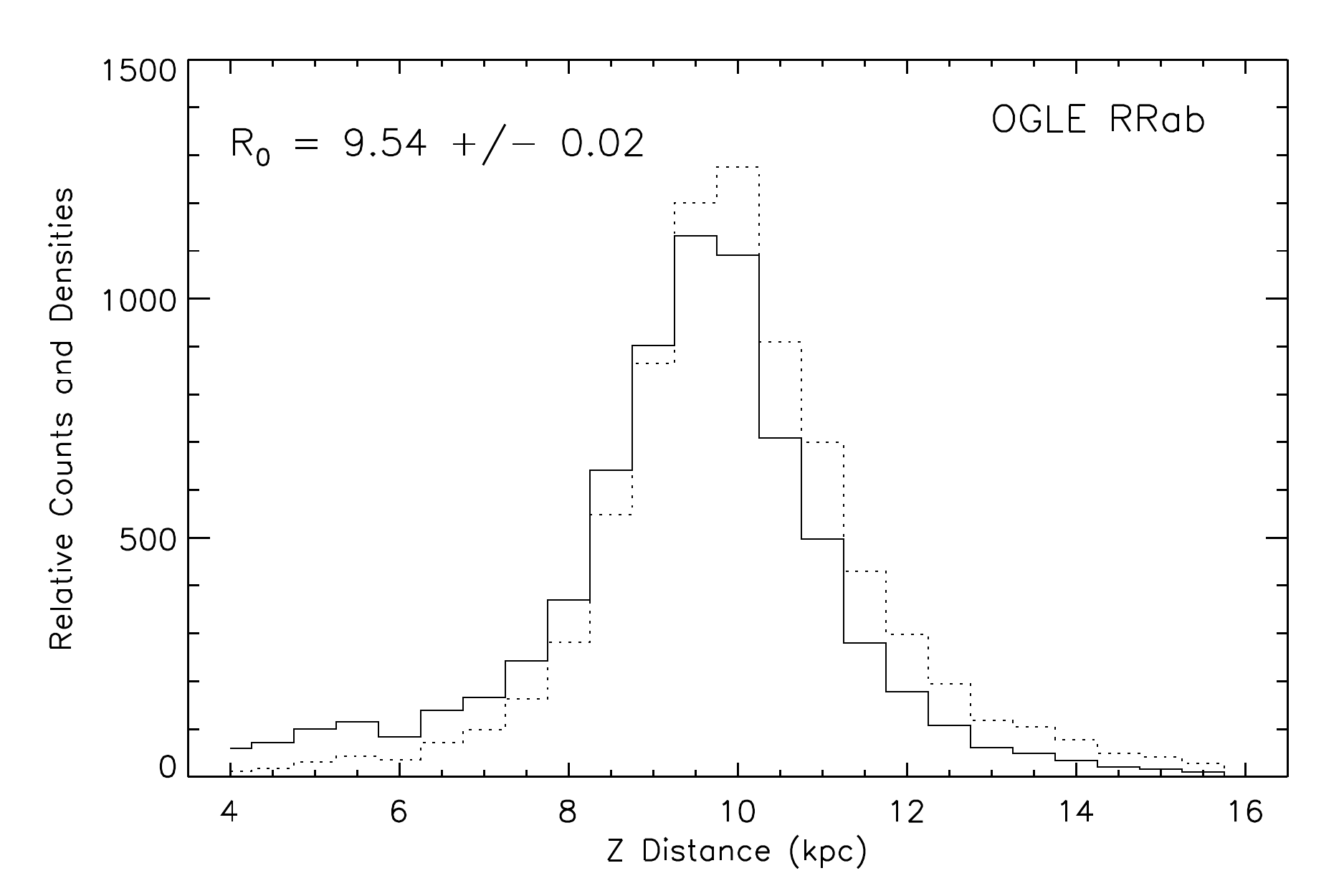}
\caption{The histogram of star counts (dashed line) and RRab relative spatial density (solid line) as a function of projected distance along the direction towards the Galactic center. See text for details.}.
\label{fig:OGdist}
\end{figure}

Limiting ourselves to Galactic longitudes $l$ bounded within $-5^{\circ} < l < +5^{\circ}$ and Galactic latitudes $b$ within $-8^{\circ} < b < 0^{\circ}$ (where 
the line-of-sight looks relatively close to the Galactic center, and the distances are not biased by RRab's in background star streams), we have 8092 RRab's from the OGLE-III catalog.  Calculating their distances using Equations~\ref{eqn:OGLEtoI0} and \ref{eqn:absVM5} we obtain the projected distance $z$ on the scale of Layden's distance to M5 using Equation~\ref{eqn:zproject}.  We construct the histogram of the RRab's for $z$
as shown by the dashed lines in Figure~\ref{fig:OGdist}.  To make the
histogram represent relative star densities, we reconstruct using a
weighted count for each star, where the weight decreases as the
inverse square of the line-of-sight distance.  The solid line in
Figure~\ref{fig:OGdist} shows this modified histogram.  Using the same
peak centroiding methods as used above, we obtain a peak density at $z
= 9.54 \pm 0.02$ kpc, where the quoted uncertainty refers only to the
centroiding error.  If the spatial distribution of the RR~Lyrae stars
is azimuthally symmetric, $z$ is a good estimator of $R_{0}$.  It
differs from Equation~\ref{eqn:rnought} by less than 1 percent. On the
one hand this agreement is only to be expected, because the same
precepts for reddening and absolute magnitudes of the RRab's have been
used for both derivations. However, on the other hand, the sub-sample
from OGLE-III used here has 40 times more stars, spread over a wider
spatial extent, so the agreement validates implicit assumptions
regarding the spatial distribution of the RR~Lyraes.  This result
however is pre-mature in detail.  \citet{Nataf16}
showed that not only is the reddening towards the bulge non-standard, but that it also varies from one line-of-sight to another within the angular scale of the bulge. We might therefore expect that Equation~\ref{eqn:DEC2OGLE}, and therefore Equation~\ref{eqn:OGLEtoI0} will be different for different lines of sight.

With the wider perspective of the OGLE  coverage, it is now possible, in principle, by applying the reddening corrections as done here, to deduce the spatial distribution of the RR~Lyraes near the Galactic center, especially the flattening of the density ellipsoid.  The gaps in coverage in $l$ and $b$, and possible incompleteness along lines of very high extinction thwart the direct calculation of relative densities, and makes this task messy: we do not attempt it here.  The caution above about variation of the reddening law itself along different sight lines also applies to most applications made possible by he OGLE RR~Lyrae data-set in the bulge.  We will be able to ascertain how much Equation~\ref{eqn:DEC2OGLE} and~\ref{eqn:OGLEtoI0} change when we present the analysis for the 5 remaining fields in our study.

\subsection{The Density Distribution of RR~Lyraes in the Galactic Bulge}

\begin{figure*}[ht!]
\epsscale{0.9}
\plotone{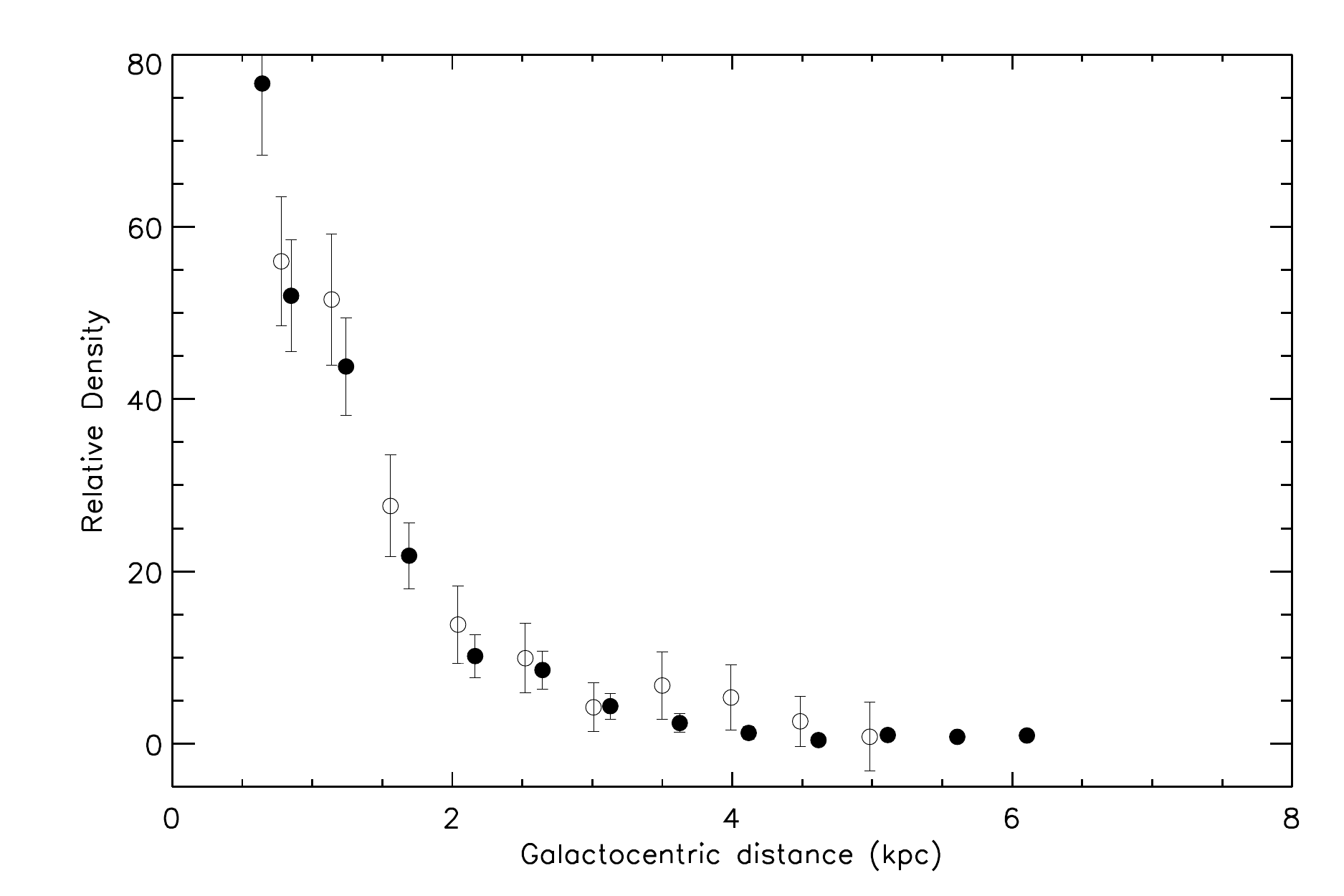}
\caption{Relative density of RR~Lyrae stars as a function of Galactocentric distance in kpc as derived from the RRab's in field B1.  Open circles denote locations where the distance $d$ from the Sun is less than $R_{0}$, and filled circles for $d > R_{0}$. The error bars are calculated using Poisson statistics of the counts of RRab in each ``bin.''}
\label{fig:rdens1}
\end{figure*}

RR~Lyrae are well known tracers of ancient stellar populations.  With over 450 RRab's in our B1 field, whose line-of-sight  passes close to the Galactic center, for which we have well derived distances, and from which we have derived a distance to the center $R_{0}$, we are poised to examine the density distribution of the parent population of ancient stars.  If the distance to an object is $d$, and its Galactic coordinates are $l$ and $b$, then the Galactocentric distance $r$ is given by:
\begin{equation}
\label{eqn:Gcentric}
r^{2} = R_{0}^{2} + d^{2} - 2dR_{0} \cos(l) \cos(b)
\end{equation}

Applying this to the count and density histograms derived in \S~\ref{sec:distance}, and using $R_{0} = 9.47~{\rm kpc}$ from Equation~\ref{eqn:rnought}, we can remap them as a function of $r$.  Figure~\ref{fig:rdens1} shows the distribution: open circles denote values for $d < R_{0}$ and filled circles for $d > R_{0}$.  The error bars are calculated using Poisson statistics of the counts of RRab in each ``bin,'' and dividing by the normalized volume for the corresponding line-of-sight distance. Since the line-of-sight is at non-zero $b$, flattening of the bulge along the polar axis can produce different density values at the same $r$, for locations on the near side, vs. that on the far side of the Galactic center.

\begin{figure*}[ht!]
\epsscale{0.9}
\plotone{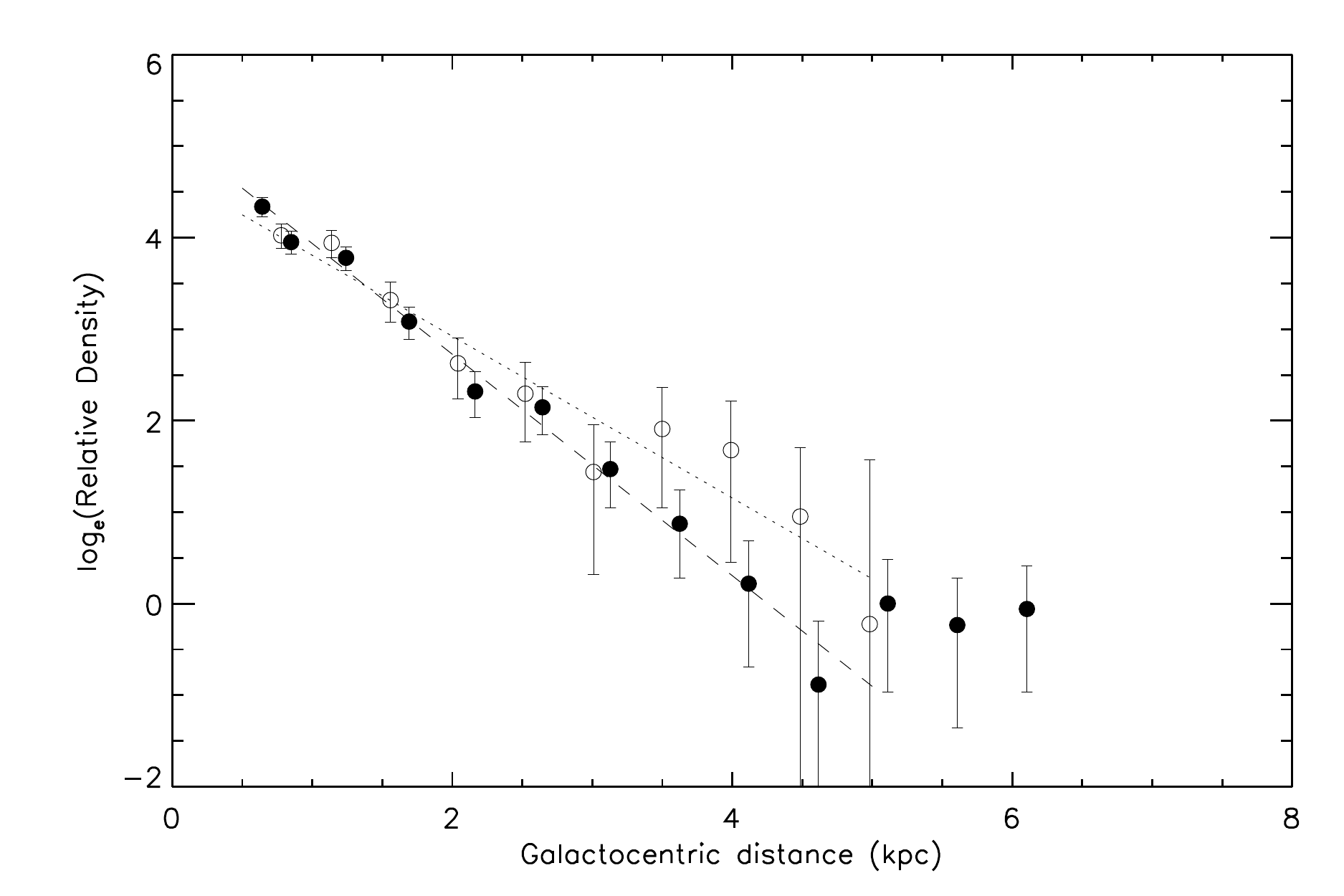}
\caption{Natural log of the relative density of RR~Lyrae stars as a function of Galactocentric distance in kpc as derived from the RRab's in field B1.  Open circles denote locations where the distance $d$ from the Sun is less than $R_{0}$, and filled circles for $d > R_{0}$. }
\label{fig:rdens2}
\end{figure*}

Figure~\ref{fig:rdens2} shows the same plot, but now with the natural log of the relative density on the ordinate.  The points out to $r <  5$ kpc appear to decrease linearly with $r$, indicating an exponential decline in the density of the form

\begin{equation}
\label{eqn:expdens}
\rho = B e^{-r/a}
\end{equation}

Formal fits to the near side (open circles) and far side (filled circles) show slightly different slopes of -0.90 and -1.21 ${\rm kpc}^{-1}$ respectively, corresponding to $a = 1.11$ kpc  and $a = 0.83$ kpc for the near and far sides respectively.  These (pseudo-)scale lengths obviously are along the line-of-sight and not along any axis of symmetry of the actual distribution (if indeed it is even elliptical).  However, we expect to learn more from the other 5 fields in this study.  On the far side  there is a rise in the RR~Lyrae density beyond $r \sim 5$ kpc.  This may indicate an encounter of the line-of-sight with features in the thick disk (such as a spiral arm), or possibly a stellar stream. For instance, 
\emph{Gaia} recently identified the remnant of a galaxy,
Gaia-Enceladus, which merged with the Milky Way approximately 10 Gyr
ago \citep{helmi18b}. This merger produced different stellar streams
that cover the entire sky and cross the Galactic disk and also have associated RR~Lyrae stars \citep[see Figure~3 of ][]{helmi18b}.

\section{Summary Discussion}
\label{sec:discussion}
 
 To date, the bulge has been probed primarily by spectroscopy of the
 brighter stars \citep[e.g.,][]{kun11}. We know from this spectroscopy that there is a huge range of metallicities, but no definitive constraints on the age distribution.  The available giants do not proportionally represent all elements of the underlying population mix.  Being mindful of the Initial Mass Function, one expects older stars to be over-represented among the giants, because their turnoff luminosities and masses are lower than their younger counterparts.  Main sequence stars in the bulge are prohibitively faint for detailed spectroscopic analysis, except when such stars are temporarily rendered brighter due to lensing events.  From a handful of main sequence stars studied in this way, \citet{bens18} argue that there are significant numbers of stars younger than 8~Gyr in the central kilo-parsec, a conclusion that is apparently at odds with that from the study of giants alone.  Given that the sample of 19 main sequence stars is unlikely to increase significantly in the foreseeable future, it appears that the issue of selective representation of sub-populations of stars by the giants is best mitigated by the synthesis of observed Hess diagrams of the bulge, using methods along the lines of \citet{dolph02} which are capable of producing a more complete picture of the star formation history and the age-metallicity relation. Analysis of the VVV generated near infrared CMD's are already underway \citep{surot19}.  The visual wavelength multi-band CMD's presented here provide higher leverage on metallicities (and hence on ages as well), but they are not yet Hess diagrams where the selection effects and completeness are fully understood and characterized.

The production of adequate quality Hess diagrams of the bulge has faced several challenges. In this paper we have demonstrated that the major problem of decoding the line-of-sight reddening with sufficient angular resolution as well as 
deriving and applying the correct reddening law is tractable through the RR~Lyrae stars. This builds on the work of \citet{Nataf13, Nataf16}, who used Red Giant Clump stars as standard candles and distance markers.  Our work here uses independently derived photometry, a different color and distance marker, and extends the reddening analysis to much bluer passbands (where the clump shows structure in color).  We have embarked on a program to search empirically for any systematic issues with RRab minimum light colors as a standard color marker, by extending the work done on the globular cluster M5 \citep{vivas17}, to a number of other RR~Lyrae bearing clusters with different metallicities and Oosterhoff types.  While the work presented here is thus a major step forward, there are two other issues that need resolving before the population synthesis mechanism can be brought to bear in full measure:

\begin{enumerate}
\item
Removal of the foreground contamination along the line-of-sight.  As discussed in \S~\ref{sec:intro}, it is only a matter of time before surveys like VVV and their derivatives 
are able to solve this issue. 

\item
Estimation of the completeness in the CMDs.  The usual procedure of deriving the completeness from artificial star tests is made more complicated by the extreme range and angular scales over which the extinction changes.  Effectively, each $30$ arcsec square bin must be evaluated independently.  Extinction not only changes the de-reddened magnitude from the brightness recovered from the image, but also affects how crowded a given patch of the field appears, and the confusion limit for detection and measurement errors.  This is not an intractable problem, but one that will be time consuming. We intend to address it after the equivalent of the study in this paper is done for all 6 of our fields.

\end{enumerate}

So how might the CMDs generated here, and the equivalent for the other fields in our study be used in the interim period before Hess diagrams with completeness estimates and foreground contamination removed can be produced?  First, one can ascertain the locations of the stars that have been spectroscopically studied already on our CMDs. Are certain areas of the CMD excluded?  A case in point are the more luminous among the ``blue loop'' stars discussed in \S~\ref{sec:CMDfeatures} and \S~\ref{sec:blueloopspec}, and the associated questions raised by the existence of that feature. Are all regions of the fan shaped giant branch structure in our CMDs represented?  Have clump stars from different parts of the extended clump structure in Figure~\ref{fig:cmd_umg} been observed?  How do stars from the blue and red ends of that extension differ in their spectroscopic characteristics?  Conversely one can use the de-reddened CMDs to pick stars from regions of particular interest to follow up spectroscopically in the next generation of surveys, such as APOGEE V.   When the equivalent CMDs for the other fields become available, we will be able to look for differences. Perhaps an empirical CMD of the foreground disk stars will emerge that will help model the foreground population in preparation for synthesizing the Hess diagrams to come.
Our plan is to proceed first with deriving and presenting the CMDs for all the remaining fields, followed by work along the lines outlined above.

We note that of the 4877 putative variables of all types we have detected, 2265 were detected independently in at least 2 of the 5 passbands, implying that these are almost certainly not false positives.  Since our time coverage is limited, and focussed on obtaining light curves for RR~Lyraes, our data are inadequate for obtaining light curves and periods for all these objects. Fortunately, many of these are also known and classified by the OGLE survey.  Our five band, de-reddened data for this set of objects are a ``training set'' for identifying variable stars from future LSST data, which will have sparse cadence, but information in panchromatic passbands.  With the exception of the ``Stripe 82'' field from SDSS \citep[e.g.,][]{sesar07, bramich08}, there are no suitable publicly available data sets with time-domain coverage in multiple passbands at this time that are suitable for developing and testing algorithms for parsing variability characteristics from LSST data. Our results provide the ability for LSST time-domain brokering projects to develop the necessary techniques and algorithms to handle the 
variable star data that LSST will generate.

%
%\
%\begin{quotation}
This project used data obtained with the Dark Energy Camera (DECam),
which was constructed by the Dark Energy Survey (DES) collaboration.
Funding for the DES Projects has been provided by 
the U.S. Department of Energy, 
the U.S. National Science Foundation, 
the Ministry of Science and Education of Spain, 
the Science and Technology Facilities Council of the United Kingdom, 
the Higher Education Funding Council for England, 
the National Center for Supercomputing Applications at the University of Illinois at Urbana-Champaign, 
the Kavli Institute of Cosmological Physics at the University of Chicago, 
the Center for Cosmology and Astro-Particle Physics at the Ohio State University, 
the Mitchell Institute for Fundamental Physics and Astronomy at Texas A\&M University, 
Financiadora de Estudos e Projetos, Funda{\c c}{\~a}o Carlos Chagas Filho de Amparo {\`a} Pesquisa do Estado do Rio de Janeiro, 
Conselho Nacional de Desenvolvimento Cient{\'i}fico e Tecnol{\'o}gico and the Minist{\'e}rio da Ci{\^e}ncia, Tecnologia e Inovac{\~a}o, 
the Deutsche Forschungsgemeinschaft, 
and the Collaborating Institutions in the Dark Energy Survey. 
The Collaborating Institutions are 
Argonne National Laboratory, 
the University of California at Santa Cruz, 
the University of Cambridge, 
Centro de Investigaciones En{\'e}rgeticas, Medioambientales y Tecnol{\'o}gicas-Madrid, 
the University of Chicago, 
University College London, 
the DES-Brazil Consortium, 
the University of Edinburgh, 
the Eidgen{\"o}ssische Technische Hoch\-schule (ETH) Z{\"u}rich, 
Fermi National Accelerator Laboratory, 
the University of Illinois at Urbana-Champaign, 
the Institut de Ci{\`e}ncies de l'Espai (IEEC/CSIC), 
the Institut de F{\'i}sica d'Altes Energies, 
Lawrence Berkeley National Laboratory, 
the Ludwig-Maximilians Universit{\"a}t M{\"u}nchen and the associated Excellence Cluster Universe, 
the University of Michigan, 
{the} National Optical Astronomy Observatory, 
the University of Nottingham, 
the Ohio State University, 
the OzDES Membership Consortium
the University of Pennsylvania, 
the University of Portsmouth, 
SLAC National Accelerator Laboratory, 
Stanford University, 
the University of Sussex, 
and Texas A\&M University.

Based on observations at Cerro Tololo Inter-American Observatory, National Optical
Astronomy Observatory (NOAO Prop. 2013A-0719 and PI A. Saha), which is operated by the Association of
Universities for Research in Astronomy (AURA) under a cooperative agreement with the
National Science Foundation.

%\end{quotation}

We thank the support staff at CTIO for their excellent support during the observations.   Olszewski was partially supported by NSF Grants AST-1313001 and AST-1815767. 
We are grateful for perceptive (and ultra-prompt!) comments from an anonymous referee which have resulted in substantive improvements to the paper.

\bibliographystyle{aasjournal}
\bibliography{ms}

\begin{thebibliography}{}
\expandafter\ifx\csname natexlab\endcsname\relax\def\natexlab#1{#1}\fi

\bibitem[{{Abolfathi} {et~al.}(2018){Abolfathi}, {Aguado}, {Aguilar}, {Allende
  Prieto}, {Almeida}, {Ananna}, {Anders}, {Anderson}, {Andrews}, {Anguiano}, \&
  et~al.}]{abol18}
{Abolfathi}, B., {Aguado}, D.~S., {Aguilar}, G., {et~al.} 2018, \apjs, 235, 42

\bibitem[{{Abuter} {et~al.}(2018){Abuter}, {Amorim}, {Anugu}, {Baub{\"o}ck},
  {Benisty}, {Berger}, {Blind}, {Bonnet}, {Brandner}, {Buron}, {Collin},
  {Chapron}, {Cl{\'e}net}, {Coud{\'e} Du Foresto}, {de Zeeuw}, {Deen},
  {Delplancke- Str{\"o}bele}, {Dembet}, {Dexter}, {Duvert}, {Eckart},
  {Eisenhauer}, {Finger}, {F{\"o}rster Schreiber}, {F{\'e}dou}, {Garcia},
  {Garcia Lopez}, {Gao}, {Gendron}, {Genzel}, {Gillessen}, {Gordo}, {Habibi},
  {Haubois}, {Haug}, {Hau{\ss}mann}, {Henning}, {Hippler}, {Horrobin},
  {Hubert}, {Hubin}, {Jimenez Rosales}, {Jochum}, {Jocou}, {Kaufer}, {Kellner},
  {Kendrew}, {Kervella}, {Kok}, {Kulas}, {Lacour}, {Lapeyr{\`e}re}, {Lazareff},
  {Le Bouquin}, {L{\'e}na}, {Lippa}, {Lenzen}, {M{\'e}rand}, {M{\"u}ler},
  {Neumann}, {Ott}, {Palanca}, {Paumard}, {Pasquini}, {Perraut}, {Perrin},
  {Pfuhl}, {Plewa}, {Rabien}, {Ram{\'\i}rez}, {Ramos}, {Rau},
  {Rodr{\'\i}guez-Coira}, {Rohloff}, {Rousset}, {Sanchez-Bermudez},
  {Scheithauer}, {Sch{\"o}ller}, {Schuler}, {Spyromilio}, {Straub},
  {Straubmeier}, {Sturm}, {Tacconi}, {Tristram}, {Vincent}, {von Fellenberg},
  {Wank}, {Waisberg}, {Widmann}, {Wieprecht}, {Wiest}, {Wiezorrek}, {Woillez},
  {Yazici}, {Ziegler}, \& {Zins}}]{abuter18}
{Abuter}, R., {Amorim}, A., {Anugu}, N., {et~al.} 2018, \aap, 615, L15

\bibitem[{{Arenou} {et~al.}(2018){Arenou}, {Luri}, {Babusiaux}, {Fabricius},
  {Helmi}, {Muraveva}, {Robin}, {Spoto}, {Vallenari}, {Antoja},
  {Cantat-Gaudin}, {Jordi}, {Leclerc}, {Reyl{\'e}}, {Romero-G{\'o}mez}, {Shih},
  {Soria}, {Barache}, {Bossini}, {Bragaglia}, {Breddels}, {Fabrizio},
  {Lambert}, {Marrese}, {Massari}, {Moitinho}, {Robichon}, {Ruiz-Dern},
  {Sordo}, {Veljanoski}, {Eyer}, {Jasniewicz}, {Pancino}, {Soubiran}, {Spagna},
  {Tanga}, {Turon}, \& {Zurbach}}]{arenou18}
{Arenou}, F., {Luri}, X., {Babusiaux}, C., {et~al.} 2018, \aap, 616, A17

\bibitem[{{Baade}(1946)}]{baade46}
{Baade}, W. 1946, \pasp, 58, 249

\bibitem[{{Barbuy} {et~al.}(2018){Barbuy}, {Chiappini}, \&
  {Gerhard}}]{barbuy18}
{Barbuy}, B., {Chiappini}, C., \& {Gerhard}, O. 2018, ArXiv e-prints,
  arXiv:1805.01142

\bibitem[{{Benedict} {et~al.}(2011){Benedict}, {McArthur}, {Feast}, {Barnes},
  {Harrison}, {Bean}, {Menzies}, {Chaboyer}, {Fossati}, {Nesvacil}, {Smith},
  {Kolenberg}, {Laney}, {Kochukhov}, {Nelan}, {Shulyak}, {Taylor}, \&
  {Freedman}}]{benedict11}
{Benedict}, G.~F., {McArthur}, B.~E., {Feast}, M.~W., {et~al.} 2011, \aj, 142,
  187

\bibitem[{{Bensby} {et~al.}(2018){Bensby}, {Feltzing}, {Gould}, {Yee},
  {Johnson}, {Asplund}, {Mel{\'e}ndez}, {Lucatello}, \& {Howes}}]{bens18}
{Bensby}, T., {Feltzing}, S., {Gould}, A., {et~al.} 2018, in IAU Symposium,
  Vol. 334, Rediscovering Our Galaxy, ed. C.~{Chiappini}, I.~{Minchev},
  E.~{Starkenburg}, \& M.~{Valentini}, 86--89

\bibitem[{{Bernard} {et~al.}(2018){Bernard}, {Schultheis}, {Di Matteo}, {Hill},
  {Haywood}, \& {Calamida}}]{bernard18}
{Bernard}, E.~J., {Schultheis}, M., {Di Matteo}, P., {et~al.} 2018, \mnras,
  477, 3507

\bibitem[{{Blanco}(1992)}]{blanco92}
{Blanco}, B.~M. 1992, \aj, 103, 1872

\bibitem[{{Blanco} \& {Blanco}(1997)}]{blancos97}
{Blanco}, B.~M., \& {Blanco}, V.~M. 1997, \aj, 114, 2596

\bibitem[{{Blum} {et~al.}(2003){Blum}, {Ram{\'{\i}}rez}, {Sellgren}, \&
  {Olsen}}]{blum03}
{Blum}, R.~D., {Ram{\'{\i}}rez}, S.~V., {Sellgren}, K., \& {Olsen}, K. 2003,
  \apj, 597, 323

\bibitem[{{Bramich} {et~al.}(2008){Bramich}, {Vidrih}, {Wyrzykowski}, {Munn},
  {Lin}, {Evans}, {Smith}, {Belokurov}, {Gilmore}, {Zucker}, {Hewett},
  {Watkins}, {Faria}, {Fellhauer}, {Miknaitis}, {Bizyaev}, {Ivezi{\'c}},
  {Schneider}, {Snedden}, {Malanushenko}, {Malanushenko}, \& {Pan}}]{bramich08}
{Bramich}, D.~M., {Vidrih}, S., {Wyrzykowski}, L., {et~al.} 2008, \mnras, 386,
  887

\bibitem[{{Brown} {et~al.}(2009){Brown}, {Sahu}, {Zoccali}, {Renzini},
  {Ferguson}, {Anderson}, {Smith}, {Bond}, {Minniti}, {Valenti}, {Casertano},
  {Livio}, {Panagia}, {Vanden Berg}, \& {Valenti}}]{brown09}
{Brown}, T.~M., {Sahu}, K., {Zoccali}, M., {et~al.} 2009, \aj, 137, 3172

\bibitem[{{Burstein} \& {Heiles}(1978)}]{burstein78}
{Burstein}, D., \& {Heiles}, C. 1978, \apj, 225, 40

\bibitem[{{Calamida} {et~al.}(2014){Calamida}, {Sahu}, {Anderson}, {Casertano},
  {Cassisi}, {Salaris}, {Brown}, {Sokol}, {Bond}, {Ferraro}, {Ferguson},
  {Livio}, {Valenti}, {Buonanno}, {Clarkson}, \& {Pietrinferni}}]{cala14}
{Calamida}, A., {Sahu}, K.~C., {Anderson}, J., {et~al.} 2014, \apj, 790, 164

\bibitem[{{Chu} {et~al.}(2018){Chu}, {Do}, {Hees}, {Ghez}, {Naoz}, {Witzel},
  {Sakai}, {Chappell}, {Gautam}, {Lu}, \& {Matthews}}]{chu18}
{Chu}, D.~S., {Do}, T., {Hees}, A., {et~al.} 2018, \apj, 854, 12

\bibitem[{{Clarkson} {et~al.}(2008){Clarkson}, {Sahu}, {Anderson}, {Smith},
  {Brown}, {Rich}, {Casertano}, {Bond}, {Livio}, {Minniti}, {Panagia},
  {Renzini}, {Valenti}, \& {Zoccali}}]{clark08}
{Clarkson}, W., {Sahu}, K., {Anderson}, J., {et~al.} 2008, \apj, 684, 1110

\bibitem[{{Clement} {et~al.}(2001){Clement}, {Muzzin}, {Dufton}, {Ponnampalam},
  {Wang}, {Burford}, {Richardson}, {Rosebery}, {Rowe}, \& {Hogg}}]{clement01}
{Clement}, C.~M., {Muzzin}, A., {Dufton}, Q., {et~al.} 2001, \aj, 122, 2587

\bibitem[{{de Grijs} \& {Bono}(2016)}]{degrijs16}
{de Grijs}, R., \& {Bono}, G. 2016, \apjs, 227, 5

\bibitem[{{Dias} {et~al.}(2016){Dias}, {Barbuy}, {Saviane}, {Held}, {Da Costa},
  {Ortolani}, {Gullieuszik}, \& {V{\'a}squez}}]{dias16}
{Dias}, B., {Barbuy}, B., {Saviane}, I., {et~al.} 2016, \aap, 590, A9

\bibitem[{{Dolphin}(2002)}]{dolph02}
{Dolphin}, A.~E. 2002, \mnras, 332, 91

\bibitem[{{Dwek} {et~al.}(1995){Dwek}, {Arendt}, {Hauser}, {Kelsall}, {Lisse},
  {Moseley}, {Silverberg}, {Sodroski}, \& {Weiland}}]{dwek95}
{Dwek}, E., {Arendt}, R.~G., {Hauser}, M.~G., {et~al.} 1995, \apj, 445, 716

\bibitem[{{Fitzpatrick}(1999)}]{fitz99}
{Fitzpatrick}, E.~L. 1999, Publications of the Astronomical Society of the
  Pacific, 111, 63

\bibitem[{{Flaugher} {et~al.}(2015){Flaugher}, {Diehl}, {Honscheid}, {Abbott},
  {Alvarez}, {Angstadt}, {Annis}, {Antonik}, {Ballester}, {Beaufore},
  {Bernstein}, {Bernstein}, {Bigelow}, {Bonati}, {Boprie}, {Brooks},
  {Buckley-Geer}, {Campa}, {Cardiel-Sas}, {Castander}, {Castilla}, {Cease},
  {Cela-Ruiz}, {Chappa}, {Chi}, {Cooper}, {da Costa}, {Dede}, {Derylo},
  {DePoy}, {de Vicente}, {Doel}, {Drlica-Wagner}, {Eiting}, {Elliott}, {Emes},
  {Estrada}, {Fausti Neto}, {Finley}, {Flores}, {Frieman}, {Gerdes},
  {Gladders}, {Gregory}, {Gutierrez}, {Hao}, {Holland}, {Holm}, {Huffman},
  {Jackson}, {James}, {Jonas}, {Karcher}, {Karliner}, {Kent}, {Kessler},
  {Kozlovsky}, {Kron}, {Kubik}, {Kuehn}, {Kuhlmann}, {Kuk}, {Lahav}, {Lathrop},
  {Lee}, {Levi}, {Lewis}, {Li}, {Mandrichenko}, {Marshall}, {Martinez},
  {Merritt}, {Miquel}, {Mu{\~n}oz}, {Neilsen}, {Nichol}, {Nord}, {Ogando},
  {Olsen}, {Palaio}, {Patton}, {Peoples}, {Plazas}, {Rauch}, {Reil}, {Rheault},
  {Roe}, {Rogers}, {Roodman}, {Sanchez}, {Scarpine}, {Schindler}, {Schmidt},
  B3~{Schmitt}, {Schubnell}, {Schultz}, {Schurter}, {Scott}, {Serrano}, {Shaw},
  {Smith}, {Soares-Santos}, {Stefanik}, {Stuermer}, {Suchyta}, {Sypniewski},
  {Tarle}, {Thaler}, {Tighe}, {Tran}, {Tucker}, {Walker}, {Wang}, {Watson},
  {Weaverdyck}, {Wester}, {Woods}, {Yanny}, \& {DES
  Collaboration}}]{flaugher15}
{Flaugher}, B., {Diehl}, H.~T., {Honscheid}, K., {et~al.} 2015, \aj, 150, 150

\bibitem[{{Fukugita} {et~al.}(1996){Fukugita}, {Ichikawa}, {Gunn}, {Doi},
  {Shimasaku}, \& {Schneider}}]{fuk96}
{Fukugita}, M., {Ichikawa}, T., {Gunn}, J.~E., {et~al.} 1996, \aj, 111, 1748

\bibitem[{{Gaia Collaboration} {et~al.}(2018){Gaia Collaboration}, {Helmi},
  {van Leeuwen}, {McMillan}, {Massari}, {Antoja}, {Robin}, {Lindegren},
  {Bastian}, {Arenou}, \& et~al.}]{helmi18}
{Gaia Collaboration}, {Helmi}, A., {van Leeuwen}, F., {et~al.} 2018, \aap, 616,
  A12

\bibitem[{{Garc{\'{\i}}a P{\'e}rez} {et~al.}(2016){Garc{\'{\i}}a P{\'e}rez},
  {Allende Prieto}, {Holtzman}, {Shetrone}, {M{\'e}sz{\'a}ros}, {Bizyaev},
  {Carrera}, {Cunha}, {Garc{\'{\i}}a-Hern{\'a}ndez}, {Johnson}, {Majewski},
  {Nidever}, {Schiavon}, {Shane}, {Smith}, {Sobeck}, {Troup}, {Zamora},
  {Weinberg}, {Bovy}, {Eisenstein}, {Feuillet}, {Frinchaboy}, {Hayden},
  {Hearty}, {Nguyen}, {O'Connell}, {Pinsonneault}, {Wilson}, \&
  {Zasowski}}]{gper16}
{Garc{\'{\i}}a P{\'e}rez}, A.~E., {Allende Prieto}, C., {Holtzman}, J.~A.,
  {et~al.} 2016, \aj, 151, 144

\bibitem[{{Garc{\'{\i}}a P{\'e}rez} {et~al.}(2018){Garc{\'{\i}}a P{\'e}rez},
  {Ness}, {Robin}, {Martinez-Valpuesta}, {Sobeck}, {Zasowski}, {Majewski},
  {Bovy}, {Allende Prieto}, {Cunha}, {Girardi}, {M{\'e}sz{\'a}ros}, {Nidever},
  {Schiavon}, {Schultheis}, {Shetrone}, \& {Smith}}]{gper18}
{Garc{\'{\i}}a P{\'e}rez}, A.~E., {Ness}, M., {Robin}, A.~C., {et~al.} 2018,
  \apj, 852, 91

\bibitem[{{Gontcharov} {et~al.}(2019){Gontcharov}, {Mosenkov}, \&
  {Khovritchev}}]{gont19}
{Gontcharov}, G.~A., {Mosenkov}, A.~V., \& {Khovritchev}, M.~Y. 2019, \mnras,
  483, 4949

\bibitem[{{Helmi} {et~al.}(2018){Helmi}, {Babusiaux}, {Koppelman}, {Massari},
  {Veljanoski}, \& {Brown}}]{helmi18b}
{Helmi}, A., {Babusiaux}, C., {Koppelman}, H.~H., {et~al.} 2018, \nat, 563, 85

\bibitem[{{Holtzman} {et~al.}(2018){Holtzman}, {Hasselquist}, {Shetrone},
  {Cunha}, {Allende Prieto}, {Anguiano}, {Bizyaev}, {Bovy}, {Casey},
  {Edvardsson}, {Johnson}, {J{\"o}nsson}, {Meszaros}, {Smith}, {Sobeck},
  {Zamora}, {Chojnowski}, {Fernandez-Trincado}, {Garcia-Hernandez}, {Majewski},
  {Pinsonneault}, {Souto}, {Stringfellow}, {Tayar}, {Troup}, \&
  {Zasowski}}]{holtz18}
{Holtzman}, J.~A., {Hasselquist}, S., {Shetrone}, M., {et~al.} 2018, \aj, 156,
  125

\bibitem[{{Iben}(1964)}]{iben64}
{Iben}, Jr., I. 1964, \apj, 140, 1631

\bibitem[{{J{\"o}nsson} {et~al.}(2018){J{\"o}nsson}, {Allende Prieto},
  {Holtzman}, {Feuillet}, {Hawkins}, {Cunha}, {M{\'e}sz{\'a}ros},
  {Hasselquist}, {Fern{\'a}ndez-Trincado}, {Garc{\'{\i}}a-Hern{\'a}ndez},
  {Bizyaev}, {Carrera}, {Majewski}, {Pinsonneault}, {Shetrone}, {Smith},
  {Sobeck}, {Souto}, {Stringfellow}, {Teske}, \& {Zamora}}]{jonss18}
{J{\"o}nsson}, H., {Allende Prieto}, C., {Holtzman}, J.~A., {et~al.} 2018, \aj,
  156, 126

\bibitem[{{Kiraga} {et~al.}(1997){Kiraga}, {Paczy{\'n}ski}, \&
  {Stanek}}]{kiraga97}
{Kiraga}, M., {Paczy{\'n}ski}, B., \& {Stanek}, K.~Z. 1997, \apj, 485, 611

\bibitem[{{Kormendy} \& {Kennicutt}(2004)}]{kormendy04}
{Kormendy}, J., \& {Kennicutt}, Jr., R.~C. 2004, \araa, 42, 603

\bibitem[{{Kunder} {et~al.}(2011){Kunder}, {de Propris}, {Rich}, {Koch},
  {Howard}, {Johnson}, {Clarkson}, {Mallery}, {Kormendy}, {Robin}, {Fux},
  {David}, {Zhao}, {Kuijken}, {Pipino}, \& {Shen}}]{kun11}
{Kunder}, A.~M., {de Propris}, R., {Rich}, M., {et~al.} 2011, in American
  Astronomical Society Meeting Abstracts, Vol. 217, American Astronomical
  Society Meeting Abstracts \#217, 241.12

\bibitem[{{Layden} {et~al.}(2005){Layden}, {Sarajedini}, {von Hippel}, \&
  {Cool}}]{layden05}
{Layden}, A.~C., {Sarajedini}, A., {von Hippel}, T., \& {Cool}, A.~M. 2005,
  \apj, 632, 266

\bibitem[{{Lindegren} {et~al.}(2018){Lindegren}, {Hern{\'a}ndez}, {Bombrun},
  {Klioner}, {Bastian}, {Ramos-Lerate}, {de Torres}, {Steidelm{\"u}ller},
  {Stephenson}, {Hobbs}, {Lammers}, {Biermann}, {Geyer}, {Hilger}, {Michalik},
  {Stampa}, {McMillan}, {Casta{\~n}eda}, {Clotet}, {Comoretto}, {Davidson},
  {Fabricius}, {Gracia}, {Hambly}, {Hutton}, {Mora}, {Portell}, {van Leeuwen},
  {Abbas}, {Abreu}, {Altmann}, {Andrei}, {Anglada}, {Balaguer-N{\'u}{\~n}ez},
  {Barache}, {Becciani}, {Bertone}, {Bianchi}, {Bouquillon}, {Bourda},
  {Br{\"u}semeister}, {Bucciarelli}, {Busonero}, {Buzzi}, {Cancelliere},
  {Carlucci}, {Charlot}, {Cheek}, {Crosta}, {Crowley}, {de Bruijne}, {de
  Felice}, {Drimmel}, {Esquej}, {Fienga}, {Fraile}, {Gai}, {Garralda},
  {Gonz{\'a}lez- Vidal}, {Guerra}, {Hauser}, {Hofmann}, {Holl}, {Jordan},
  {Lattanzi}, {Lenhardt}, {Liao}, {Licata}, {Lister}, {L{\"o}ffler},
  {Marchant}, {Martin-Fleitas}, {Messineo}, {Mignard}, {Morbidelli}, {Poggio},
  {Riva}, {Rowell}, {Salguero}, {Sarasso}, {Sciacca}, {Siddiqui}, {Smart},
  {Spagna}, {Steele}, {Taris}, {Torra}, {van Elteren}, {van Reeven}, \&
  {Vecchiato}}]{lindegren18}
{Lindegren}, L., {Hern{\'a}ndez}, J., {Bombrun}, A., {et~al.} 2018, \aap, 616,
  A2

\bibitem[{{Majewski} {et~al.}(2017){Majewski}, {Schiavon}, {Frinchaboy},
  {Allende Prieto}, {Barkhouser}, {Bizyaev}, {Blank}, {Brunner}, {Burton},
  {Carrera}, {Chojnowski}, {Cunha}, {Epstein}, {Fitzgerald}, {Garc{\'{\i}}a
  P{\'e}rez}, {Hearty}, {Henderson}, {Holtzman}, {Johnson}, {Lam}, {Lawler},
  {Maseman}, {M{\'e}sz{\'a}ros}, {Nelson}, {Nguyen}, {Nidever}, {Pinsonneault},
  {Shetrone}, {Smee}, {Smith}, {Stolberg}, {Skrutskie}, {Walker}, {Wilson},
  {Zasowski}, {Anders}, {Basu}, {Beland}, {Blanton}, {Bovy}, {Brownstein},
  {Carlberg}, {Chaplin}, {Chiappini}, {Eisenstein}, {Elsworth}, {Feuillet},
  {Fleming}, {Galbraith-Frew}, {Garc{\'{\i}}a}, {Garc{\'{\i}}a-Hern{\'a}ndez},
  {Gillespie}, {Girardi}, {Gunn}, {Hasselquist}, {Hayden}, {Hekker}, {Ivans},
  {Kinemuchi}, {Klaene}, {Mahadevan}, {Mathur}, {Mosser}, {Muna}, {Munn},
  {Nichol}, {O'Connell}, {Parejko}, {Robin}, {Rocha-Pinto}, {Schultheis},
  {Serenelli}, {Shane}, {Silva Aguirre}, {Sobeck}, {Thompson}, {Troup},
  {Weinberg}, \& {Zamora}}]{maj17}
{Majewski}, S.~R., {Schiavon}, R.~P., {Frinchaboy}, P.~M., {et~al.} 2017, \aj,
  154, 94

\bibitem[{{Martig} {et~al.}(2016){Martig}, {Fouesneau}, {Rix}, {Ness},
  {M{\'e}sz{\'a}ros}, {Garc{\'{\i}}a-Hern{\'a}ndez}, {Pinsonneault},
  {Serenelli}, {Silva Aguirre}, \& {Zamora}}]{mart16}
{Martig}, M., {Fouesneau}, M., {Rix}, H.-W., {et~al.} 2016, \mnras, 456, 3655

\bibitem[{{Minniti} {et~al.}(2010){Minniti}, {Lucas}, {Emerson}, {Saito},
  {Hempel}, {Pietrukowicz}, {Ahumada}, {Alonso}, {Alonso-Garcia}, {Arias},
  {Bandyopadhyay}, {Barb{\'a}}, {Barbuy}, {Bedin}, {Bica}, {Borissova},
  {Bronfman}, {Carraro}, {Catelan}, {Clari{\'a}}, {Cross}, {de Grijs},
  {D{\'e}k{\'a}ny}, {Drew}, {Fari{\~n}a}, {Feinstein}, {Fern{\'a}ndez
  Laj{\'u}s}, {Gamen}, {Geisler}, {Gieren}, {Goldman}, {Gonzalez}, {Gunthardt},
  {Gurovich}, {Hambly}, {Irwin}, {Ivanov}, {Jord{\'a}n}, {Kerins}, {Kinemuchi},
  {Kurtev}, {L{\'o}pez-Corredoira}, {Maccarone}, {Masetti}, {Merlo},
  {Messineo}, {Mirabel}, {Monaco}, {Morelli}, {Padilla}, {Palma}, {Parisi},
  {Pignata}, {Rejkuba}, {Roman-Lopes}, {Sale}, {Schreiber}, {Schr{\"o}der},
  {Smith}, {}, {Soto}, {Tamura}, {Tappert}, {Thompson}, {Toledo}, {Zoccali}, \&
  {Pietrzynski}}]{minniti10}
{Minniti}, D., {Lucas}, P.~W., {Emerson}, J.~P., {et~al.} 2010, \na, 15, 433

\bibitem[{{Minniti} {et~al.}(2017){Minniti}, {D{\'e}k{\'a}ny}, {Majaess},
  {Palma}, {Pullen}, {Rejkuba}, {Alonso-Garc{\'{\i}}a}, {Catelan}, {Contreras
  Ramos}, {Gonzalez}, {Hempel}, {Irwin}, {Lucas}, {Saito}, {Tissera},
  {Valenti}, \& {Zoccali}}]{minniti17}
{Minniti}, D., {D{\'e}k{\'a}ny}, I., {Majaess}, D., {et~al.} 2017, \aj, 153,
  179

\bibitem[{{Muraveva} {et~al.}(2018){Muraveva}, {Delgado}, {Clementini},
  {Sarro}, \& {Garofalo}}]{mura18}
{Muraveva}, T., {Delgado}, H.~E., {Clementini}, G., {Sarro}, L.~M., \&
  {Garofalo}, A. 2018, ArXiv e-prints, arXiv:1805.08742

\bibitem[{{Narayan} {et~al.}(2016){Narayan}, {Axelrod}, {Holberg}, {Matheson},
  {Saha}, {Olszewski}, {Claver}, {Stubbs}, {Bohlin}, {Deustua}, \&
  {Rest}}]{nar16}
{Narayan}, G., {Axelrod}, T., {Holberg}, J.~B., {et~al.} 2016, \apj, 822, 67

\bibitem[{{Nataf} {et~al.}(2013){Nataf}, {Gould}, {Fouqu{\'e}}, {Gonzalez},
  {Johnson}, {Skowron}, {Udalski}, {Szyma{\'n}ski}, {Kubiak},
  {Pietrzy{\'n}ski}, {Soszy{\'n}ski}, {Ulaczyk}, {Wyrzykowski}, \&
  {Poleski}}]{Nataf13}
{Nataf}, D.~M., {Gould}, A., {Fouqu{\'e}}, P., {et~al.} 2013, \apj, 769, 88

\bibitem[{{Nataf} {et~al.}(2016){Nataf}, {Gonzalez}, {Casagrande}, {Zasowski},
  {Wegg}, {Wolf}, {Kunder}, {Alonso-Garcia}, {Minniti}, {Rejkuba}, {Saito},
  {Valenti}, {Zoccali}, {Poleski}, {Pietrzy{\'n}ski}, {Skowron},
  {Soszy{\'n}ski}, {Szyma{\'n}ski}, {Udalski}, {Ulaczyk}, \&
  {Wyrzykowski}}]{Nataf16}
{Nataf}, D.~M., {Gonzalez}, O.~A., {Casagrande}, L., {et~al.} 2016, \mnras,
  456, 2692

\bibitem[{{O'Donnell}(1994)}]{odonnell94}
{O'Donnell}, J.~E. 1994, \apj, 422, 158

\bibitem[{{Pietrukowicz} {et~al.}(2012){Pietrukowicz}, {Udalski},
  {Soszy{\'n}ski}, {Nataf}, {Wyrzykowski}, {Poleski}, {Koz{\l}owski},
  {Szyma{\'n}ski}, {Kubiak}, {Pietrzy{\'n}ski}, \& {Ulaczyk}}]{Piet12}
{Pietrukowicz}, P., {Udalski}, A., {Soszy{\'n}ski}, I., {et~al.} 2012, \apj,
  750, 169

\bibitem[{{Pinsonneault} {et~al.}(2014){Pinsonneault}, {Elsworth}, {Epstein},
  {Hekker}, {M{\'e}sz{\'a}ros}, {Chaplin}, {Johnson}, {Garc{\'{\i}}a},
  {Holtzman}, {Mathur}, {Garc{\'{\i}}a P{\'e}rez}, {Silva Aguirre}, {Girardi},
  {Basu}, {Shetrone}, {Stello}, {Allende Prieto}, {An}, {Beck}, {Beers},
  {Bizyaev}, {Bloemen}, {Bovy}, {Cunha}, {De Ridder}, {Frinchaboy},
  {Garc{\'{\i}}a-Hern{\'a}ndez}, {Gilliland}, {Harding}, {Hearty}, {Huber},
  {Ivans}, {Kallinger}, {Majewski}, {Metcalfe}, {Miglio}, {Mosser}, {Muna},
  {Nidever}, {Schneider}, {Serenelli}, {Smith}, {Tayar}, {Zamora}, \&
  {Zasowski}}]{pins14}
{Pinsonneault}, M.~H., {Elsworth}, Y., {Epstein}, C., {et~al.} 2014, \apjs,
  215, 19

\bibitem[{{Saha} \& {Hoessel}(1990)}]{saha90}
{Saha}, A., \& {Hoessel}, J.~G. 1990, \aj, 99, 97

\bibitem[{{Saha} \& {Vivas}(2017)}]{saha17}
{Saha}, A., \& {Vivas}, A.~K. 2017, \aj, 154, 231

\bibitem[{{Saha} {et~al.}(2010){Saha}, {Olszewski}, {Brondel}, {Olsen},
  {Knezek}, {Harris}, {Smith}, {Subramaniam}, {Claver}, {Rest}, {Seitzer},
  {Cook}, {Minniti}, \& {Suntzeff}}]{saha10}
{Saha}, A., {Olszewski}, E.~W., {Brondel}, B., {et~al.} 2010, \aj, 140, 1719

\bibitem[{{Sandage}(1990)}]{sandage90}
{Sandage}, A. 1990, \apj, 350, 631

\bibitem[{{Schechter} {et~al.}(1993){Schechter}, {Mateo}, \&
  {Saha}}]{schechter93}
{Schechter}, P.~L., {Mateo}, M., \& {Saha}, A. 1993, \pasp, 105, 1342

\bibitem[{{Schlegel} {et~al.}(1998){Schlegel}, {Finkbeiner}, \&
  {Davis}}]{sfd98}
{Schlegel}, D.~J., {Finkbeiner}, D.~P., \& {Davis}, M. 1998, \apj, 500, 525

\bibitem[{{Schultheis} {et~al.}(2017){Schultheis}, {Rojas-Arriagada},
  {Garc{\'{\i}}a P{\'e}rez}, {J{\"o}nsson}, {Hayden}, {Nandakumar}, {Cunha},
  {Allende Prieto}, {Holtzman}, {Beers}, {Bizyaev}, {Brinkmann}, {Carrera},
  {Cohen}, {Geisler}, {Hearty}, {Fernandez-Tricado}, {Maraston}, {Minnitti},
  {Nitschelm}, {Roman-Lopes}, {Schneider}, {Tang}, {Villanova}, {Zasowski}, \&
  {Majewski}}]{schult17}
{Schultheis}, M., {Rojas-Arriagada}, A., {Garc{\'{\i}}a P{\'e}rez}, A.~E.,
  {et~al.} 2017, \aap, 600, A14

\bibitem[{{Sesar} {et~al.}(2007){Sesar}, {Ivezi{\'c}}, {Lupton}, {Juri{\'c}},
  {Gunn}, {Knapp}, {DeLee}, {Smith}, {Miknaitis}, {Lin}, {Tucker}, {Doi},
  {Tanaka}, {Fukugita}, {Holtzman}, {Kent}, {Yanny}, {Schlegel}, {Finkbeiner},
  {Padmanabhan}, {Rockosi}, {Bond}, {Lee}, {Stoughton}, {Jester}, {Harris},
  {Harding}, {Brinkmann}, {Schneider}, {York}, {Richmond}, \& {Vanden
  Berk}}]{sesar07}
{Sesar}, B., {Ivezi{\'c}}, {\v Z}., {Lupton}, R.~H., {et~al.} 2007, \aj, 134,
  2236

\bibitem[{{Sesar} {et~al.}(2010){Sesar}, {Ivezi{\'c}}, {Grammer}, {Morgan},
  {Becker}, {Juri{\'c}}, {De Lee}, {Annis}, {Beers}, {Fan}, {Lupton}, {Gunn},
  {Knapp}, {Jiang}, {Jester}, {Johnston}, \& {Lampeitl}}]{sesar10}
{Sesar}, B., {Ivezi{\'c}}, {\v Z}., {Grammer}, S.~H., {et~al.} 2010, \apj, 708,
  717

\bibitem[{{Skottfelt} {et~al.}(2015){Skottfelt}, {Bramich}, {Figuera Jaimes},
  {J{\o}rgensen}, {Kains}, {Arellano Ferro}, {Alsubai}, {Bozza}, {Calchi
  Novati}, {Ciceri}, {D'Ago}, {Dominik}, {Galianni}, {Gu}, {Harps{\o}e},
  {Haugb{\o}lle}, {Hinse}, {Hundertmark}, {Juncher}, {Korhonen}, {Liebig},
  {Mancini}, {Popovas}, {Rabus}, {Rahvar}, {Scarpetta}, {Schmidt}, {Snodgrass},
  {Southworth}, {Starkey}, {Street}, {Surdej}, {Wang}, \& {Wertz (The Mindstep
  Consortium)}}]{skot15}
{Skottfelt}, J., {Bramich}, D.~M., {Figuera Jaimes}, R., {et~al.} 2015, \aap,
  573, A103

\bibitem[{{Soszy{\'n}ski} {et~al.}(2014){Soszy{\'n}ski}, {Udalski},
  {Szyma{\'n}ski}, {Pietrukowicz}, {Mr{\'o}z}, {Skowron}, {Koz{\l}owski},
  {Poleski}, {Skowron}, {Pietrzy{\'n}ski}, {Wyrzykowski}, {Ulaczyk}, \&
  {Kubiak}}]{sosz14}
{Soszy{\'n}ski}, I., {Udalski}, A., {Szyma{\'n}ski}, M.~K., {et~al.} 2014,
  \actaa, 64, 177

\bibitem[{{Stanek}(1996)}]{stanek96}
{Stanek}, K.~Z. 1996, \apjl, 460, L37

\bibitem[{{Sturch}(1966)}]{sturch66}
{Sturch}, C. 1966, \apj, 143, 774

\bibitem[{{Surot} {et~al.}(2019){Surot}, {Valenti}, {Hidalgo}, {Zoccali},
  {S{\"o}kmen}, {Rejkuba}, {Minniti}, {Gonzalez}, {Cassisi}, {Renzini}, \&
  {Weiss}}]{surot19}
{Surot}, F., {Valenti}, E., {Hidalgo}, S.~L., {et~al.} 2019, arXiv e-prints,
  arXiv:1902.01695

\bibitem[{{Valdes} {et~al.}(2014){Valdes}, {Gruendl}, \& {DES
  Project}}]{Valdes14}
{Valdes}, F., {Gruendl}, R., \& {DES Project}. 2014, in Astronomical Society of
  the Pacific Conference Series, Vol. 485, Astronomical Data Analysis Software
  and Systems XXIII, ed. N.~{Manset} \& P.~{Forshay}, 379

\bibitem[{{van Albada} \& {Baker}(1971)}]{vanalbada71}
{van Albada}, T.~S., \& {Baker}, N. 1971, \apj, 169, 311

\bibitem[{{Vivas} {et~al.}(2017){Vivas}, {Saha}, {Olsen}, {Blum}, {Olszewski},
  {Claver}, {Valdes}, {Axelrod}, {Kaleida}, {Kunder}, {Narayan}, {Matheson}, \&
  {Walker}}]{vivas17}
{Vivas}, A.~K., {Saha}, A., {Olsen}, K., {et~al.} 2017, \aj, 154, 85

\bibitem[{{Walker} \& {Terndrup}(1991)}]{walker91}
{Walker}, A.~R., \& {Terndrup}, D.~M. 1991, \apj, 378, 119

\bibitem[{{Zasowski} {et~al.}(2019){Zasowski}, {Schultheis}, {Hasselquist},
  {Cunha}, {Sobeck}, {Johnson}, {Rojas-Arriagada}, {Majewski}, {Andrews},
  {J{\"o}nsson}, {Beers}, {Chojnowski}, {Frinchaboy}, {Holtzman}, {Minniti},
  {Nidever}, \& {Nitschelm}}]{zas19}
{Zasowski}, G., {Schultheis}, M., {Hasselquist}, S., {et~al.} 2019, \apj, 870,
  138

\end{thebibliography}

\end{document}